\documentclass[prd,showpacs,nofootinbib,preprintnumbers,preprint]{revtex4}

\usepackage{amsmath}
\usepackage{amsfonts}
\usepackage{amssymb}
\usepackage{mathrsfs}
\usepackage{graphicx}
\usepackage{hyperref}
\usepackage[caption=false]{subfig}
\usepackage{blindtext, subfig}

\graphicspath{{TTS/}}

\begin{document}

\title{ 
Completing characterization of photon orbits in Kerr and Kerr-Newman metrics}
\author{D. V. Gal'tsov} \email{galtsov@phys.msu.ru}
\affiliation{Faculty of Physics,
Moscow State University, 119899, Moscow, Russia,\\Kazan Federal University, 420008 Kazan, Russia}
\author{K.V.~Kobialko} \email{kobyalkokv@yandex.ru}
\affiliation{Faculty of Physics, Moscow State University, 119899, Moscow, Russia}

\begin{abstract}  Recently, several new characteristics have been introduced to describe null geodesic structure of stationary spacetimes,  such as photon regions (PR) and transversely trapping surfaces (TTS). The former are three-dimensional domains confining the spherical photon orbits, while the latter are closed two-surface of spherical topology which fill other regions called TTR.   It is argued that in generic stationary axisymmetric spacetime it is natural to consider also the non-closed TTSs of the geometry of spherical cups, satisfying the same conditions (``partial'' TTS or PTTS), which fill the three-dimensional regions, PTTR. We then show that PR, TTR and PTTR together with the corresponding anti-trapping regions constitute the complete set of regions filling the entire three-space (where timelike surfaces are defined) of Kerr-like spacetimes. This construction provides a novel optical description of such spacetimes without recurring to explicit solution of the geodesic equations. Applying this analysis to Kerr-Newman metrics (including the overspinning ones) we reveal four different optical types for different sets of the rotation and charge parameters. To illustrate their  properties we extend Synge analysis of photon escape in the Schwarzschild metric to stationary spacetimes and construct density graphs describing escape of photons from all the above regions.

\end{abstract}

\maketitle


\setcounter{page}{2}

\setcounter{equation}{0}

\section{Introduction}
{\em Photon spheres} are important characteristics of spherically symmetric spacetimes. They are closely related to gravitational lensing \cite{Virbhadra:1999nm} and the BH shadows \cite {Synge}.   
In the case of the Schwarzschild metric studied in detail by Virbhadra and Ellis \cite{Virbhadra:1999nm}, closed circular photon orbits (often called photon rings) are located at $ r = 3M $ and form a  photon sphere  due to spherical symmetry. It is essential that the photon rings cover the photon sphere densely, i.e. the photons can have any direction tangent to  the sphere. The photon rings in Schwarzchild spacetime are unstable (so do the photon spheres); their Lyapunov exponents   define the   high frequency spectrum of the quasi-normal modes \cite{Cardoso:2008bp} (see also \cite{Khanna:2016yow,Konoplya:2017wot}). The photon sphere also has an interesting property to be a separatrice where the centrifugal force acting on an orbiting body changes sign, becoming inward directed inside it \cite{Hasse:2001by,Abramowicz:1990cg}. In spacetime, a photon sphere form a timelike hypersurface
marking  the closest distance of approach of a scattered light ray where
the  bending angle become infinitely large.  An existence of a
photon sphere thus may indicate on emergency of 
multiple images. (In what follows we use the same term PS for a two-dimensional sphere in space and the corresponding three-dimensional hypersurface in spacetime, assuming that the correct meaning is clear from the context.)

Obviously, existence of the photon sphere is related to spherical symmetry of spacetime. It is worth noting, that the photon sphere is not destroyed by the Newman-Unti-Tamburino (NUT) parameter, in which case the $so(3)$ algebra still holds  locally, though metric is already non-static. With this exception, stationary metrics with true rotation do not admits photon spheres or more general photon surfaces. In static spacetime 
various uniqueness theorems were formulated in which an assumption of the existence of a regular horizon was replaced by an assumption of  existence of a photon sphere \cite{Cederbaum,Yazadjiev:2015hda,Yazadjiev:2015mta,Yazadjiev:2015jza,Rogatko,Cederbaumo}. No such general results are available for stationary spacetimes. So the problem of optical characterization of stationary metrics which we discuss in this paper remains actual. Mention an interesting relation between the separability of spacetime and properties of the circular and the spherical photon orbits discovered recently. Namely, a spacetime is non-separable, if there exist an equatorial circular orbit and, at the same time, there are no spherical orbits beyond the equatorial plane \cite{Pappas:2018opz,Glampedakis:2018blj}. This property may serve a diagnostic of the non-Kerr nature of spacetime.

In certain cases the circular photon orbits may be stable. Normally this does not happen in the BH spacetimes outside the event horizon, but often happens inside the inner horizon or in the field of naked singularities and wormholes. In spherically symmetric static spacetimes, stable circular geodesics form "anti-photon" surfaces, a term suggested by Cvetic, Gibbons and Pope \cite{Cvetic:2016bxi}. In a spacetime with separable geodesic equations, the stable photon orbits correspond to a minimum of the effective radial potential. Above the minimum, the  bounded light orbits with two radial turning points are located. Their existence serves an indication that the solution may be unstable \cite{Keir:2013jga,Cardoso:2014sna,Dolan:2016bxj,Cunha:2017eoe}. Stable photon rings are often encountered near (hypothetical) horizonless ultra-compact objects, thus indicating on their instability \cite{Cunha:2017qtt}. They were shown to exist also on the horizons of extreme static black holes \cite{Khoo:2016xqv,Tang:2017enb}.
 
In non-spherical static spacetimes, properties of the photon spheres are shared by the {\em photon surfaces} of non-spherical form.
In \cite {Claudel:2000yi}, the photon surface is defined as a closed timelike hypersurface $ S $, such that any null geodesic initially tangent to $ S $ continues to be included in $ S $. Several examples of spacetime have been found that allow non-spherical  photon  surfaces, which are not necessarily asymptotically flat
(vacuum C-metric, Melvin's solution of Einstein-Maxwell theory and its generalizations including the dilaton field \cite {Gibbons}). 

Mathematically, an important property of the photon  surfaces is established by the theorem asserting that these  are  conformally invariant and totally umbilical hypersurfaces in spacetime \cite{Okumura,Senovilla:2011np}. This means that their second fundamental form is pure trace, i.e. is proportional to the induced metric. This property is especially useful in the cases when the geodesic equations are non-separable, so no analytic solution can be found, while the umbilical hypersurfaces still can be described analytically.  

Recently two other interesting characteristic surfaces in the strong gravitational field were suggested. One is  the {\em loosely trapped surface} (LTS) \cite{Shiromizu:2017ego}. As was shown long ago by Synge \cite{Synge}, the region between the horizon and the photon sphere in the Schwarzschild metric is approximately ``half'' trapped: if  photons are isotropically distributed in this region, more than half of them will be absorbed by the black hole. In a sense, these photons are ``loosely'' trapped. Once the horizon is approached, they become fully trapped. Considering two-spheres between the horizon and the photon sphere one can associate the radial derivative of the trace of their extrinsic curvature as a measure of the gravity strength: while this derivative is negative outside the photon sphere, it becomes positive and when moving towards the horizon. This gives rise to the definition of the loosely trapped surface in a non-vacuum spacetime as a compact 2-surface with positive trace of its extrinsic curvature and the non-negative radial derivative of the trace. Then for such a surface a Penrose-like inequality holds involving the area of the photon surface, provided the scalar curvature is positive \cite{Shiromizu:2017ego}.

In rotating spacetimes the photon orbits with constant Boyer-Lindquist radius
may exist as well (e.g. spherical orbits in Kerr  \cite{Wilkins:1972rs,Teo}), but they do not fill densely the photon spheres, since their existence requires certain relation between the constants of motion. Such orbits fill the three-dimensional volumes --- the {\em photon regions} \cite{Grenzebach,Grenzebach:2015oea}.
In more general spacetimes with two commuting Killing vectors, the orbits which fill some compact region  were called {\em fundamental photon orbits} \cite{Cunha:2017eoe}. In the Kerr case such orbits (outside the horizon) are unstable. But in more general case they may also be stable --- a signal of spacetime instability.

Another generalization of the notion of the photon surface is the {\em transversely trapped surface} (TTS)  \cite{Yoshino1}. Its relation to LTS is discussed in  \cite{Yoshino1}. TTS generalize PS allowing for the initially tangent photons to leave the closed two-surface but only in the inward direction. Similarly to PS,  the TTS can be defined in geometric terms using inequalities involving the second quadratic form of the corresponding hypersurfaces in spacetime. This can be also useful in the cases when the geodesic equations are non-separable, such as certain Weyl spacetimes. Here we argue that TTS may be reasonably generalized further to {\em partial} TTS, or PTTS, allowing for non-closed two-sections. In the Boyer-Lindquist coordinates such two-surfaces are spherical caps. The PTTS can be left by photons in both directions, but only at certain restricted angles. We hope that this concept may be useful for analysis of strong gravitational lensing. Finally, to complete the list of characteristic surfaces in a way to ensure foliation of most of the spacetime manifold,
we introduce the anti-TTS (ATTS) and anti-PTTS (APTTS) hypersurfaces such that their two-sections can be left by the initially tangential photons in the outward direction. Using the Kerr-Newman metric as an example, it will be shown that the set of surfaces presented above is sufficient to fully characterize the null geodesic structure of Kerr-like spacetimes.

The plan of the paper is as follows. In Section~2, we define the characteristic hypersurfaces in space-time based on the properties of the second fundamental form. In Section~3, the necessary and sufficient conditions for the surfaces PS, TTS and ATTS are obtained for stationary space-times parametrized by Boyer-Lindquist type coordinates. Section~4 is devoted to the applications of our formalism to the Kerr-Newman  spacetime.
The critical values of the rotation parameter and the charge separating the four optical types of KN are found. In Section~5, we study the output of photons from different characteristic regions to spatial infinity depending on the angle of Singe and present numerous density graphs illustrating the picture of the escape from different regions. In conclusion, the main results are summarized and some perspectives are discussed. 

\setcounter{equation}{0}

\section{The setup}

Recall the basic concepts of the geometry of hypersurfaces  \cite{Okumura}. Let $\hat{M}$ be a $4$-dimensional spacetime. Consider an embedded three-dimensional timelike hypersurface $M$ represented in the parametric form $F(M): M\rightarrow \hat{M}$ 
as 
\begin{equation}
x^{\mu}=f^{\mu}(\sigma^A), \qquad \mu=0,...,3, \qquad A=0,...,2,
\label{a1}
\end{equation}
where $\sigma^A$  are the local coordinates on $M$. We denote by $f^\mu_A=\partial f^\mu/\partial \sigma^A$
the  linearly independent tangent vectors at each point of $F(M)$.
The induced metric $g_{AB}$ and the unit spacelike normal vector field $n^{\mu}$ on $M$  are defined by
\begin{equation}
g_{AB}=\hat{g}_{\mu\nu}f^{\mu}_Af^{\nu}_B, \quad \hat{g}_{\mu\nu}f^{\mu}_An^{\nu} =0, \quad \hat{g}_{\mu\nu}n^{\mu}n^{\nu}=1.
\label{a2}
\end{equation}
The components of the second fundamental tensor $H_{AB}$  are obtained from the Gauss decomposition:
\begin{equation}
H_{AB}=\hat{g}_{\mu\nu}(\hat{\nabla}_{f_A}f^{\mu}_{B})n^{\nu},
\label{a3}
\end{equation}
where $\hat{\nabla}$ covariant derivative in $\hat{M}$.

Consider a null affinely parameterized geodesic $\hat{\gamma}$ with the tangent vector $\dot{\hat{\gamma}}^\mu(s)$ emitted tangentially to $M$ at some point  
 $F(P)$ on $F(M)$:  
\begin{equation}
\hat{\nabla}_{\dot{\hat{\gamma}}}\dot{\hat{\gamma}}^\mu(s)=0.
\label{a4} 
\end{equation} 
 Let us introduce another null curve $\gamma$ starting from the point $P$ with the tangent vector $\dot{\gamma}^A(s)$, which is assumed to be a null geodesic on the hypersurface $M$. Suppose that at an initial moment the tangent vectors to both geodesics coincide. Thus at the point $P$, we can choose  $\dot{\hat{\gamma}}^\mu(0)=\dot{\gamma}^\mu(0)\equiv f^{\mu}_A\dot{\gamma}^A(0)$. Rewriting the equation $\nabla_{\dot{\gamma}}\dot{\gamma}^A(s)=0$ for   $\gamma$ in terms of the four-dimensional quantities, we find
\begin{equation}
\hat{\nabla}_{\dot{\gamma}}\dot{\gamma}^\mu(s)=(H_{AB}\dot{\gamma}^A(s)\dot{\gamma}^B(s))n^{\mu}.
\label{a6}
\end{equation} 
There are the following basic possibilities.
The first is that the two trajectories  $\hat{\gamma}$ and  $\gamma$ locally coincide,   in which case $H_{AB}\dot{\gamma}^A(s)\dot{\gamma}^B(s)=0$. Suppose that for some  timelike hypersurface this condition is satisfied for every null tangent vector $\dot{\sigma}^A$ in $M$:  
\begin{equation}
H_{AB}\dot{\sigma}^A\dot{\sigma}^B=0.
\label{a7}
\end{equation} 
Then any initially tangent null geodesic remains in the hypersurface at any time. This is a well-known property of a {\em photon sphere} and its generalization --  a {\em photon surface} (PS) \cite{Claudel:2000yi}. Equivalently, the Eq. (\ref{a7}) can be rewritten as statement that the surface is totally umbilic  \cite{Claudel:2000yi,Okumura}:
\begin{equation}
H_{AB}=\frac13 Hg_{AB}.
\label{a8}
\end{equation}
The last statement is fairly restrictive both on the geometry of the hypersurface and the spacetime allowing for its existence.  Photon surfaces may exist in the gravitational field of black holes, wormholes, naked singularities and other ultracompact objects, being a useful tool allowing to discriminate their different optical behavior \cite{Virbhadra:2002ju}. It was proved that in the vacuum asymptotically flat case the Schwarzschild PS at $r=3M$ can not be deformed, though it is not forbidden in presence of matter \cite{Yoshino:2016kgi}.

However, PS do not exist in rotating spacetimes.  Though the so-called spherical orbits in Kerr metric exist with constant value of the Boyer-Lindquist radial coordinate $r$  \cite{Wilkins:1972rs,Teo,Paganini:2016pct}, they correspond to a discrete set of tangential directions on the sphere $r={\rm const}$. Spherical orbits with different $r$ then fill the three-dimensional domain -- the {\em photon region} (PR) \cite{Grenzebach} which is an important feature of rotating spacetimes. 

Both PS and PR may be  {\em unstable} (UPR) or {\em stable} (SPR) \cite{Grenzebach}, in the latter case signalling on instability of spacetime \cite{Keir:2013jga,Cardoso:2014sna,Dolan:2016bxj,Cunha:2017eoe}. The photon region can be considered purely geometrically in a manner similar to the photon surface. In fact, it is possible to use the condition (\ref{a7}), but demanding its validity only for some null vectors and, correspondingly, for some  directions in the tangent space \cite{Teo}. In the case of rotating space, from this we can derive exactly the photon region conditions. For more general spaces, the photon regions do  not exist, but some generalizations like fundamental photon orbits are possible \cite{Cunha:2017eoe}.
  
A completely different situation arises when the trajectories
$\hat{\gamma}$ and  $\gamma$ do not coincide. The curves $\hat{\gamma}$ propagate opposite to the outward normal  $n^\mu$ if and only if $H_{AB}\dot{\gamma}^A(s)\dot{\gamma}^B(s)>0$. 
If this condition is satisfied for each null tangent vector on some entire closed surface, then it is a trapping surface, allowing tangential photons to leave it only in the inward direction. This property underlies the notion of the {\em transversely trapping surface} (TTS) \cite{Yoshino1}. The necessary and sufficient condition for $ M $ to be a TTS is that it is timelike, and for each point on $ M $ the condition
\begin{equation}
H_{AB}\dot{\sigma}^A\dot{\sigma}^B\geq0,
\label{a9}
\end{equation} 
holds  for any null tangent vector $\dot{\sigma}^A$ on $M$. For an axially symmetric space, using $3 + 1$ splitting, this condition can be formulated purely  in  terms of the 2-dimensional external curvature and the covariant derivative of the lapse function \cite{Yoshino1}. (Note that in \cite{Yoshino1}  the sign  of the inequality  is opposite because of different conventions.) 

In the original definition of \cite{Yoshino1} this surface was assumed to be closed.  However, it turns out that already Kerr-Newman spacetime    can not be foliated by closed TTS only, since the solution of the Eq. (\ref{a9} exist with non-closed two-section. To achieve a more complete characterization of the null geodesic structure in a stationary gravitational field it is thus useful to extend the above definition of TTS to non-closed two-surfaces. Such incomplete TTS will be called {\em partial transversely trapping surfaces} (PTTS). They  have similar  confining property (\ref{a9}) locally, but they still allow for the initially tangent photons to escape from the inner part in some restricted range of directions passing across the boundary.  
However, as we will see later, in Kerr-Newman space the initially tangent to PTTS null geodesics cannot escape to  spatial infinity, so the trapping properties of  TTS and  PTTS are not much different globally. But to prove this, the separability of the geodesic equations is essential, and one can expect that in non-separable spacetimes the situation can be different. 

The third possibility is that the initially tangent trajectories $\hat{\gamma}$ propagate {\em onward} with respect  to the outer normal $n^\mu$.  The corresponding condition is 
\begin{equation}
H_{AB}\dot{\sigma}^A\dot{\sigma}^B\leq0.
\label{a10}
\end{equation} 
We will call such surfaces {\em  anti} -TTS (ATTS). In the Schwarzchild metric the ATTSs fill the region outside the photon surface up to infinity. But ATTSs may occur also inside the inner horizon describing the region which can be left by photons
into an analytically continued sector of spacetime. The antitrapping surfaces can also be non-closed, in which case we call them APTTS.
All the surfaces considered above are not isolated, but densely fill the three-dimensional volume regions which we will call the {\em transversely trapping region} (TTR), the {\em anti transversely trapping region} (ATTR) and, in the case of non-closed two-surfaces, the PTTR and the APTTR. Together with the photon regions PR (if it is defined), they all cover an entire or most of the space (where null geodesics and timelike hypersurfaces are defined) opening a way to give a complete optical characterization of many known exact solutions.

\begin{figure}[tb]
\centering
\subfloat[][]{
  \includegraphics[scale=0.4]{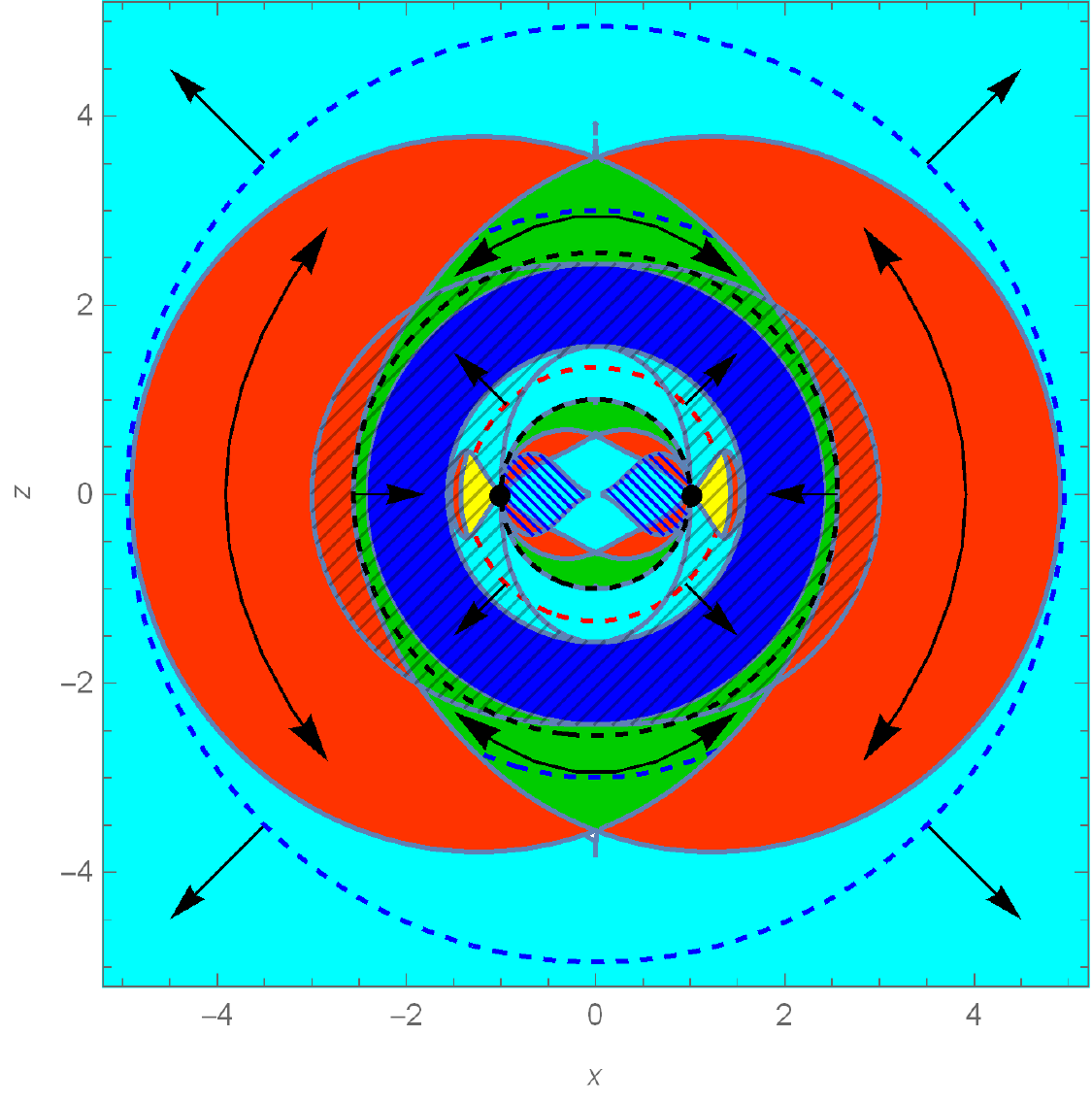} \label{KerrEXAa}
 }
 \quad
\subfloat[][]{
	\includegraphics[scale=0.41]{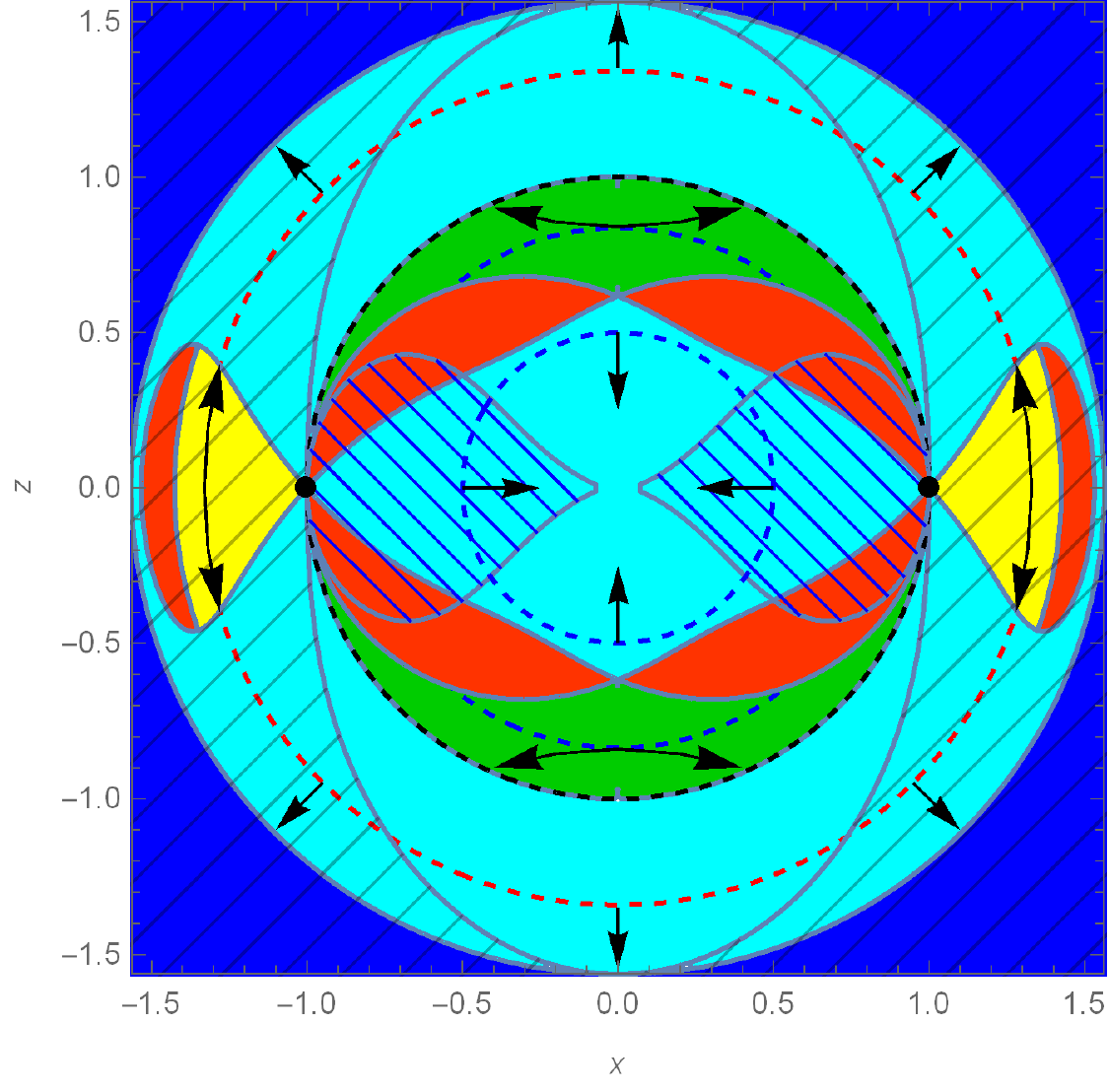} \label{KerrEXAb}	
 }
\caption{Characteristic regions in Kerr spacetime.  Red color -- the unstable photon regions, yellow -- the stable photon regions, dark blue -- the region between the horizons $\Delta_r\leq0$,  mesh -- the ergoregion, blue mesh -- the causality violating region $g_{\varphi\varphi} <0$, dash denote the throat  at $r = 0$, green - the (P)TTR, aqua -- the A(P)TTR. The dotted circles with arrows are examples of (P)TTS.  The arrows indicate directions of outgoing photons.}
\label{Kerr}
\end{figure}

In Fig.~(\ref{Kerr}) we illustrate this for the Kerr spacetime. Our picture can be compared with the picture of photon regions only in \cite{Grenzebach}. The plots represent slices $y=0$ in Cartesian type coordinates. The regions are filled with different colors; the arrows indicate directions of the outgoing photons. 

\setcounter{equation}{0}

\section{Stationary axially symmetric spacetimes}

To proceed further, we now specify the general conditions  (\ref{a7}-\ref{a10}) to the case of spacetimes with two commuting Killing vectors $\partial_t,\;\partial_\varphi$. We  take $\hat{M}$ to be a $4$-dimensional axially symmetric stationary spacetime endowed with the Boyer-Lindquist type coordinates $x^\mu=(t,r,\theta,\phi)$  defined as follows:
\begin{equation}
ds^2=-\alpha(dt-\omega d \phi)^2+\lambda dr^2 +\beta d\theta^2+\gamma d\phi^2,
\label{b1}
\end{equation}
where $\alpha,\, \omega,\, \lambda,\, \beta, \,\gamma$ are functions of $r,\,\theta$ only.  

Let's consider the  hypersurface $r=r_T={\rm const}$. As a consequence of the uniqueness theorems of \cite{Yoshino1}, the closed surfaces with  topology different from spherical do not exist.  This hypersurface is timelike, and its outer normal for $r>0$ 
\begin{equation}
n^\mu=(0,1/\sqrt{\lambda},0,0), 
\label{b2}
\end{equation}
is well defined outside the horizon. Note that in the analytically continued region $r<0$ the outer normal is $-n^\mu$, this leads to interchange of the definitions of (P)TTS and A(P)TTS.  

From the Eq. (\ref{a3}) we obtain: 
\begin{align}
& 2\sqrt{\lambda}H_{AB}\dot{\sigma}^A\dot{\sigma}^B=\partial_r\alpha(\dot{t}-\omega \dot{\phi})^2-\partial_r\beta\dot{\theta}^2-\partial_r\gamma\dot{\phi}^2  
-2\alpha\partial_r\omega(\dot{t}-\omega \dot{\phi})\dot{\phi}, \\
&g_{AB}\dot{\sigma}^A\dot{\sigma}^B=-\alpha(\dot{t}-\omega \dot{\phi})^2+\beta\dot{\theta}^2+\gamma\dot{\phi}^2=0.
\end{align}
Like in \cite{Yoshino1}, the component $\dot{\theta}$ can be excluded via the normalization condition:
\begin{align}
&\dot{\theta}^2=\frac{\alpha}{\beta}(\dot{t}-\omega \dot{\phi})^2-\frac{\gamma}{\beta}\dot{\phi}^2\geq0
\label{b3a}, \\
&2\sqrt{\lambda}H_{AB}\dot{\sigma}^A\dot{\sigma}^B= 
\alpha(\dot{t}-\omega \dot{\phi})^2 \left(\frac{\partial_r\alpha}{\alpha}-\frac{\partial_r\beta}{\beta}\right)+\dot{\phi}^2\left(\frac{\partial_r\beta}{\beta}-\frac{\partial_r\gamma}{\gamma}\right)-2\alpha\dot{\phi}(\dot{t}-\omega \dot{\phi})\partial_r\omega.
\label{b3b}
\end{align}
Then dividing the second expression by $ \alpha(\dot{t}-\omega \dot{\phi})^2$, we find from  (\ref{a9})  the following necessary and sufficient conditions for (P)TTS:
\begin{align}
&\beta>0, \quad \alpha>0, \quad \gamma>0, \quad 
\xi^2=\frac{\gamma\dot{\phi}^2}{\alpha(\dot{t}-\omega \dot{\phi})^2}\leq1,\\
&\xi^2\left(\frac{\partial_r\beta}{\beta}-\frac{\partial_r\gamma}{\gamma}\right)-2\sqrt{\frac{\alpha}{\gamma}}\partial_r\omega\xi+\left(\frac{\partial_r\alpha}{\alpha}-\frac{\partial_r\beta}{\beta}\right)\geq0.
\label{b3c}
\end{align}
or else for ergo region
\begin{align}
&\beta>0, \quad \alpha<0, \quad \gamma<0, \quad 
\xi^2=\frac{\gamma\dot{\phi}^2}{\alpha(\dot{t}-\omega \dot{\phi})^2}\geq1,\\
&\xi^2\left(\frac{\partial_r\beta}{\beta}-\frac{\partial_r\gamma}{\gamma}\right)-2\sqrt{\frac{\alpha}{\gamma}}\partial_r\omega\xi+\left(\frac{\partial_r\alpha}{\alpha}-\frac{\partial_r\beta}{\beta}\right)\leq0.
\label{b3d}
\end{align}

In a similar way one obtains the necessary and sufficient conditions for A(P)TTS from (\ref{a10}). Thus, the problem has been reduced to analyzing the negative/positive definitness  of the quadratic functions (\ref{b3c}, \ref{b3d}) on a given interval $\xi^2\leq1$ or $\xi^2\geq1$. The analysis is not complicated, but rather cumbersome. It is possible to obtain explicit restrictions on the coefficients of quadratic forms (see appendix A). Some more details of the computation can be found in \cite{Yoshino1} (the Eqs. 38a-c, appendix B). The only difference is that here we do not use the absolute values in the final  inequalities but add  the appropriate linear conditions. 

\begin{table}[h]
\caption{(P)TTS, A(P)TTS and PR conditions}
\begin{center}
\begin{tabular}{|c|c|c|c|}
\hline
Type & Region & Quadratic conditions & Linear conditions \\
\hline
TTS &  Outer($\alpha>0$, $\gamma>0$)  & ($i$) & $b>c$ \\
 &   & ($ii$) & $b>c$, $a>c$ \\
 &   & ($iii$) & $b\leq c$, $a>c$ \\
 &  Ergo($\alpha<0$, $\gamma<0$)  & ($i$) & $b<c$ \\
 &   & ($ii^*$) & $b<c$, $a<c$ \\
\hline
ATTS &  Outer($\alpha>0$, $\gamma>0$)  & ($i$) & $b<c$ \\
 &   & ($ii$) & $b<c$, $a<c$ \\
 &   & ($iii$) & $b\geq c$, $a<c$ \\
 &  Ergo($\alpha<0$, $\gamma<0$)  & ($i$) & $b>c$ \\
 &   & ($ii^*$) & $b >  c$, $a>c$ \\
\hline
 PR & - & $(-1)\times(iii)$ & - \\
\hline
\end{tabular}
\end{center}
\label{Con}
\end{table}

The table \ref{Con} represents the detailed form of conditions on (P)TTS and A(P)TTS
 (\ref{a9}, \ref{a10}) in the outer and the ergo regions of spacetime,  where
 \begin{align}
&a\equiv\partial_r\alpha/\alpha, \quad &b\equiv\partial_r\beta/\beta, \quad &c\equiv\partial_r\gamma/\gamma, \quad &d\equiv\sqrt{\alpha/\gamma}\partial_r\omega,
\label{b4}
\end{align}
and the quadratic inequalities are defined as
\begin{align}
&(i)   \quad (a-b)(b-c)\geq d^2, \\
&(ii)  \quad (a-c)^2\geq4 d^2>4(b-c)^2, \\ 
&(ii^*)  \quad (a-c)^2\geq4 d^2<4(b-c)^2,\\
&(iii)   \quad (a-c)^2\geq4 d^2. 
\label{b3}
\end{align}
 
It is worth noting that the conditions for PS (and PR)  for Plebanski-Demianski class of solutions \cite{Grenzebach} are based on the same quadratic condition  (iii) as (P)TTS (with opposite sign):
\begin{align}
(a-c)^2\leq4 d^2.
\label{b5}
\end{align}
For (P)TTS, however, one has to take into account the linear conditions together with  (iii). 

This is not a coincidence, but can be obtained directly from the Eq. (\ref{a7}). For this, as already noticed, it is necessary to require the fulfillment of
the Eq. (\ref{a7}) only for some isotropic vectors. The easiest way to do this is to express the null vector components in terms of the energy and the azimuthal orbital momentum and substitute them into th Eq.(\ref{a7}). In the case of Plebanski-Demianski space one then finds a $\theta$-independent relation between the energy and the angular momentum. Let's do this for the Kerr solution (see below (\ref{e1}), (\ref{f1a}-\ref{f1b})). From (\ref{a7}) and (\ref{b3b}) we get equality 
\begin{align}
Er^2(r-3m)+a^2E(r+m)+aL(r-m)=0,
\end{align}
accurate to nonzero multipliers. Obviously, the relationship between $E$ and $L$ does not depend on the $\theta$. Solving this equation for $L$ and substituting in (\ref{b3a}) we find exactly the condition  (\ref{b5}).   

Note that from the Table \ref{Con} it is clear that the quadratic inequalities distinguish between A(P)TTS and (P)TTS regions. Moreover, every region of Kerr-like space (where timelike surfaces $r={\rm const}$ are defined) to be related to one or another optical type.
In the initial formulation, the TTS and the PTTS have different optical properties  in the sense that light can leave a compact region after touching a PTTS across its boundary. However, as we will see in the case of spaces admitting separation of variables in the geodesic equations, this difference is diminished by the existence of a radial potential. Still, absence of the  closed TTSs, suggests that the strong field region is more accessible for observation. With this in mind, we present in the next section  new classification of   optical types of Kerr-Newman solution depending on parameters.
 
Note that in non-separable spacetimes, the photon regions most probably do  not exist, while the   (P)TTSs and A(P)TTSs do exist. Instead of the photon region one may encounter  a region where the the projection (\ref{a7}) of the external curvature does not have a definite sign.

\setcounter{equation}{0}

\section{Kerr-Newman optical types}
 
In this section, we illustrate how it is possible to classify optical types of the Kerr-Newman space based on our complete set of characteristic surfaces.   This classification can be considered as generalization
to stationary spaces of the scheme of \cite{Virbhadra:2002ju} applicable only in the static case.

In what follows we classify both the black holes and the naked singularity configurations like overextreme Kerr-Newman space. We make the following definitions. 
We call the type $I$ such spaces in which the closed TTSs exist in the outer domain. The second type $II$ includes solutions with the presence of only PTTSs and photon regions.  The third type $III$ includes metrics in which there are only photon regions but no PTTSs. Finally, in spaces $IV$, none of the above structures is present. The naked singularity of the type $IV$ is called strongly naked \cite{Virbhadra:2002ju}. These definitions are summarized in the (table \ref{Cla}). The optical meaning of this classification will be explained in   section~5.
 
 \begin{table}[tb]
\caption{Optical types}
\begin{center}
\begin{tabular}{|c|c|c|c|c|}
\hline
Type & TTR & PTTR & (PS)PR \\
\hline
$I$ &  +  & $\pm$ & + \\
 $II$ & $-$  & + & + \\
$III$ &  $-$ & $-$ & +  \\
$IV$  &  $-$ & $-$ & $-$\\
\hline
\end{tabular}
\end{center}
\label{Cla}
\end{table}
Now let's analyze a concrete example of a Kerr-Newman solution using Boyer-Lindquist coordinates.
The metric has the form  (\ref{b1}) with the following metric functions (see \cite{Griffiths1,Griffiths}):
\begin{align}
&\alpha=  \frac{\Delta_r -a^2\sin^2\theta}{\Sigma}, \quad \omega=-\frac{a(2m r-q^2)\sin^2\theta}{\Delta_r -a^2\sin^2\theta},\\ 
&\lambda=\frac{\Sigma}{\Delta_r}, \quad \beta=\Sigma, \quad\gamma=\frac{\Delta_r\Sigma\sin^2\theta}{\Delta_r -a^2\sin^2\theta},\\
&\Delta_r=r^2-2mr +q^2+a^2,  \quad \Sigma=r^2+a^2\cos^2\theta.
\label{e1}
\end{align}
Here $a$ is the rotation parameter, $q^2=e^2+g^2$  comprises the electric and magnetic charges. Basically, the coordinates $t$ and $r$ may range over the whole $\mathbb R$, while $\theta$ and $\phi$ are the
standard coordinates on the unit two-sphere. The horizons ($\Delta_r=0$), the ergosphere ($\alpha=0$) and the ring singularity in the equatorial plane $\theta=\pi/2$ are located at
\begin{equation}
r_{h\pm}=m\pm\sqrt{m^2-q^2-a^2},\quad r_{e\pm}=m\pm\sqrt{m^2-q^2-a^2\cos^2\theta},  \quad r_s=0.
\end{equation} 
The horizon disappears if $|a|>a_e=\sqrt{m^2-q^2}$. Since the normal vector $n^{\mu}$ to the hypersurface $r=r_T$ is spacelike only for $\Delta_r>0$ we are interested only in this area.  

First of all, we find that the condition ($i$) has  a simple form
\begin{equation}
-\frac{4a^2r^2\sin^2\theta}{\Delta_r\Sigma^2}\geq0,
\label{e2}
\end{equation}  
and obviously can not hold  if $\Delta_r>0$. Although the calculation is more tedious, one can show that the second conditions in ($ii$) and ($ii^*$) is satisfied automatically in outer and ergo regions respectively (see \cite{Yoshino1}) if other ones holds. The condition ($iii$) and the first one in ($ii$) and ($ii^*$) reads:
\begin{equation}
16 r^2a^2\Delta_r\sin^2\theta\leq(4r\Delta_r-\Sigma\Delta_r')^2.
\label{e3}
\end{equation} 
Recall that for the photon region, this condition has a reversed sign (\ref{b5}). A detailed analysis of the photon regions as well as their stability in the subextreme and overextreme Plebanski-Demianski spaces can be found in  \cite{Grenzebach,Charbulak:2018wzb}. The latter also provides classification of solutions based on the properties of the spherical photon orbits. Previous analysis of the trapping sets in the Kerr metric can be also found in \cite{Paganini:2016pct}. 

In the case of closed TTSs or ATTSs, the Eq. (\ref{e3}) should be satisfied for all $\theta$ what leads to the condition 
\begin{equation}
\left(r^2-3mr+2q^2\right)^2\geq4a^2(m r-q^2).
\label{e4}
\end{equation} 
In order to distinguish between TTS and ATTS, we must  add the linear conditions (table \ref{Con}). For example, the linear (P)TTS conditions ($a>c$) reduce  to the strongest one for $\theta=0$: 
\begin{equation}
r(r^2-3mr+2q^2)+a^2(r+m)<0,
\label{e6}
\end{equation} 
which determines, in particular, the existence of the polar PTTS. In our case, it simply selects one of the regions defined by the inequality (\ref{e4}). In the case of a sub-extremal space, an expression describing the boundary of a closed TTR can be obtained using the Cardano formulas: 
\begin{align}
&2r^{max}_{T}=3m+\sqrt{F}-\sqrt{v-F-\frac{8a^2m}{\sqrt{F}}}, \quad 3F=v+\frac{v^2-24a^2 u}{Q^{1/3}}+Q^{1/3},\nonumber \\
&Q=216a^4m^2+v^3-36a^2v u +24\sqrt{3}a^2\sqrt{(m^2-a^2-q^2)(v^2q^2-27a^2m^4)}, \\
&v=9m^2-8q^2, \quad u=3m^2-2q^2. \nonumber
\end{align} 
In the particular case of a vanishing charge $q=0$, this was found in  \cite{Yoshino1} and \cite{Paganini:2016pct} as the lower boundary of the trapping set. Solutions that allow the existence of a closed TTS in our classification will be called an optical type $I$. Such solutions contain a compact region of a strong gravitational field, completely covered with transversely trapping surfaces and, accordingly, strongly hidden for an external observer. At the same time, the closed photon region, which is also present in this case, can create a set of relativistic images \cite{Virbhadra:1999nm}. 

\begin{figure}[tb]
\centering
\subfloat[][$q=0$]{
  	\includegraphics[scale=0.3]{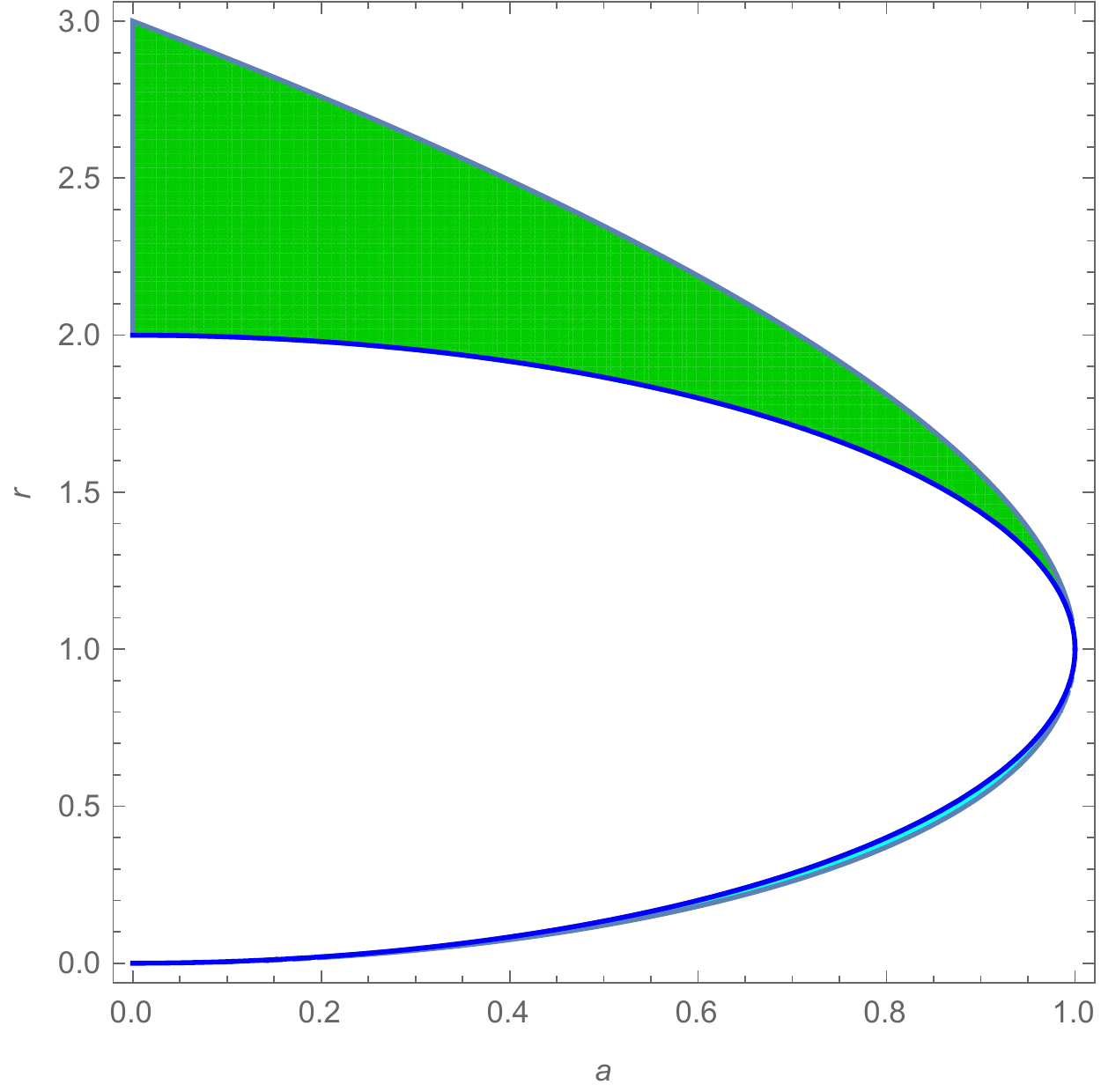} 
  	\label{EXT1}	
 }
 \quad
\subfloat[][$q<q_C$]{
\includegraphics[scale=0.3]{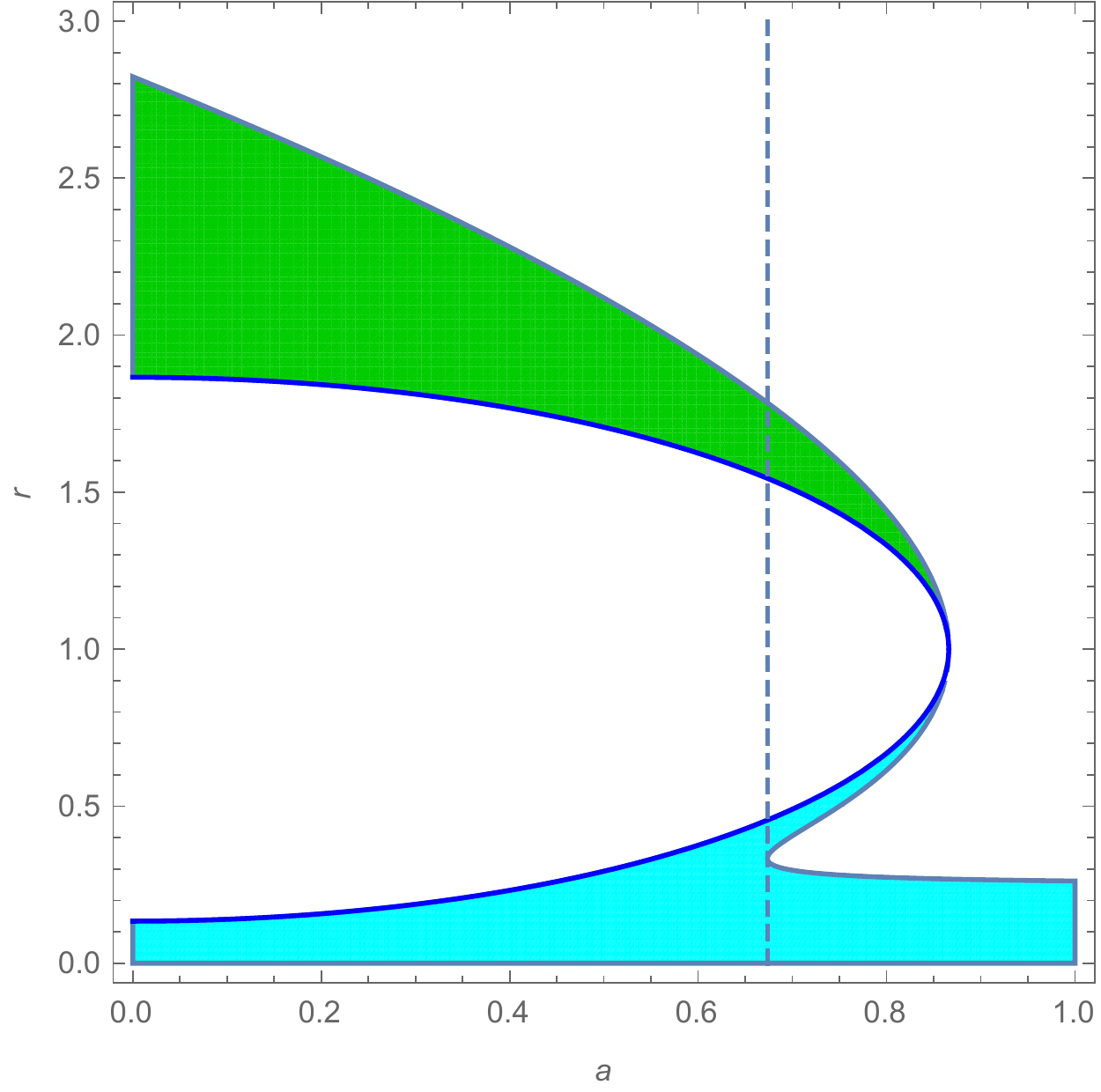}
	\label{EXT2}
 }
 \quad
 \subfloat[][$q>q_C$]{
\includegraphics[scale=0.3]{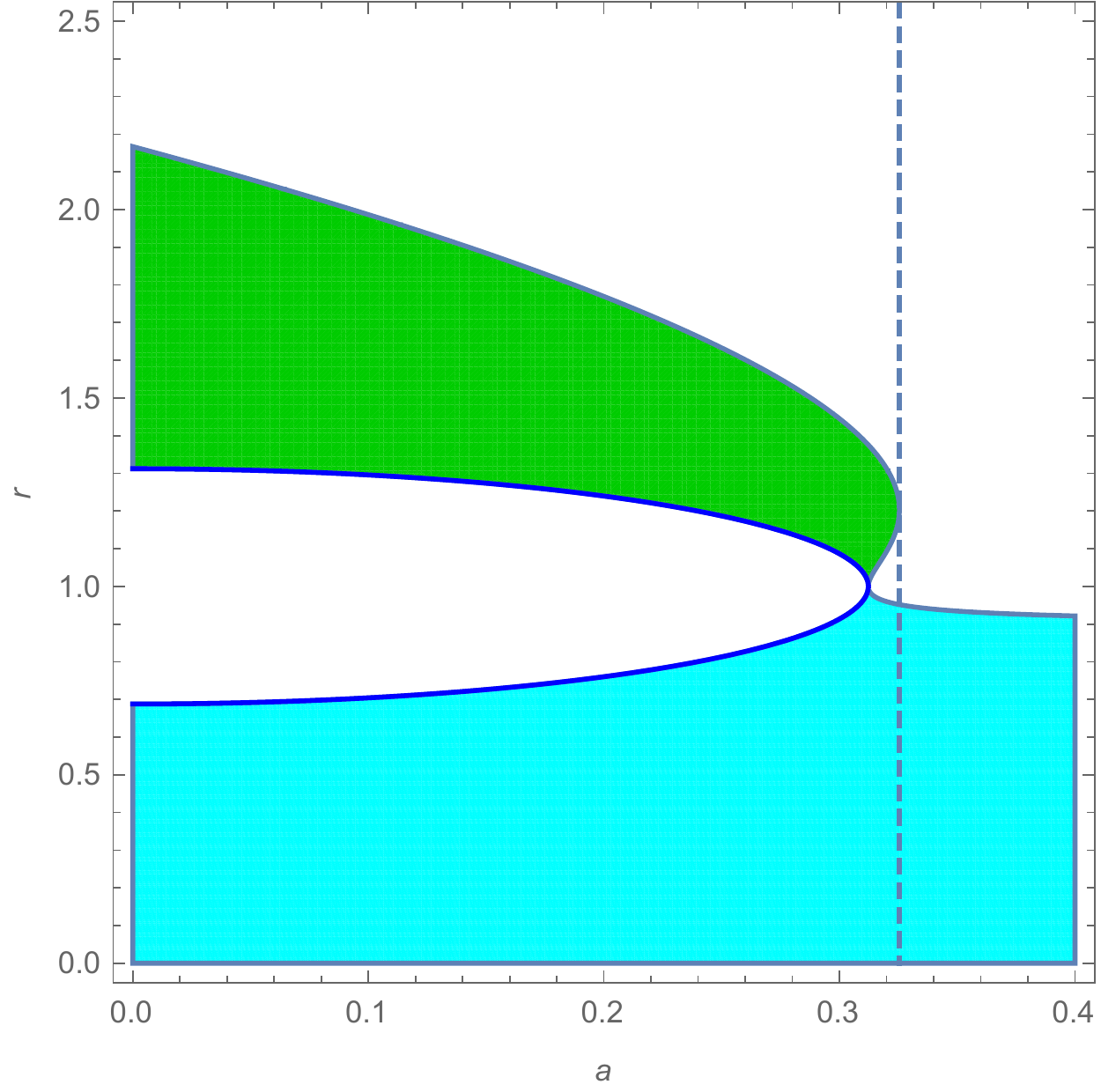}
	\label{EXT3}
 }
\caption{Closed TTR and ATTR for various $a$. Green color -- TTR, aqua -- ATTR, blue -- the horizon. Dotted lines -- $a_C$}
\label{Closed }
\end{figure}

Inequalities (\ref{e4}, \ref{e6}) can have a common solution only for certain values of the parameters ($a, q$). Accordingly, the closed TTS domain exists  exists only for the following values of $a$ and $q$:
\begin{align}
&|a|<a_{C}=\sqrt{3}q\left(1-\frac{8q^2}{9m^2}\right),  \quad a_C>a_e,   \quad q_m>q>q_C, \label{e5a}\\
&|a|\leq a_{e},   \quad  q\leq q_C,   \quad q^2_C=3/4m^2, \quad q^2_m=9/8m^2.
\label{e5b}
\end{align} 
In this expression, an important quantity is the critical charge $q_C$. For values $q\leq q_C$, the solution resembles pure Kerr. In particular,  closed TTSs do not exist in the overextreme case $|a|>a_e$ \cite{Yoshino1}. Respectively a region containing the singularity and a stable photon region becomes observable along the majority of null geodesics coming from infinity.  In the opposite case,   $q>q_C$, closed TTSs exist  even in the superextreme case,  limited by the critical value of the rotation parameter $|a|<a_C$, and, accordingly the area is hidden for most isotropic geodesics.  In the plots (\ref{EXT1}-\ref{EXT3})  we illustrate the size of the TTR depending on the magnitude of the rotation.   In the last image, there is a small TTR  bump in superextreme case, corresponding to what was said.

The structure of a closed ATTR can determine the presence of a stable photon region \cite{Grenzebach}. It exists if
\begin{align}
&|a|>a_{C}=\sqrt{3}q\left(1-\frac{8q^2}{9m^2}\right), \quad a_C<a_e,   \quad q<q_C, \label{e51b} \\
&|a|>a_{e},   \quad   q_m>q>q_C,  \quad  q^2_C=3/4m^2,  \quad q^2_m=9/8m^2.
\label{e51a}
\end{align}
In particular, a stable photon region inside the inner horizon exists only at sufficiently small values of the charge $q<q_C$. 

If the Eq. (\ref{e6}) is fulfilled for some values of the metric parameters, while the conditions of existence of  closed TTS (\ref{e5a}, \ref{e5b}) are violated, one can still have the pure polar PTTS defined by the inequalities (\ref{e3}, \ref{e6}). Such configurations belong to an optical type $II$ in our classification. In this case, the inner region of space containing the singularity is still hidden by the closed photon region, but not by the TTS, and, accordingly, is observable for a certain set of isotropic geodesics coming from infinity. Such configurations are still capable to create a set of relativistic images \cite{Virbhadra:1999nm,Virbhadra:2002ju}.

However, with certain values of the metric parameters, even the polar PTTS disappear. We present a constraint on the charge and the rotation parameter for type $II$ solution explicitly:
\begin{align}
&a^2<a^2_c=6m\sqrt{5m^2+2q^2}\sin(\vartheta+\pi/6)-2(3m^2+q^2), \quad q^2<q^2_m, \nonumber \\
&\vartheta=\frac{1}{3}\arccos\left(\frac{22m^4+14m^2q^2+q^4}{2m(5m^2+q^2)^{3/2}}\right), \quad q^2_m=9/8m^2.
\label{e7}
\end{align} 
We note that the critical value of the rotation parameter also determines the condition for the existence of polar spherical photon orbits  \cite{Charbulak:2018wzb}, which allows us to relate various classifications of   the overextreme spaces. In the particular case $q=0$, we get the previously known value:
\begin{equation}
a_c=m\sqrt{6\sqrt{3}-9}.
\end{equation}

The spaces with $|a|>a_c$ contain only a photon region which become non-closed at the poles, and goes into APTTS rather than PTTS to which PR was connected for smaller $|a|$.  Accordingly, they belong to an optical type $III$. In these spaces, there is no well-defined compact trapping region. The singularity   is strongly naked for geodesic passing near the poles. However, it is still hidden for some geodesics passing in the vicinity of the equatorial plane from infinity. 

The value $q_m$ is maximal in the sense that for larger charges $q>q_m$ both the photon regions and the PTTSs completely disappear. So $q_m$ defines the strong naked singularity condition. Such solutions have radically different optical properties, in particular, they cannot create relativistic images \cite{Virbhadra:2002ju}. These represent the fourth class of solutions, $IV$.   Interestingly, this condition accurately duplicates a similar condition for the Reissner-Nordstrom space \cite{Claudel:2000yi}. 

\begin{figure}[tb]
\centering
\subfloat[][$a=0$, $q=0$]{
  	\includegraphics[scale=0.32]{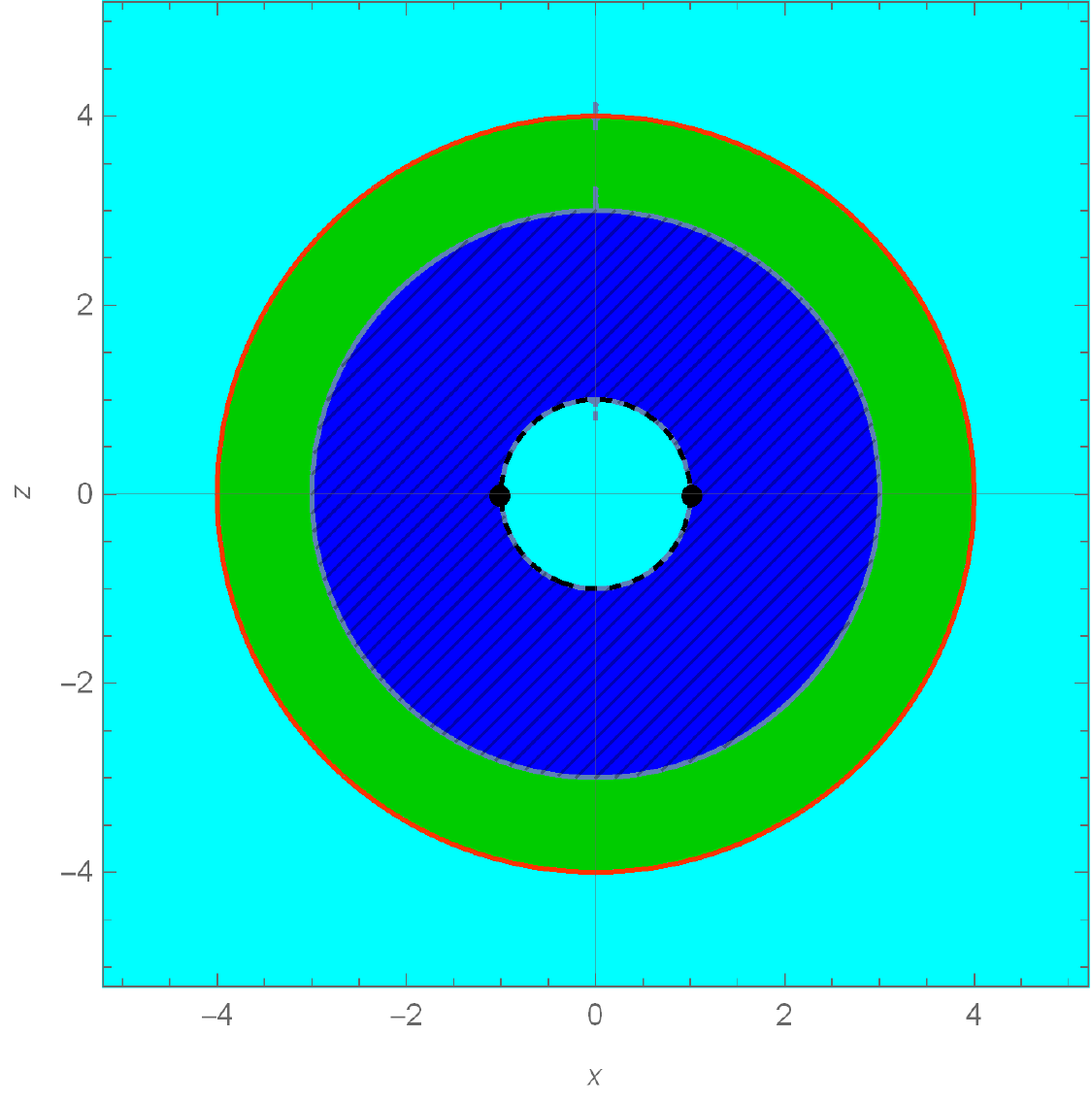}
		\label{Sha}
 }
 \quad
\subfloat[][$a=0$, $q<q_C$]{
	\includegraphics[scale=0.32]{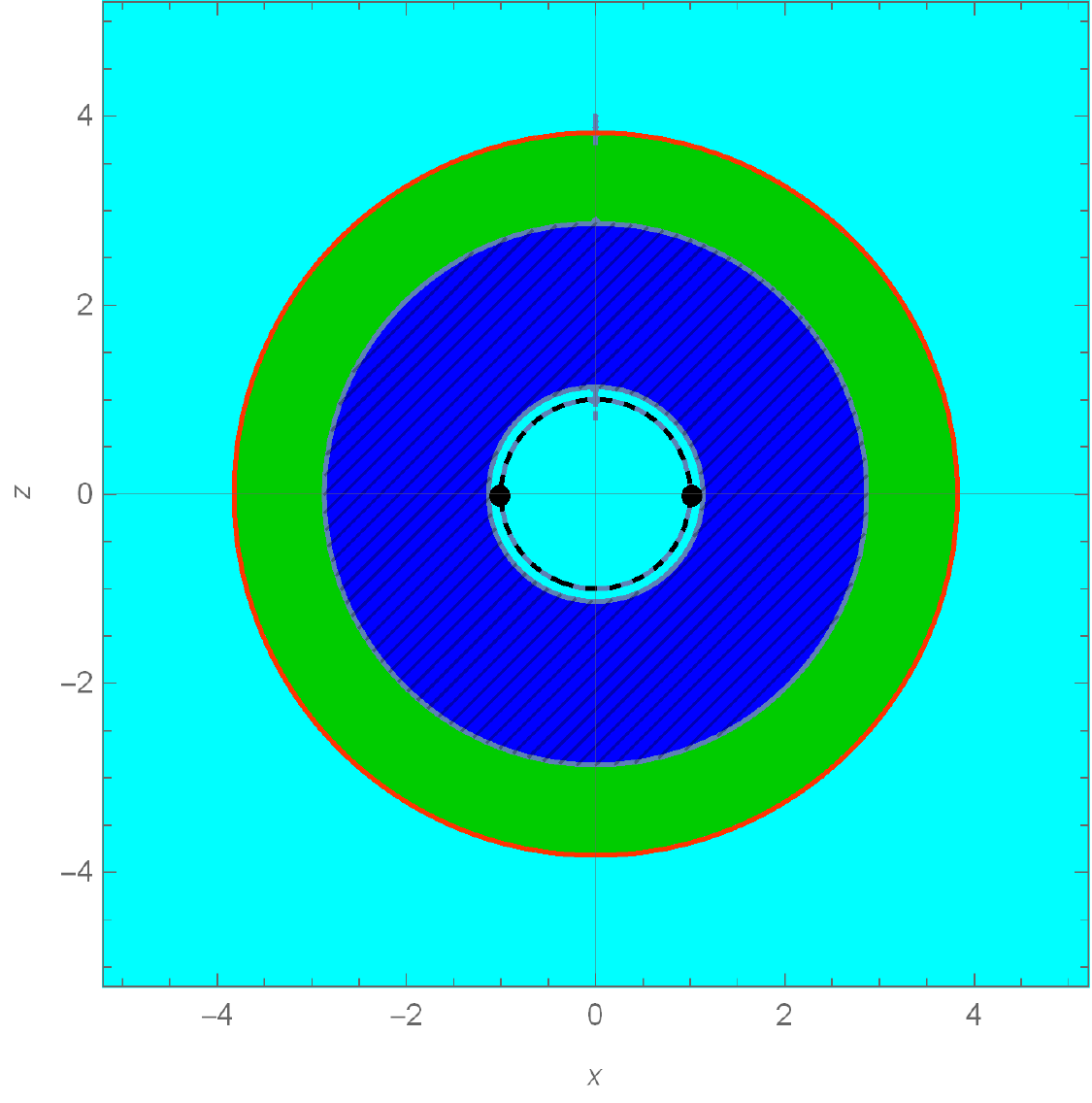}
		\label{ShN11a}
 }
 \quad
 \subfloat[][$a=0$, $q>q_C$]{
	\includegraphics[scale=0.32]{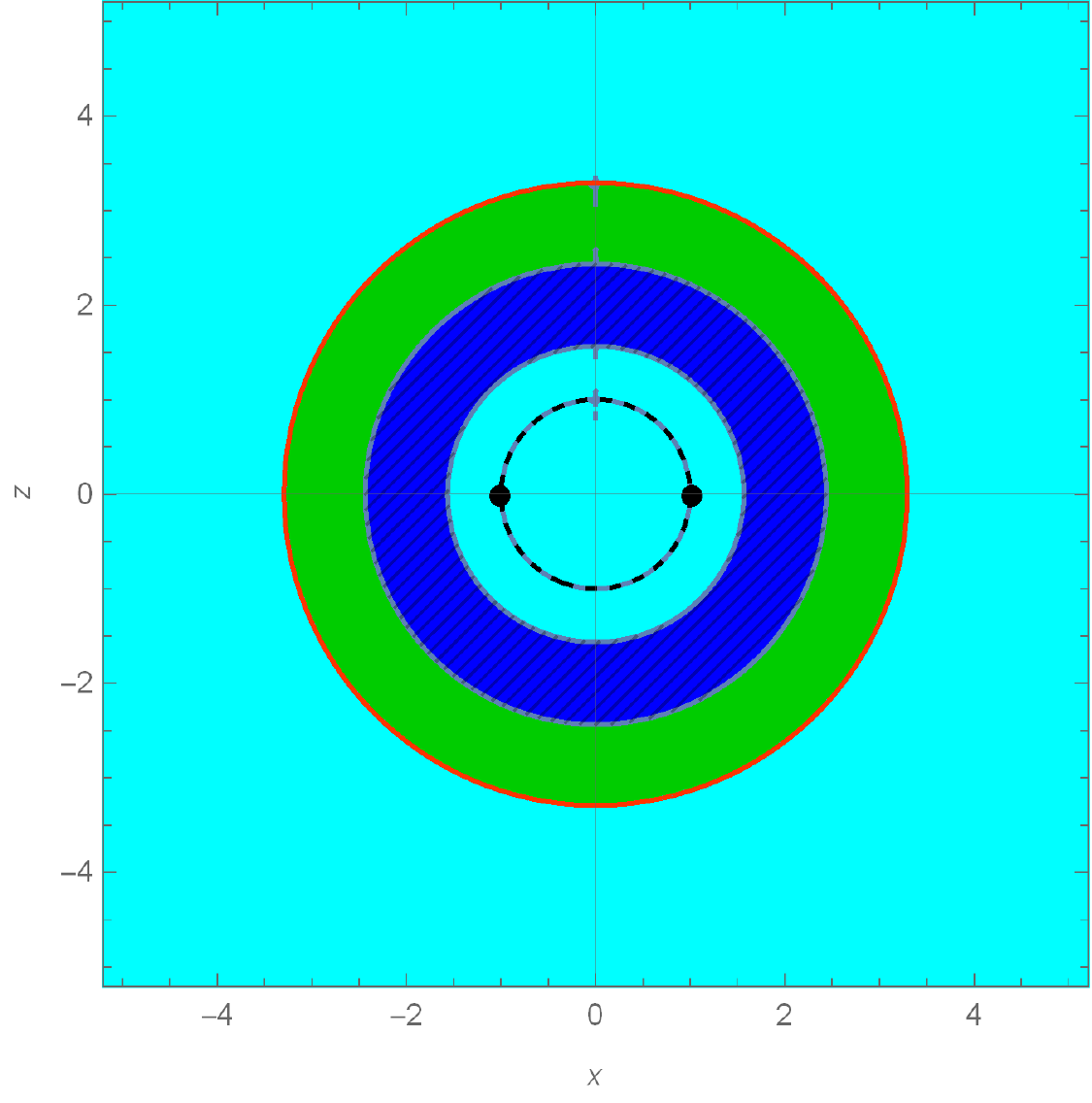}
		\label{SHN21a}
 }
 \\
 \subfloat[][$0<a<a_e$, $q=0$]{
  		\includegraphics[scale=0.32]{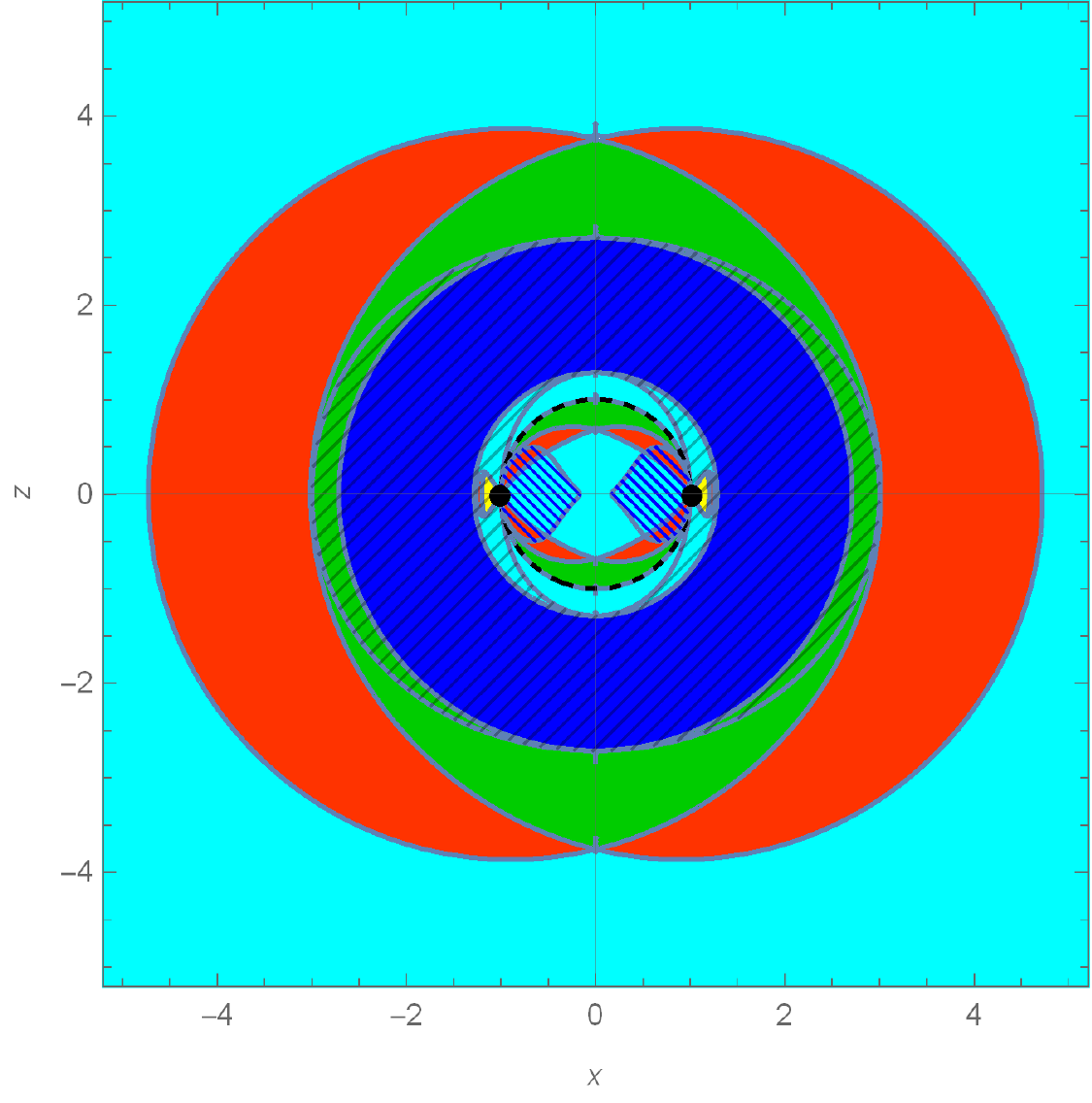}
		\label{Kerr2a}
 }
 \quad
\subfloat[][$0<a<a_C$, $q<q_C$]{
	\includegraphics[scale=0.32]{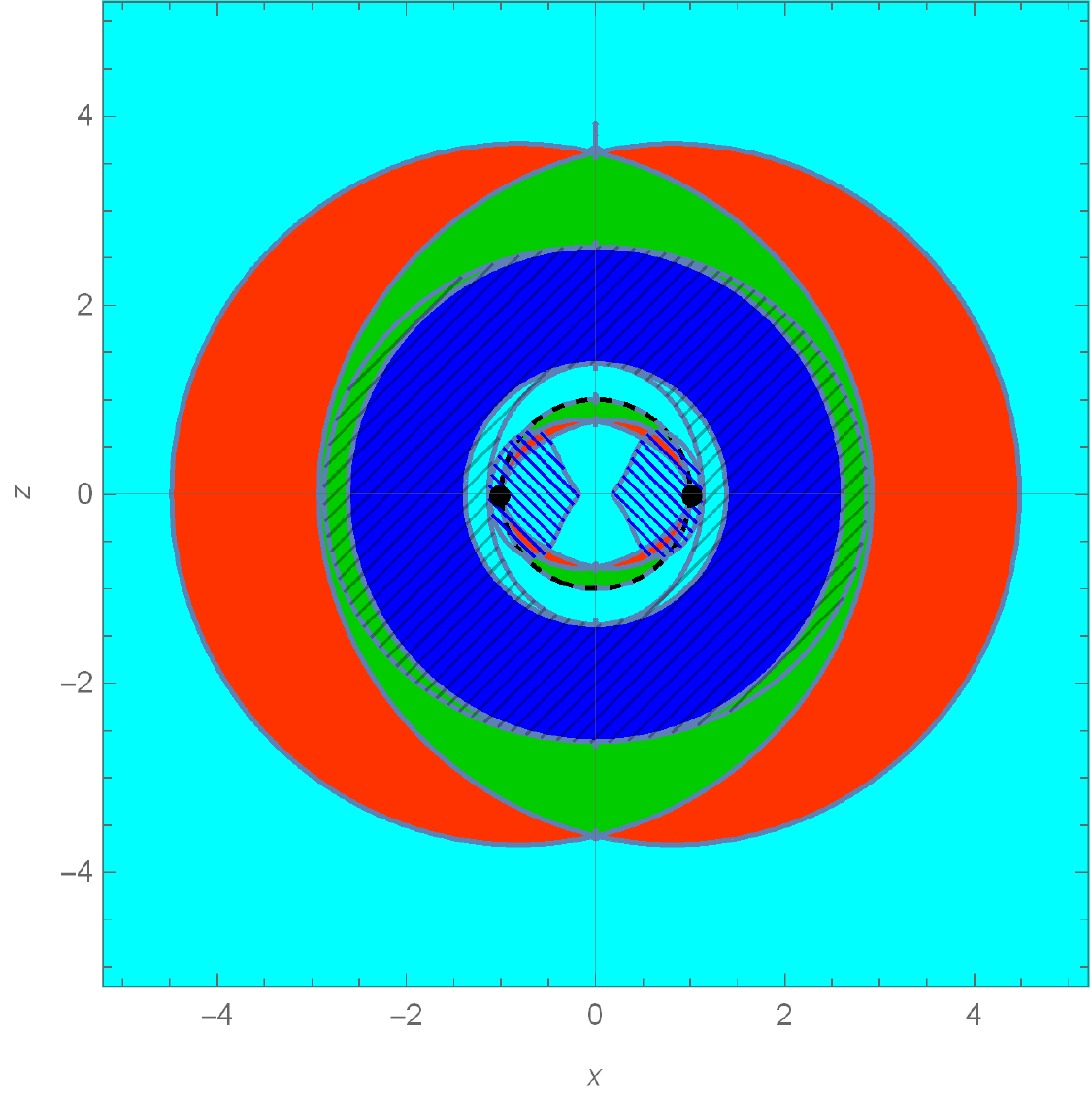}
		\label{KerrN12a}
 }
 \quad
 \subfloat[][$0<a<a_e$, $q>q_C$]{
		\includegraphics[scale=0.32]{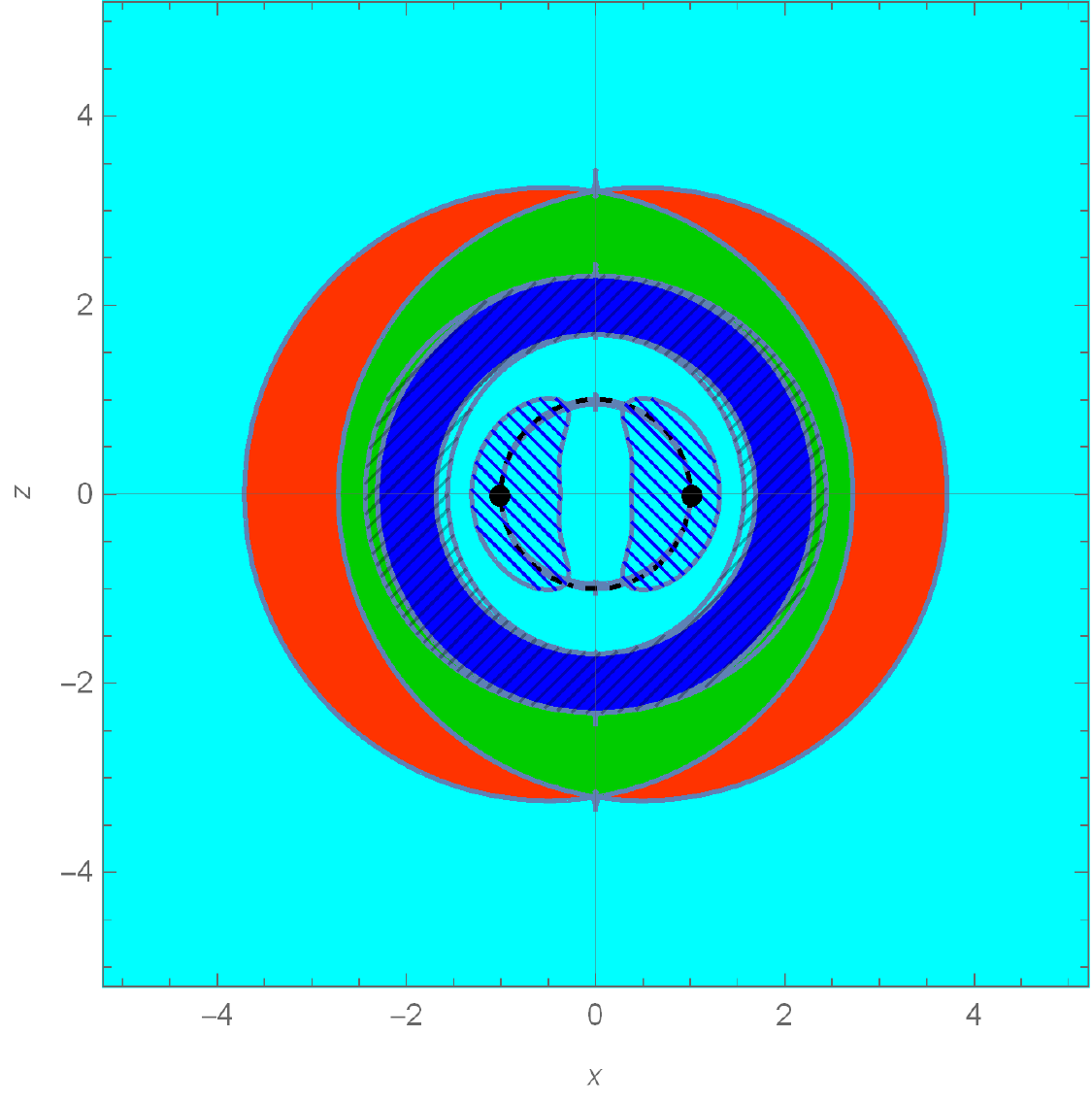}
		\label{KerrN22a}
 }
 \\
 \subfloat[][$0<a<a_e$, $q=0$]{
  		\includegraphics[scale=0.32]{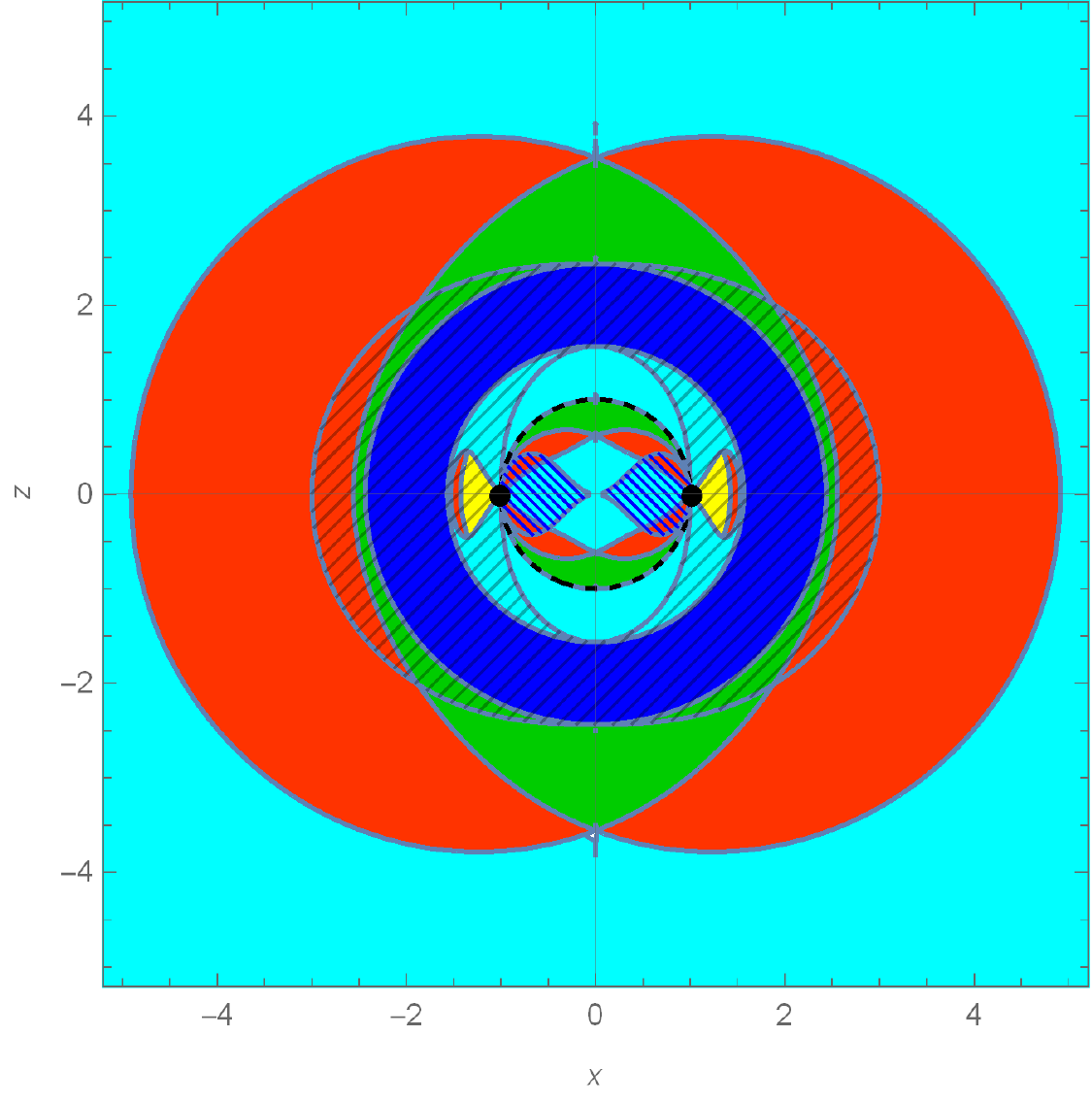}
		\label{Kerr3a}	
 }
 \quad
\subfloat[][$a_C<a<a_{e}$, $q<q_C$]{
		\includegraphics[scale=0.32]{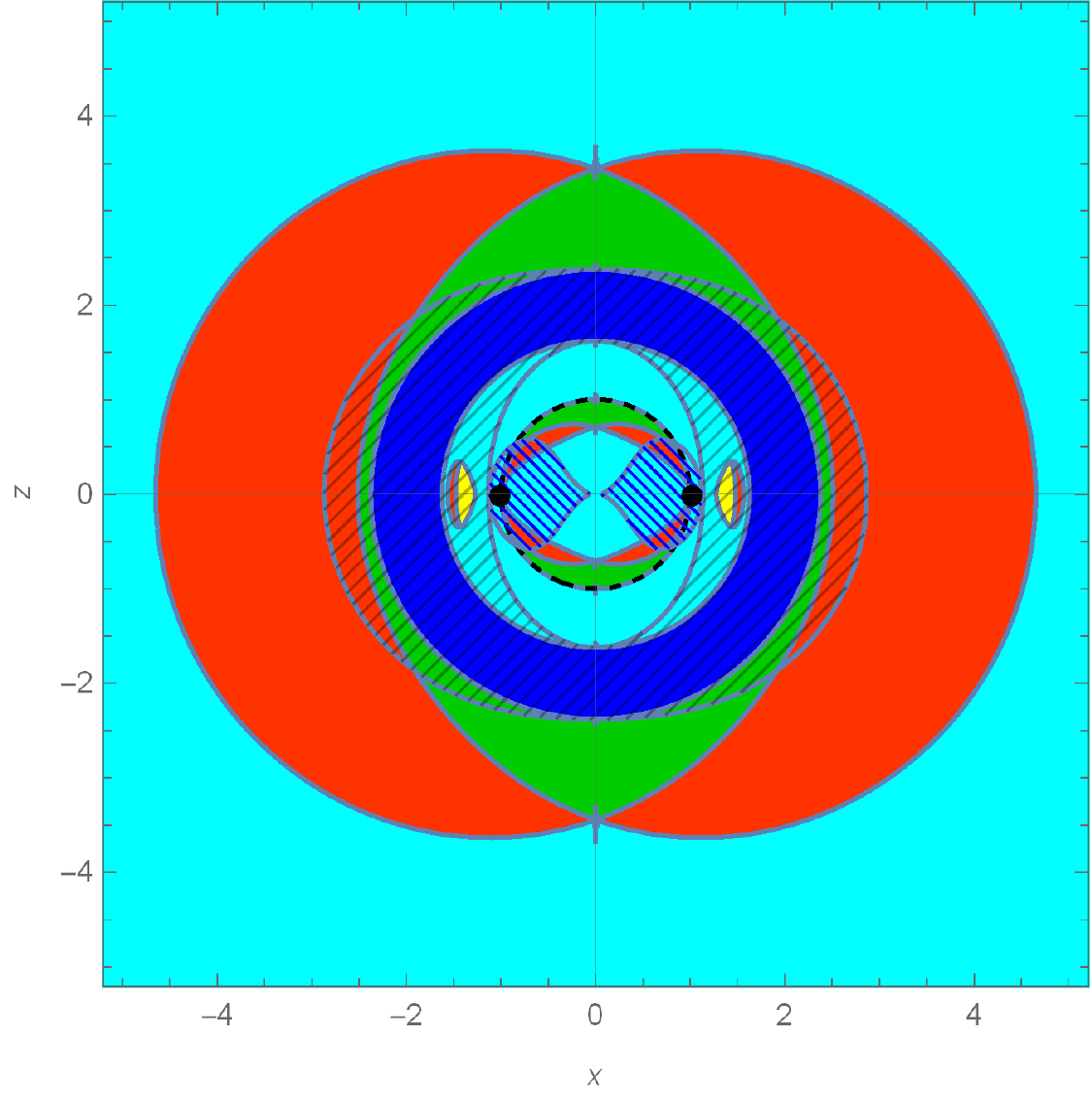}
		\label{KerrN13a}
 }
 \quad
 \subfloat[][$0<a<a_e$, $q>q_C$]{
		\includegraphics[scale=0.32]{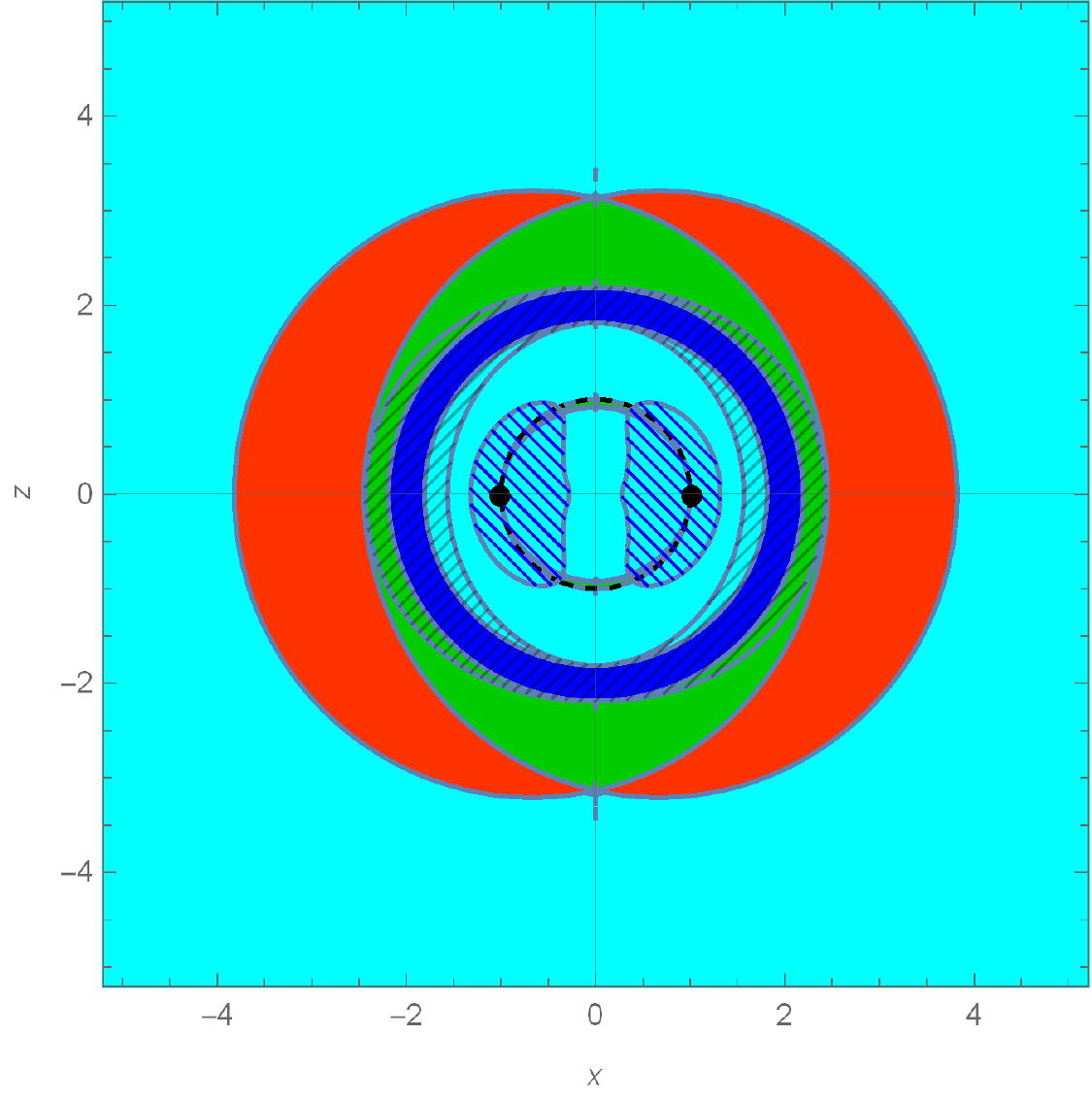}
		\label{KerrN23a}
 }
  \\
 \subfloat[][$a=a_{e}$, $q=0$]{
  		\includegraphics[scale=0.32]{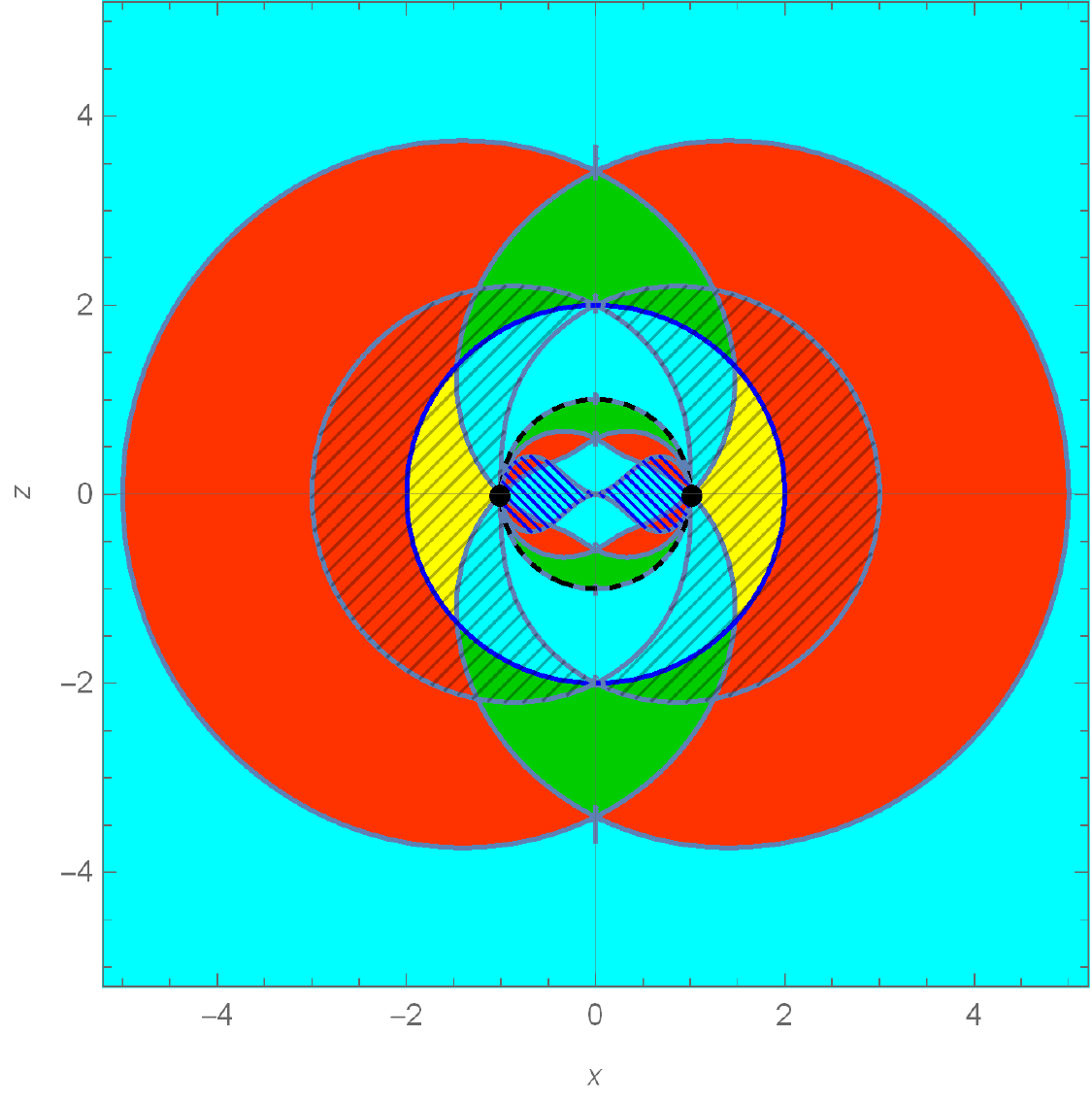}
	\label{Kerr4a}
 }
 \quad
\subfloat[][$a=a_{e}$, $q<q_C$]{
		\includegraphics[scale=0.32]{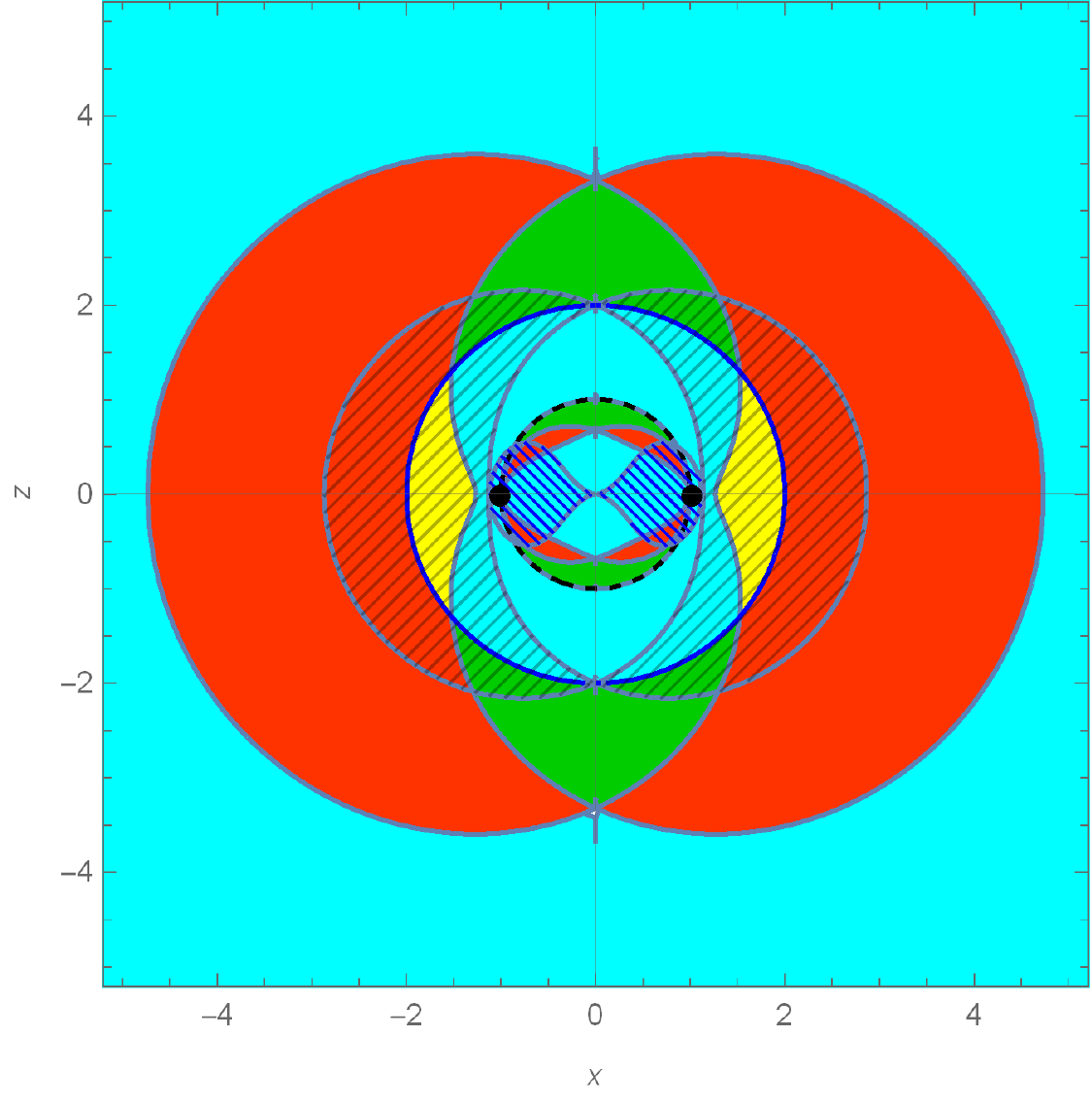}
		\label{KerrN14a}
 }
 \quad
 \subfloat[][$a=a_{e}$, $q>q_C$]{
		\includegraphics[scale=0.32]{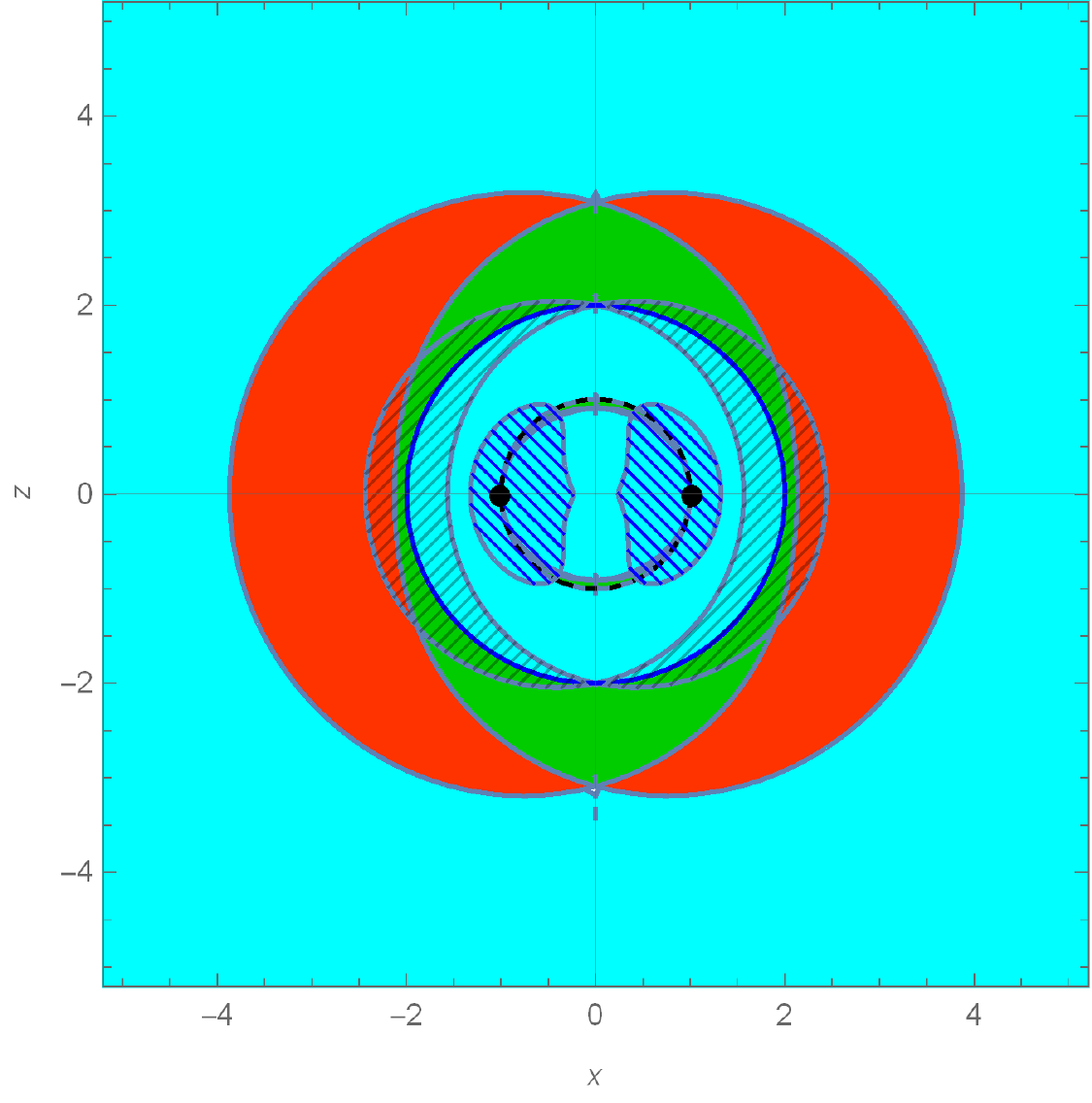}
		\label{KerrN24a}	
 }
\caption{Optical types of Kerr and Kerr-Newman spaces. Red color -- unstable photon region, yellow -- stable photon region, dark blue -- region with $\Delta_r\leq0$, dashed -- throat at $r = 0$, mesh -- ergoregion, blue mesh -- causality violation,  green -- (P)TTR, aqua -- A(P)TTR.}
\label{Kerr1}
\end{figure}

\begin{figure}[tb]
\centering
\subfloat[][$a=0$, $q=0$]{
  	\includegraphics[scale=0.32]{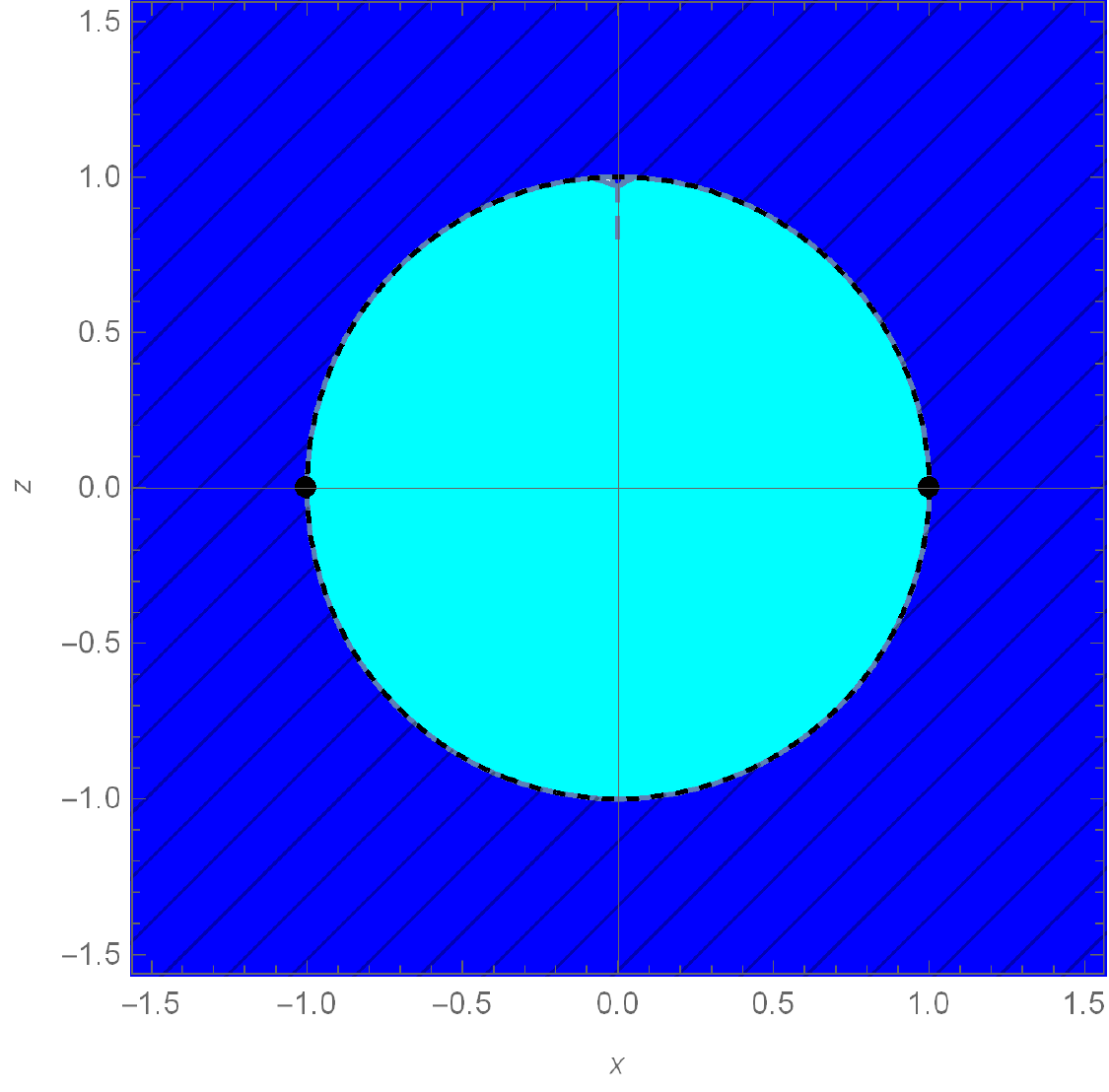}
		\label{Shb}
 }
 \quad
\subfloat[][$a=0$, $q<q_C$]{
	\includegraphics[scale=0.32]{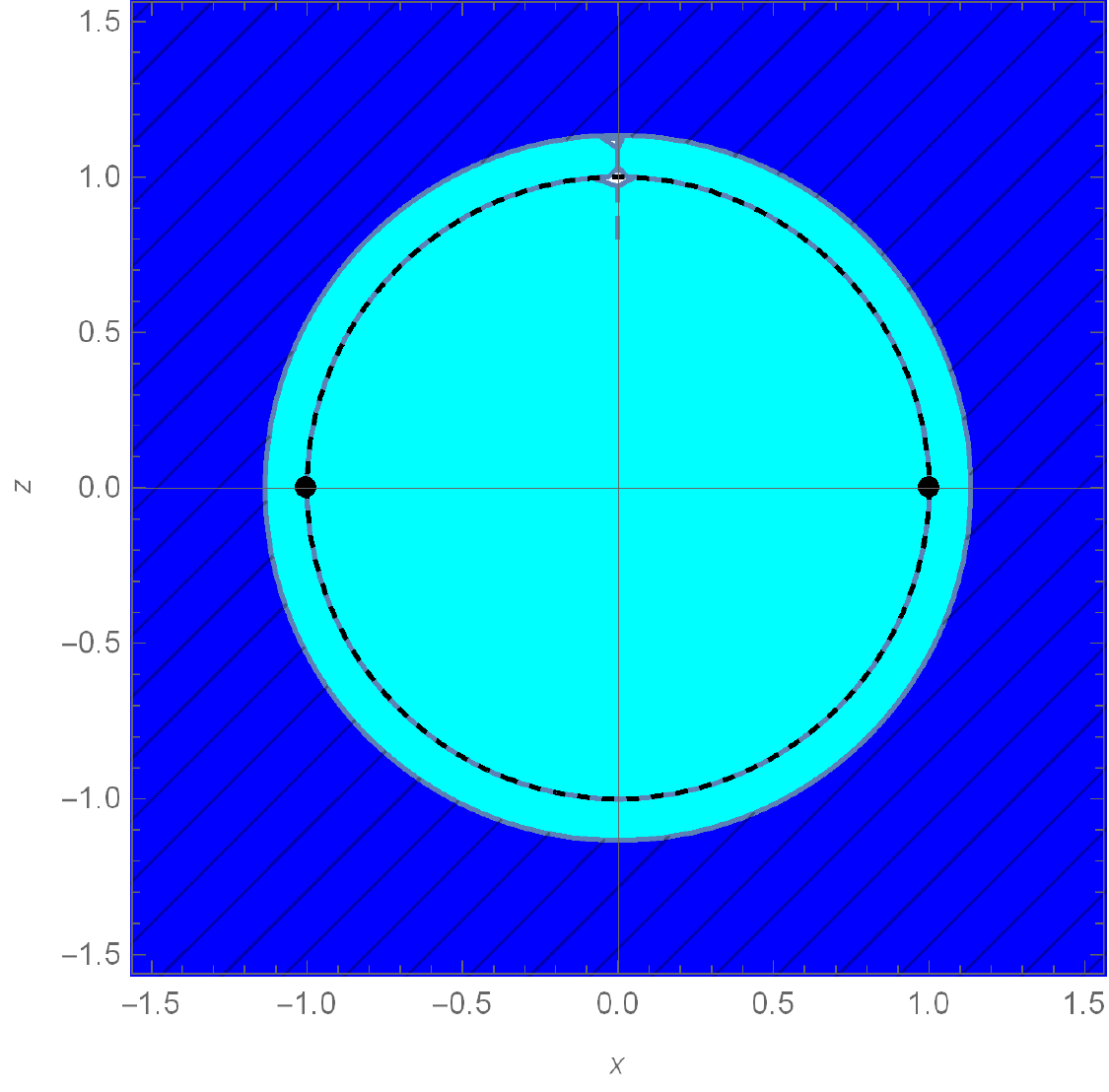}
		\label{ShN11b}
 }
 \quad
 \subfloat[][$a=0$, $q>q_C$]{
	\includegraphics[scale=0.32]{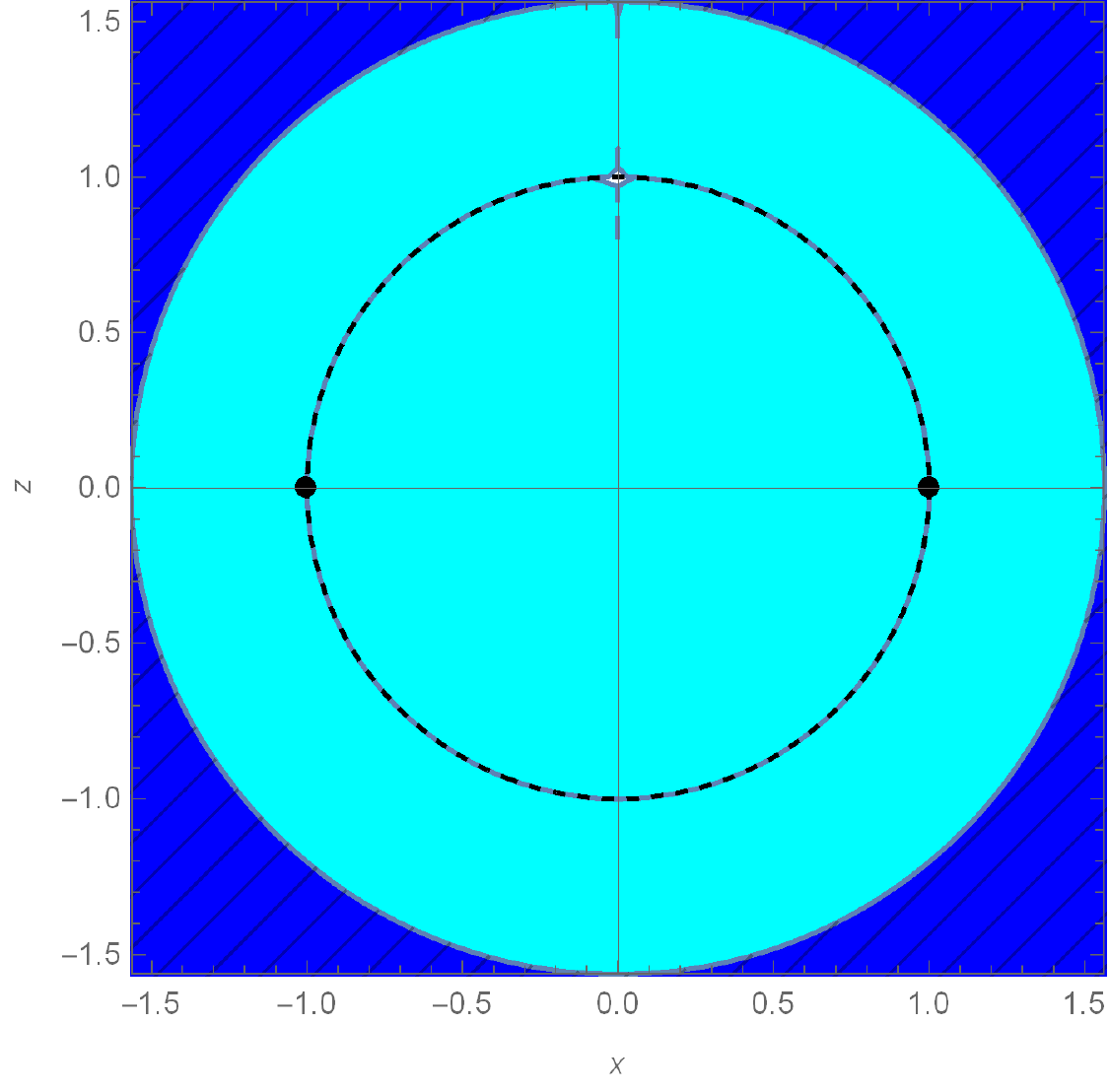}
		\label{SHN21b}
 }
 \\
 \subfloat[][$0<a<a_e$, $q=0$]{
  		\includegraphics[scale=0.32]{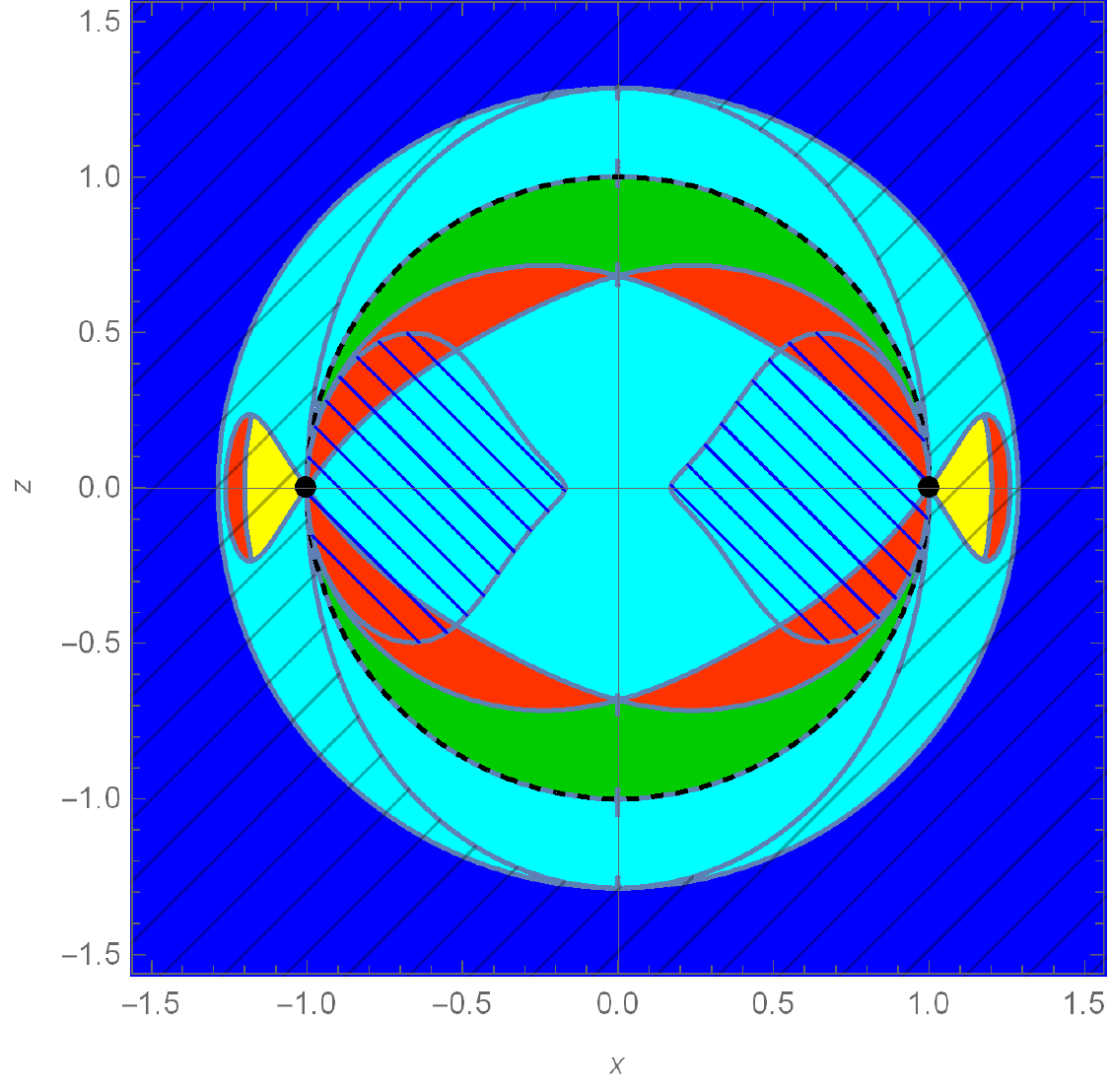}
		\label{Kerr2b}
 }
 \quad
\subfloat[][$0<a<a_C$, $q<q_C$]{
	\includegraphics[scale=0.32]{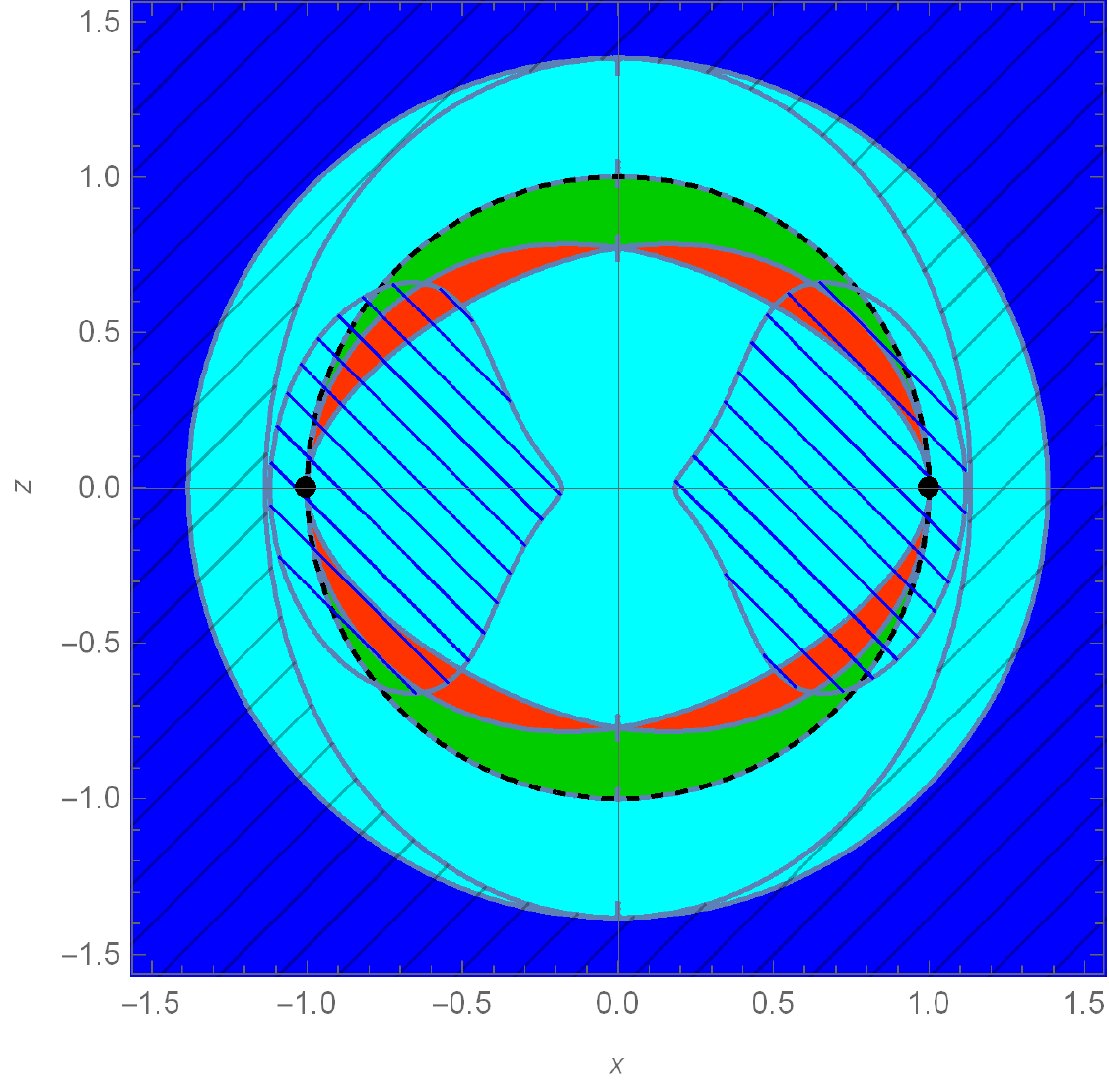}
		\label{KerrN12b}
 }
 \quad
 \subfloat[][$0<a<a_e$, $q>q_C$]{
		\includegraphics[scale=0.32]{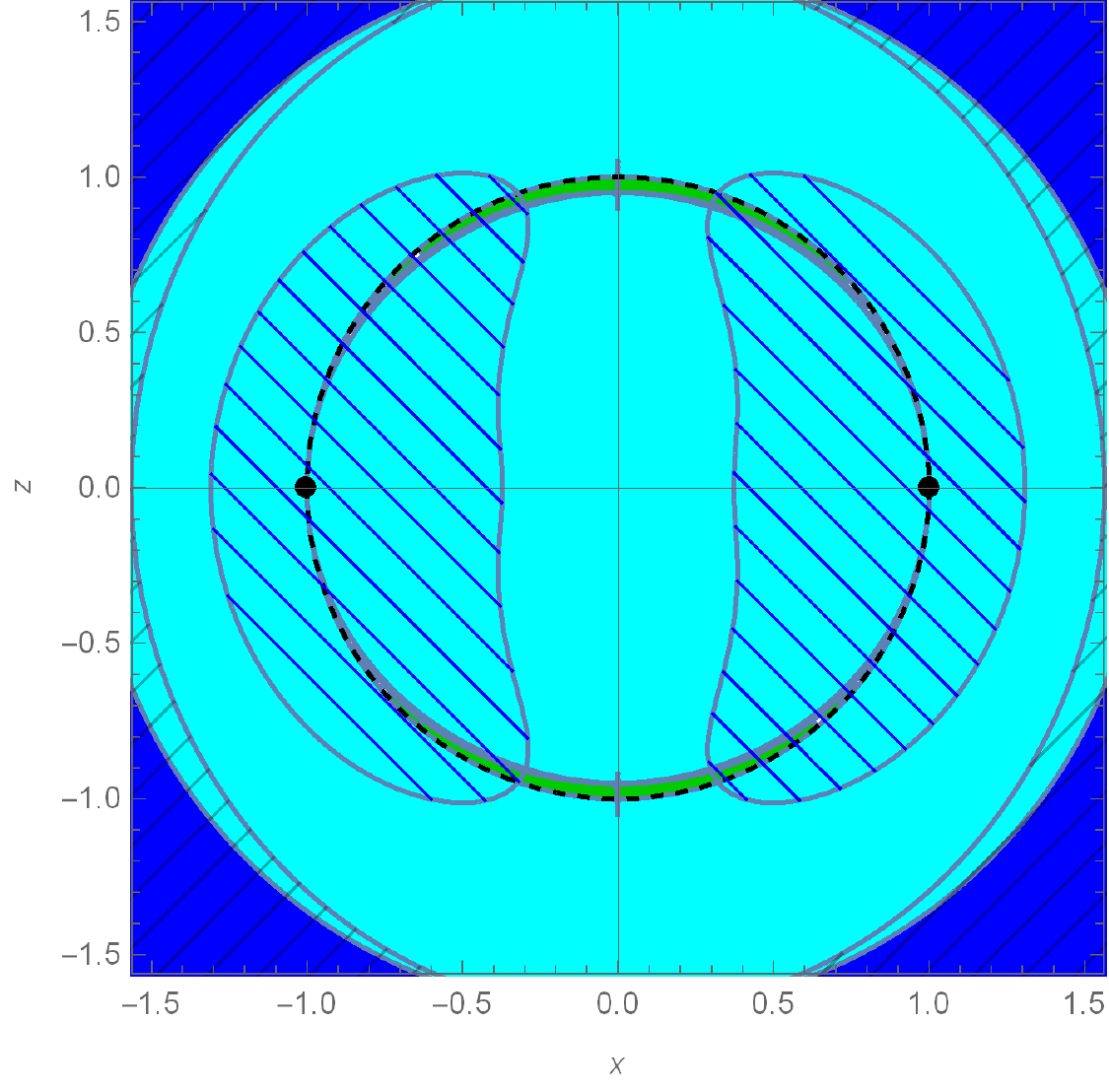}
		\label{KerrN22b}
 }
 \\
 \subfloat[][$0<a<a_e$, $q=0$]{
  		\includegraphics[scale=0.32]{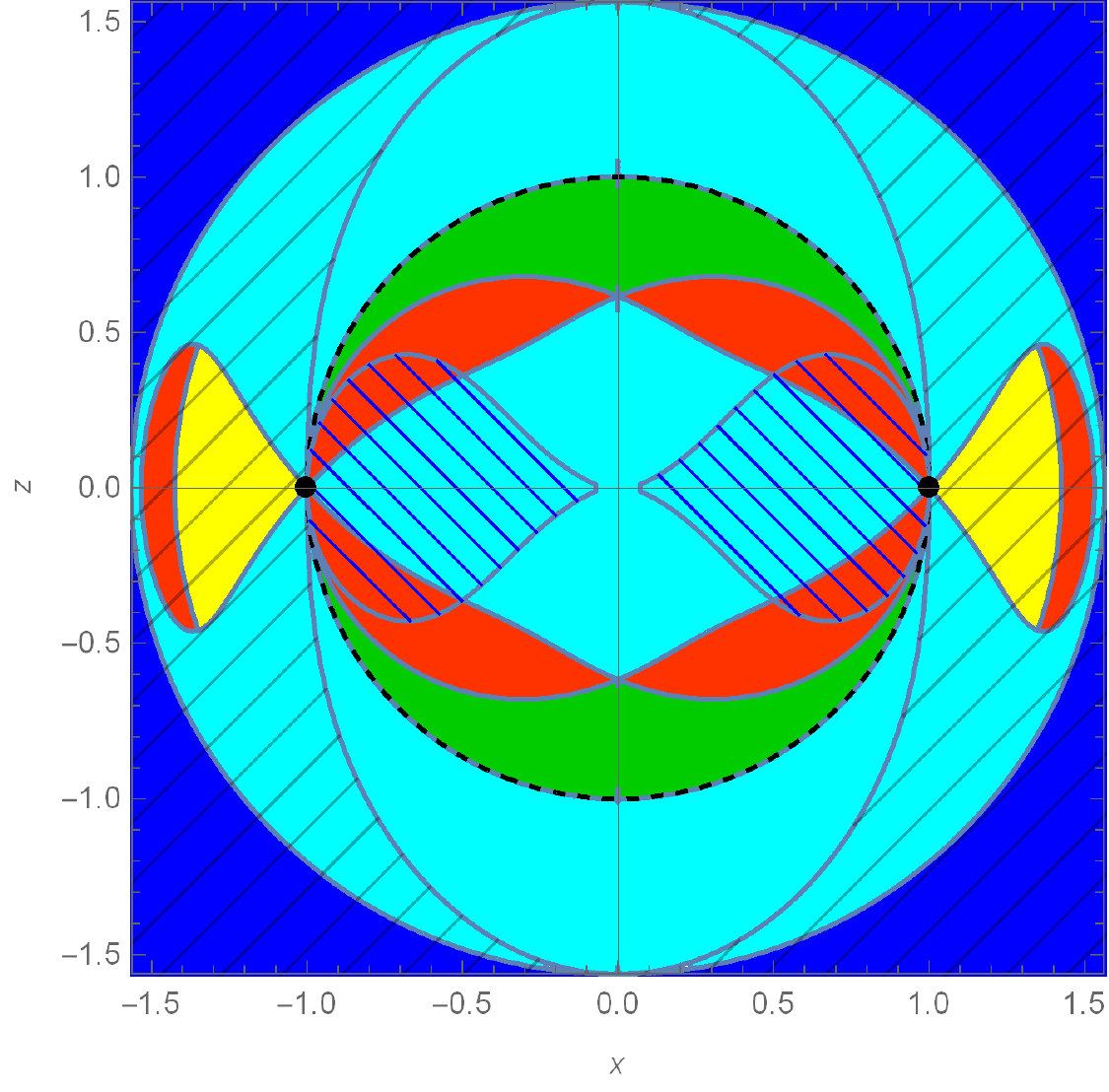}
		\label{Kerr3b}	
 }
 \quad
\subfloat[][$a_C<a<a_{e}$, $q<q_C$]{
		\includegraphics[scale=0.32]{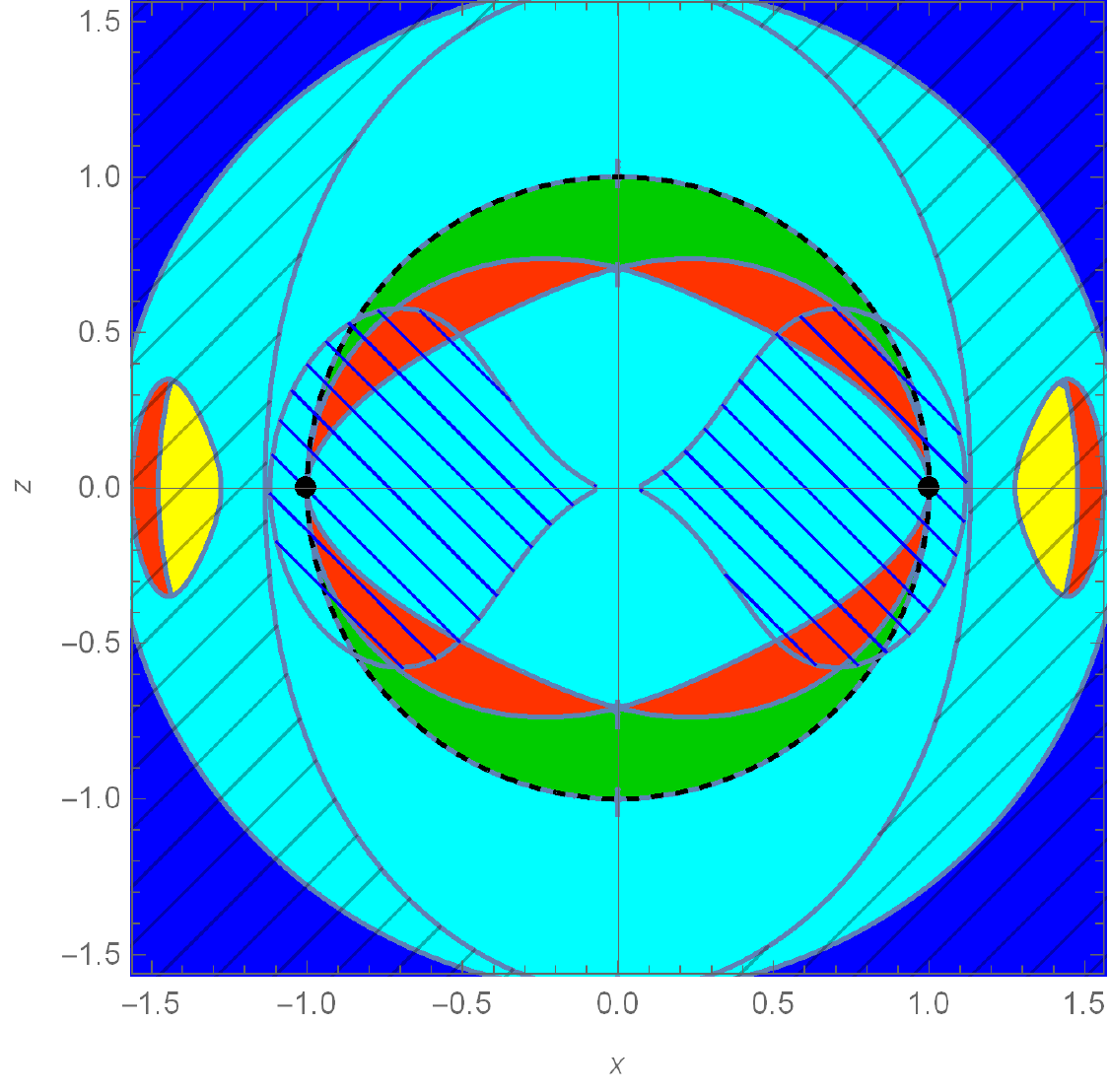}
		\label{KerrN13b}
 }
 \quad
 \subfloat[][$0<a<a_e$, $q>q_C$]{
		\includegraphics[scale=0.32]{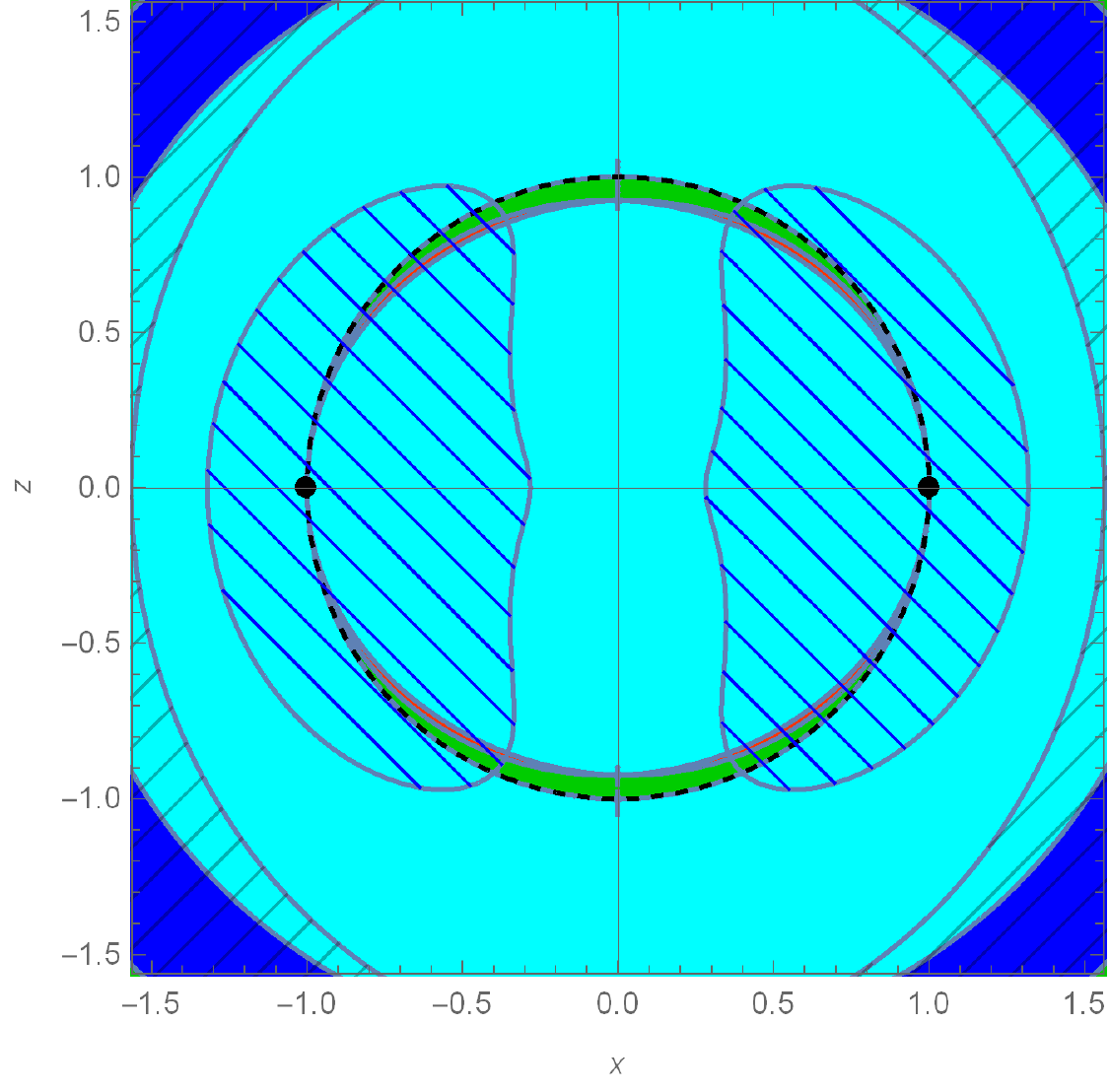}
		\label{KerrN23b}
 }
  \\
 \subfloat[][$a=a_{e}$, $q=0$]{
  		\includegraphics[scale=0.32]{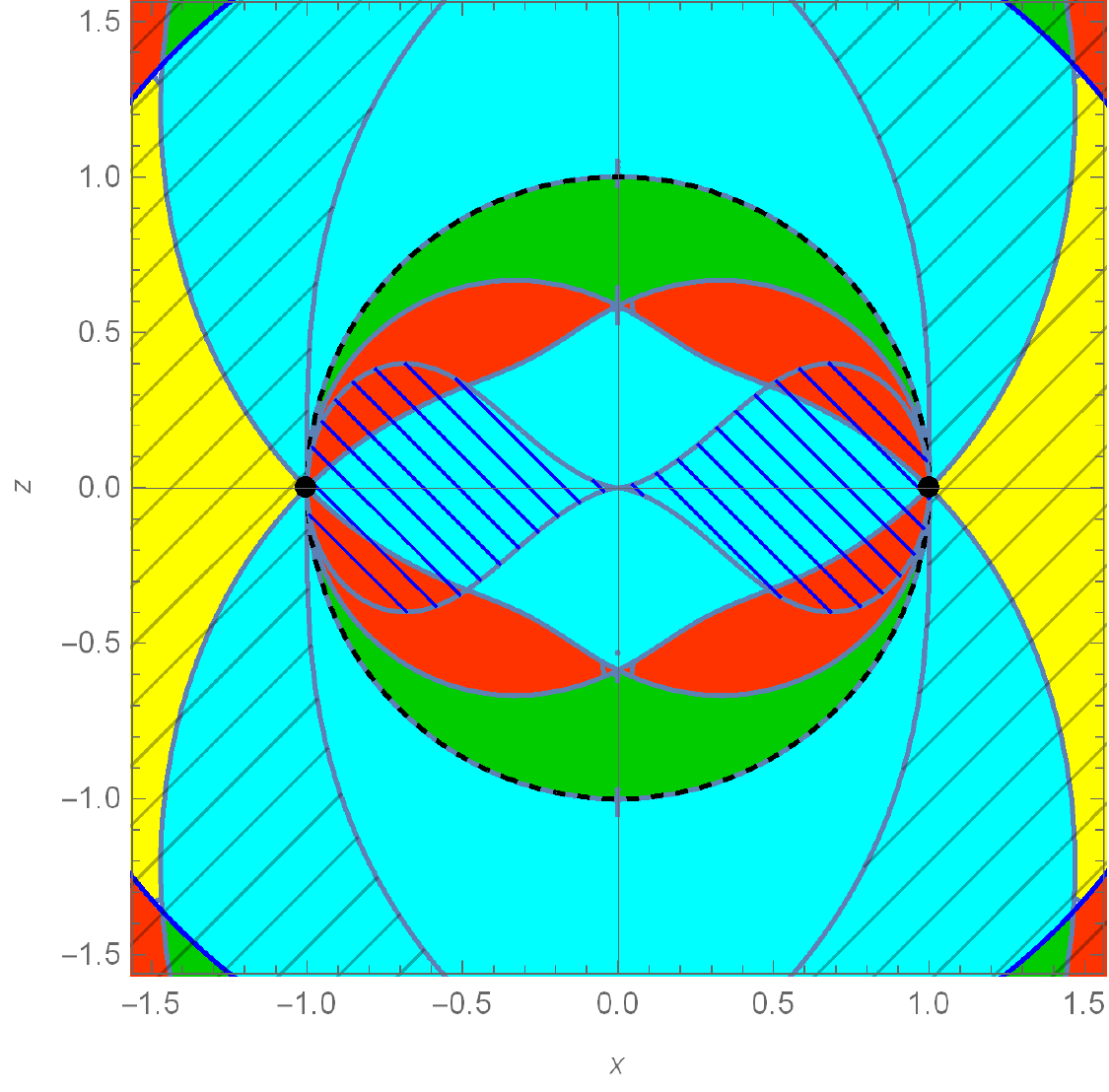}
	\label{Kerr4b}
 }
 \quad
\subfloat[][$a=a_{e}$, $q<q_C$]{
		\includegraphics[scale=0.32]{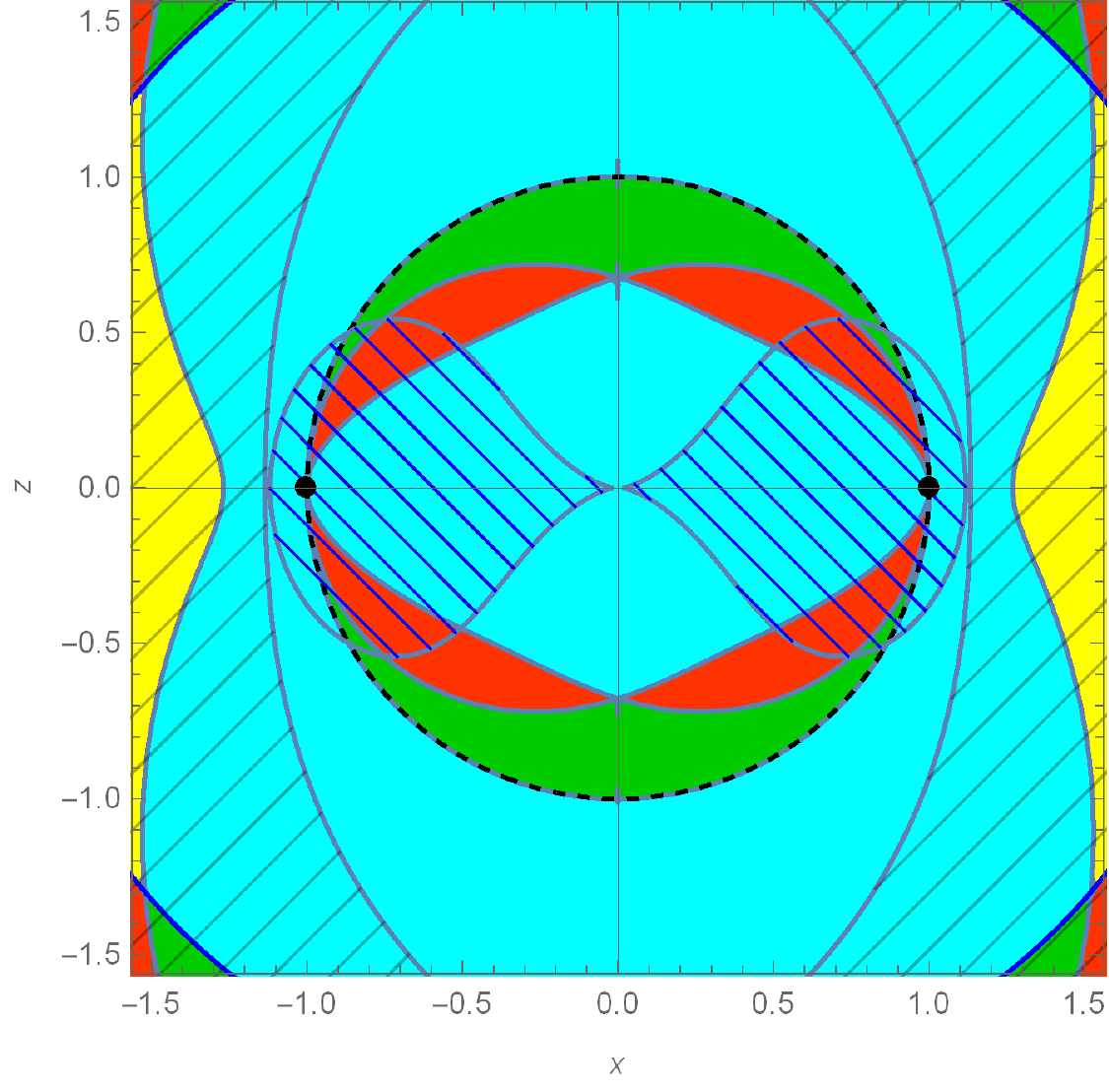}
		\label{KerrN14b}
 }
 \quad
 \subfloat[][$a=a_{e}$, $q>q_C$]{
		\includegraphics[scale=0.32]{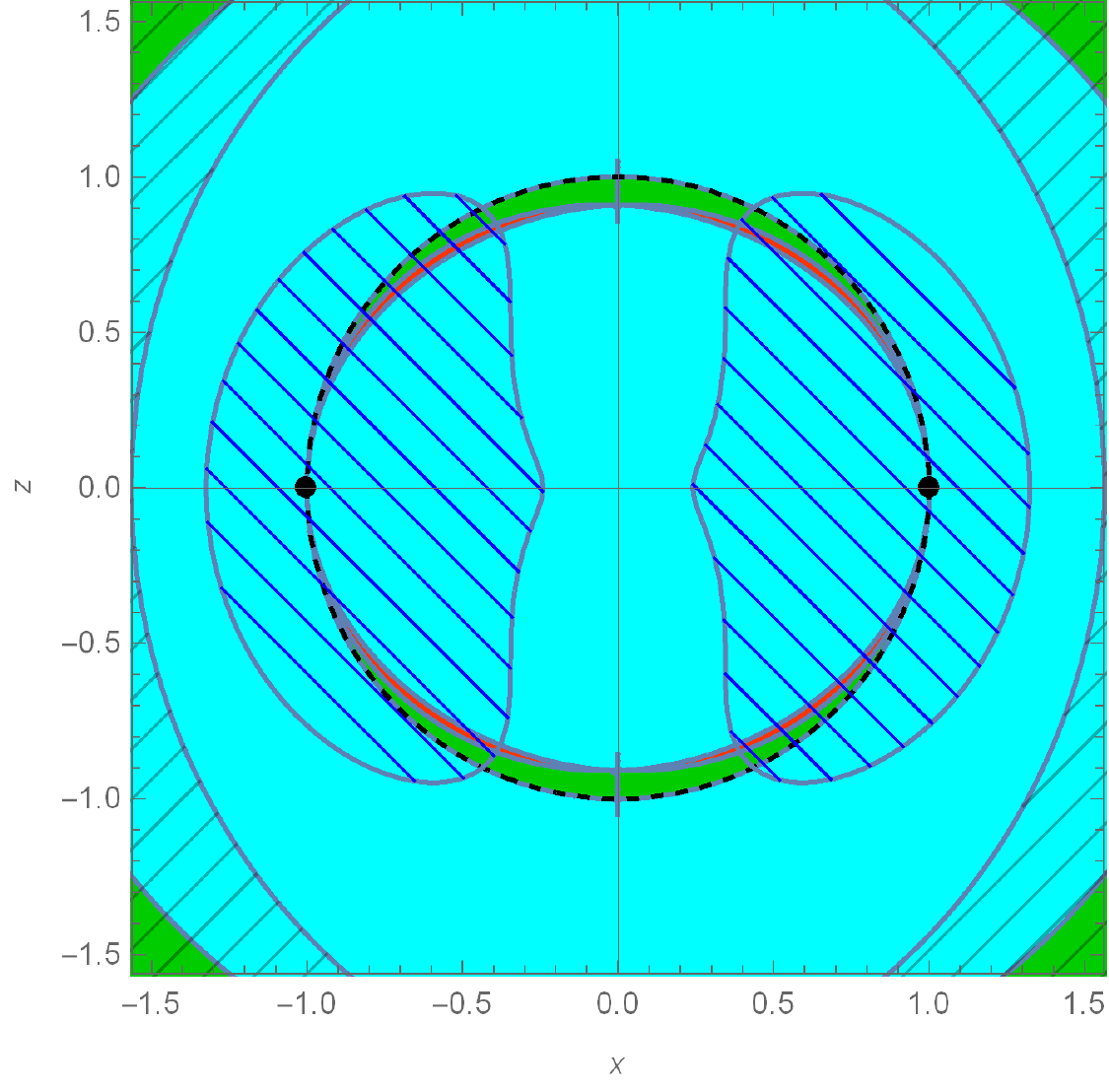}
		\label{KerrN24b}	
 }
\caption{The inner part of (\ref{Kerr1}).  Red color -- unstable photon region, yellow -- stable photon region, dark blue -- region with $\Delta_r\leq0$, dashed -- throat at $r = 0$, mesh -- ergoregion, blue mesh -- causality violation,  green -- (P)TTR, aqua -- A(P)TTR.}
\label{Kerr2}
\end{figure}

\begin{figure}[tb]
\centering
\subfloat[][$a=0$, $q=q_C$]{
  	\includegraphics[scale=0.32]{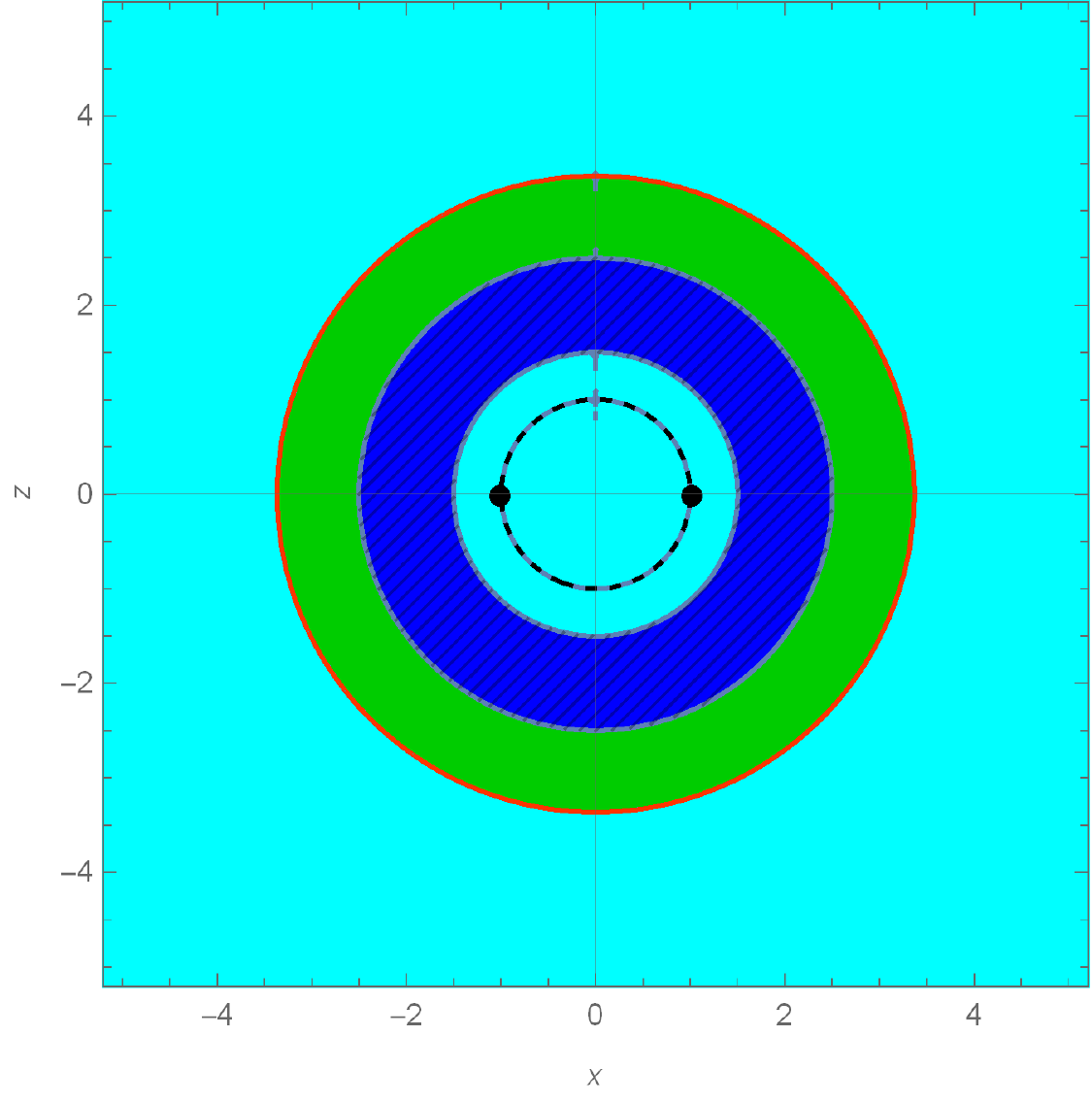}
		\label{ShN53a}	
 }
 \quad
\subfloat[][$a=a_C$, $q=q_C$]{
\includegraphics[scale=0.32]{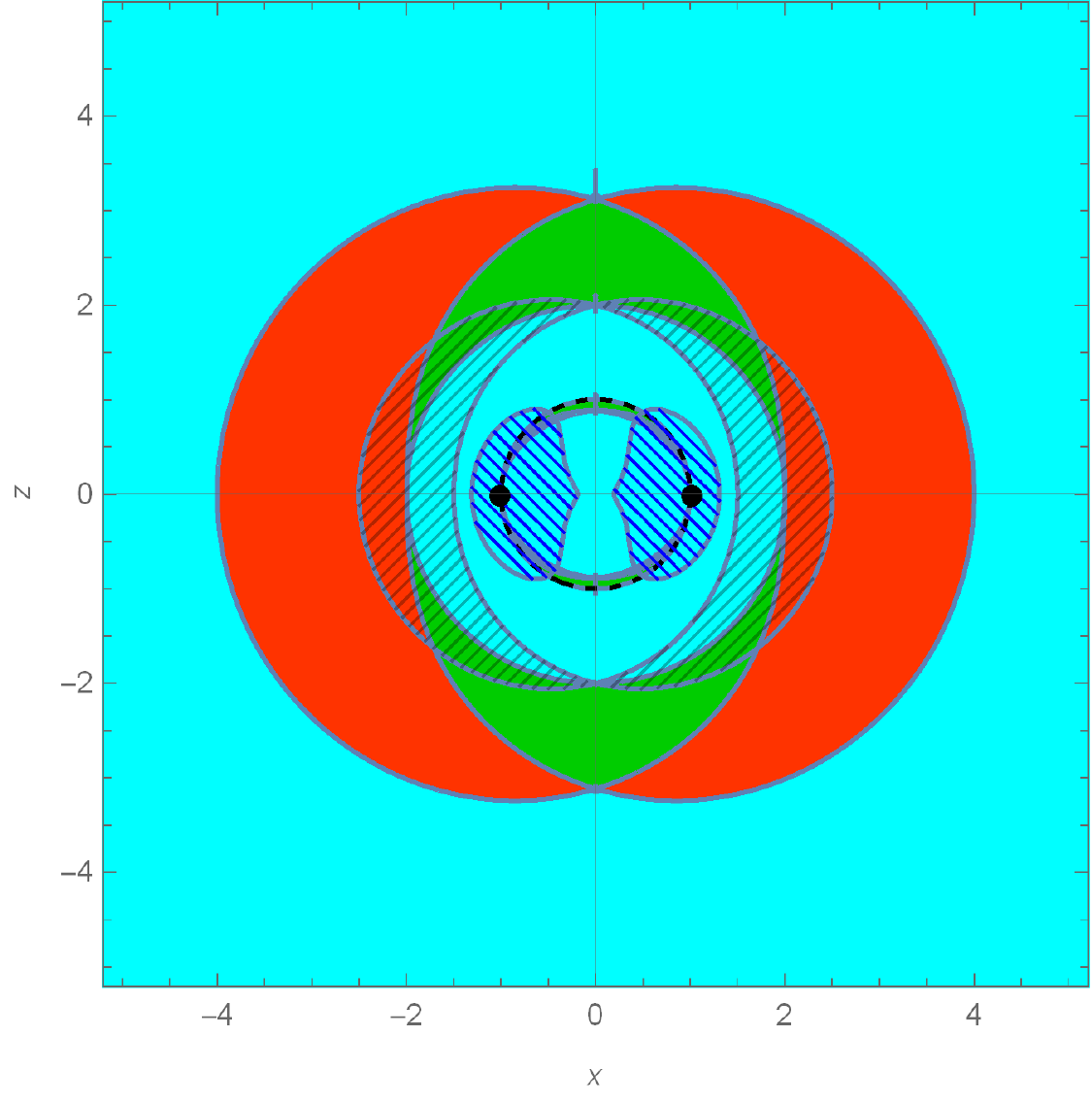}
		\label{KerrN52a}
 }
 \quad
 \subfloat[][$a=a_c$, $q=q_C$]{
		\includegraphics[scale=0.32]{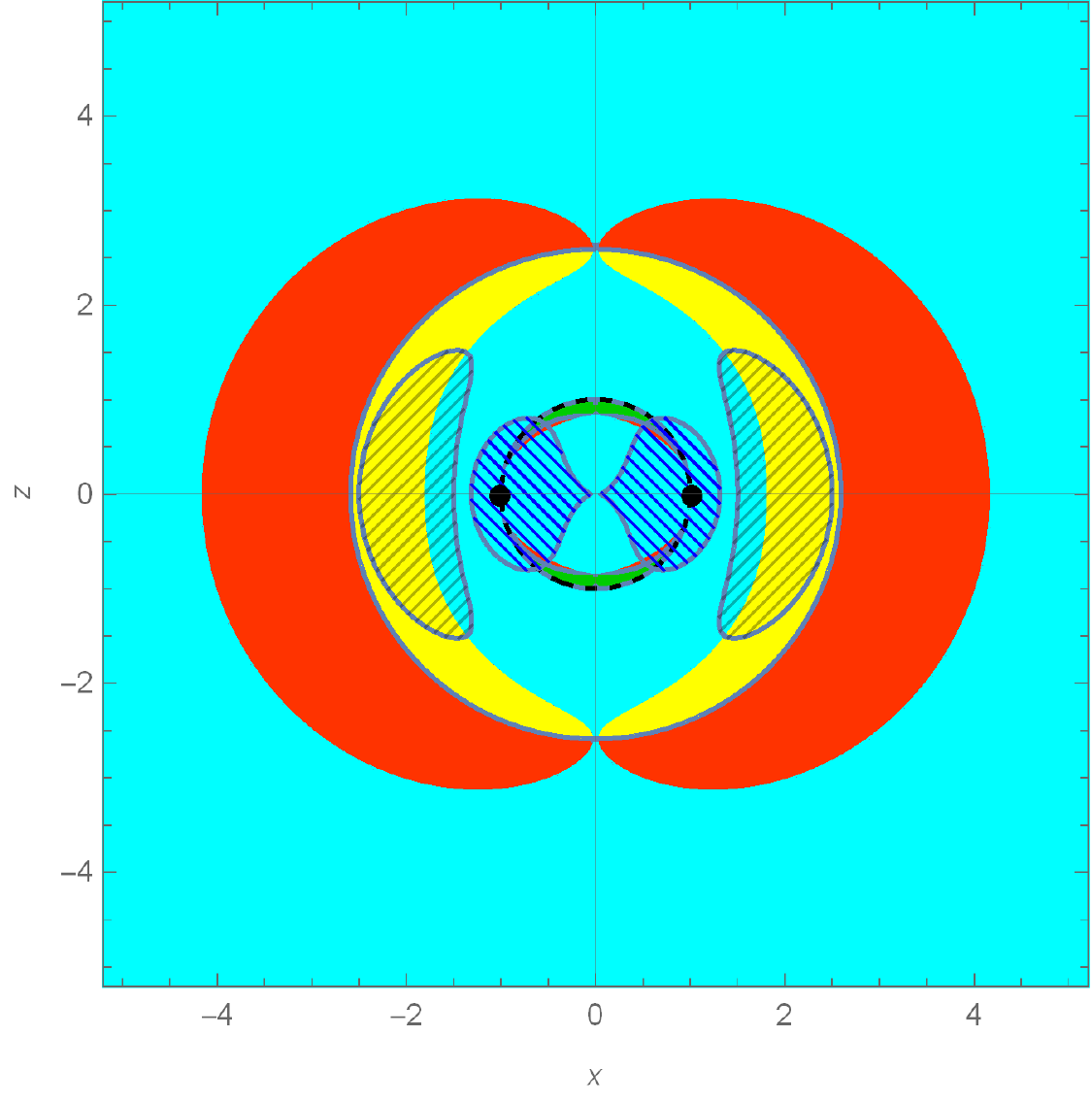}
		\label{KerrN53a}	
 }
 \\
 \subfloat[][$a=0$, $q>q_C$]{
  		\includegraphics[scale=0.32]{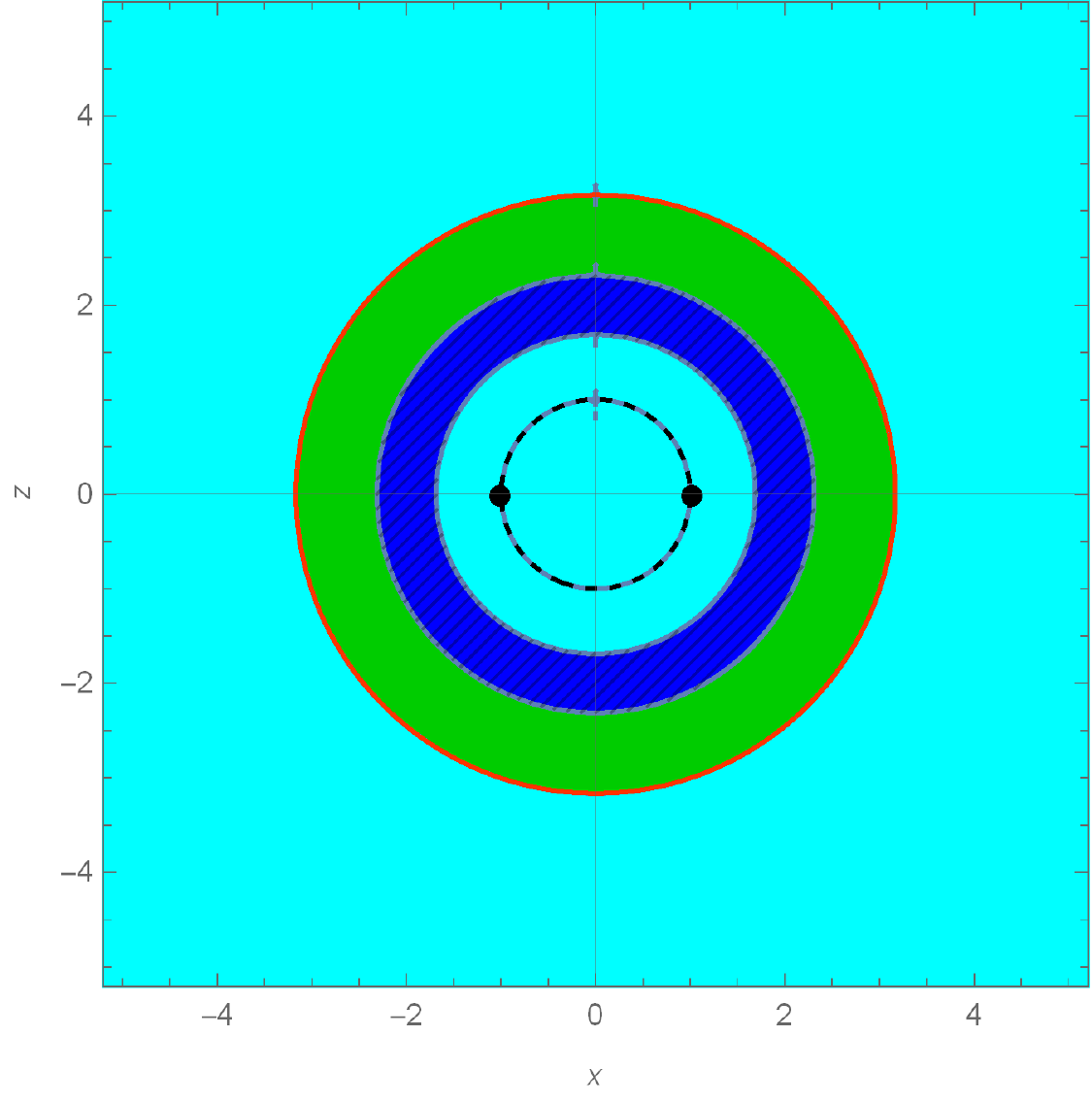}
		\label{ShN51a}
 }
 \quad
\subfloat[][$a=a_C$, $q>q_C$]{
	\includegraphics[scale=0.32]{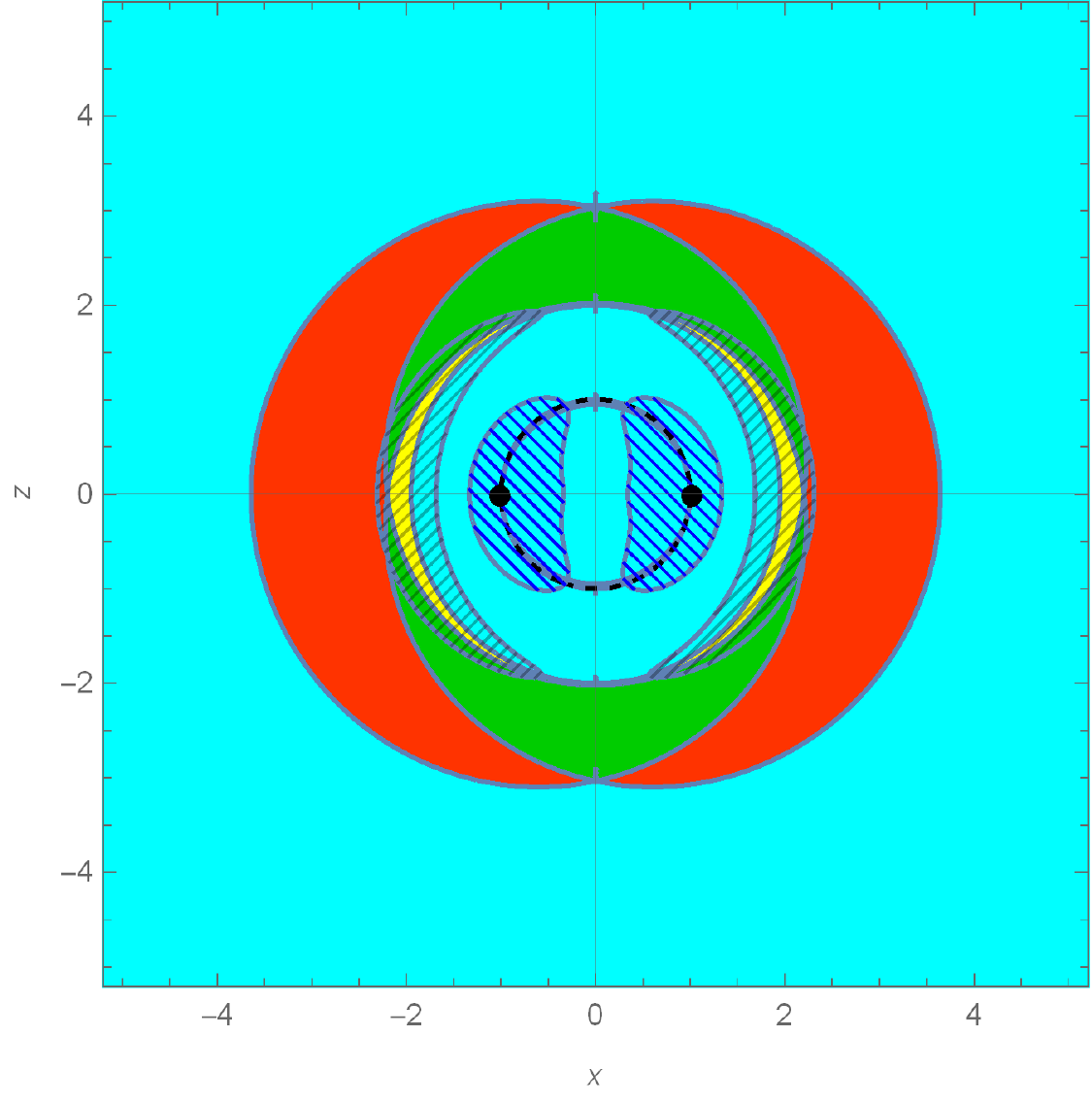}
		\label{KerrN25a} 
 }
 \quad
 \subfloat[][$a=a_c$, $q>q_C$]{
		\includegraphics[scale=0.32]{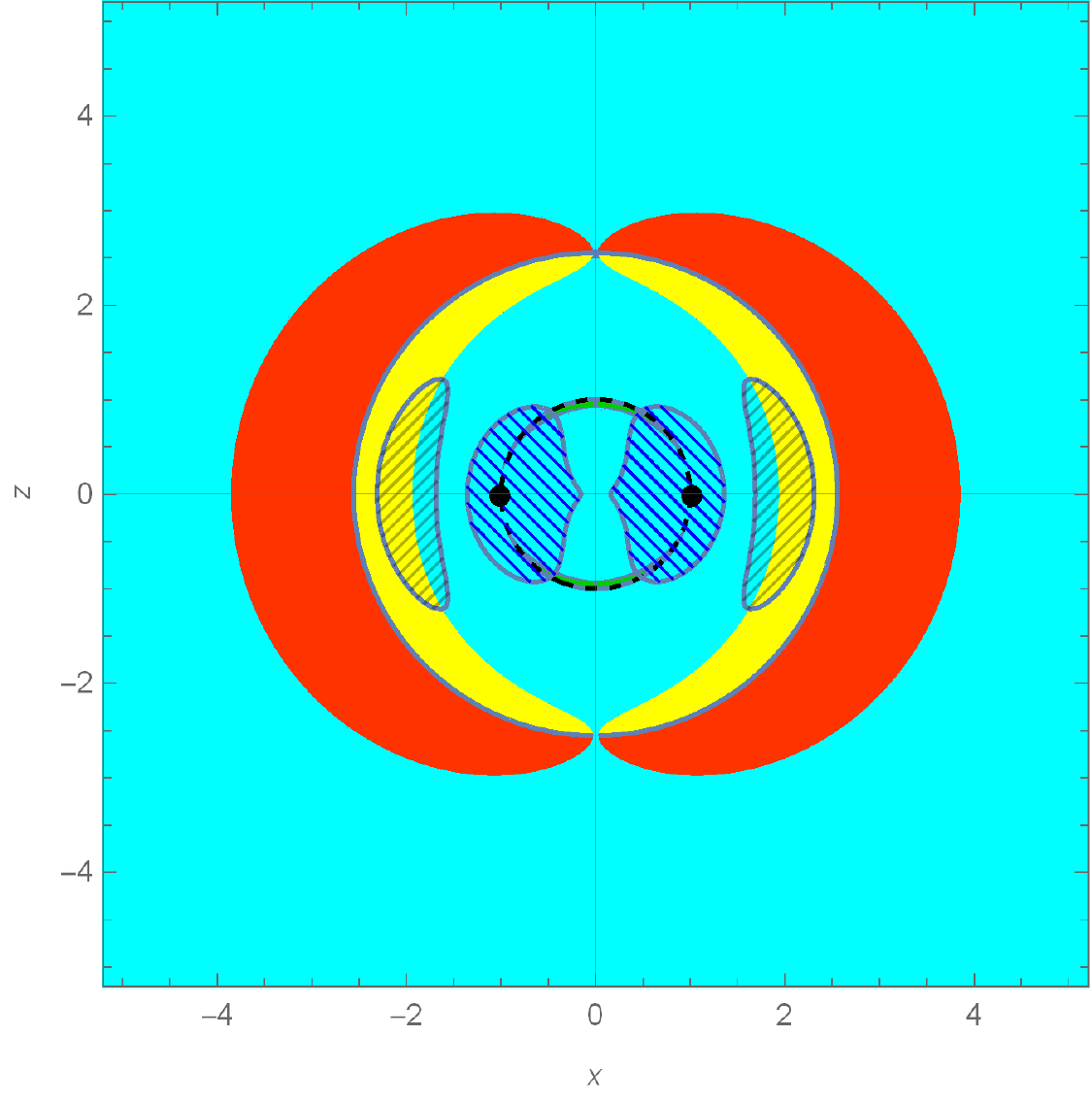}
		\label{KerrN26a}		
 }
 \\
 \subfloat[][$a=0$, $q=q_e$]{
  	\includegraphics[scale=0.32]{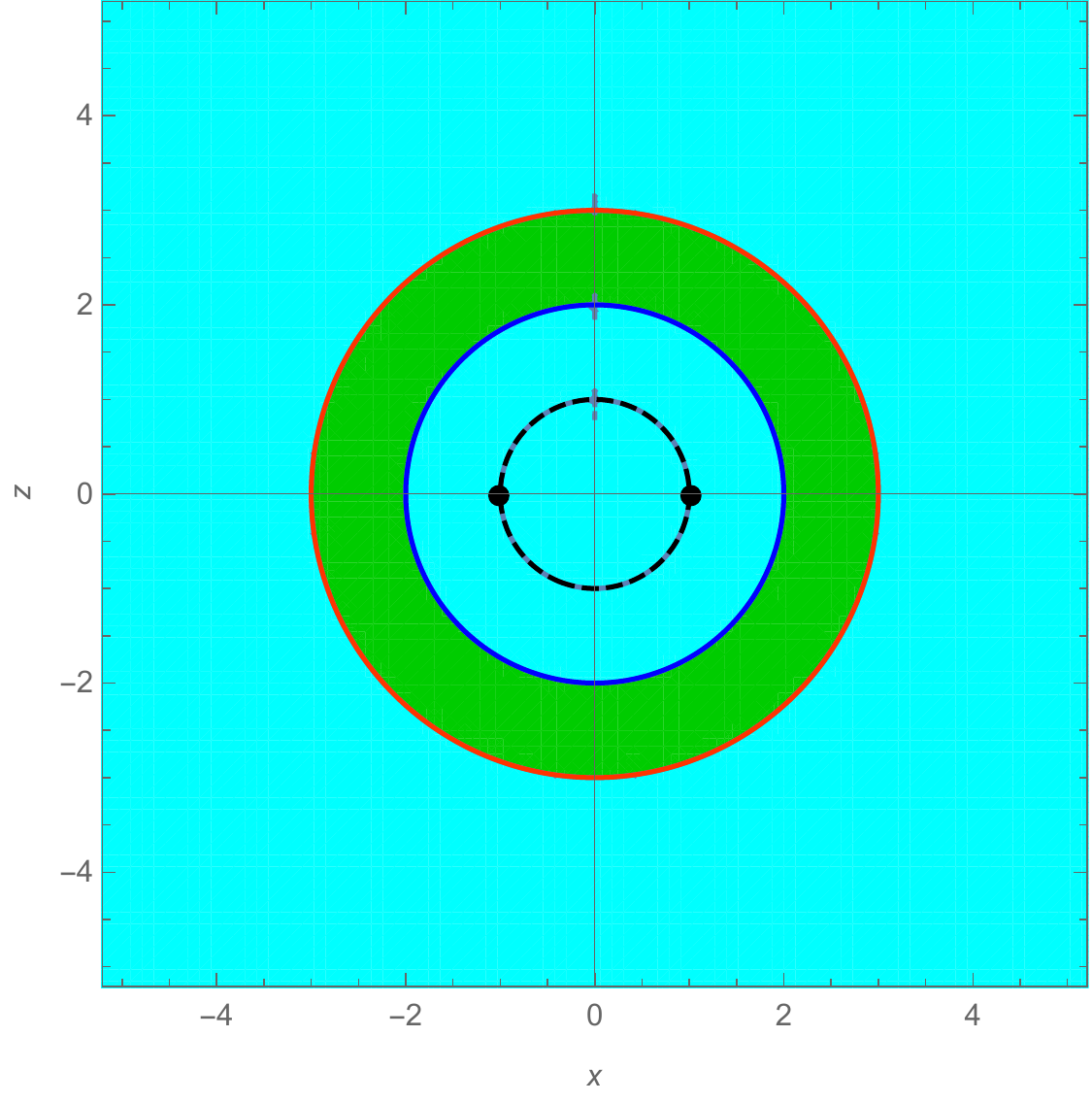}
		\label{ShN31a}	
 }
 \quad
\subfloat[][$a=a_C$, $q=q_e$]{
		\includegraphics[scale=0.32]{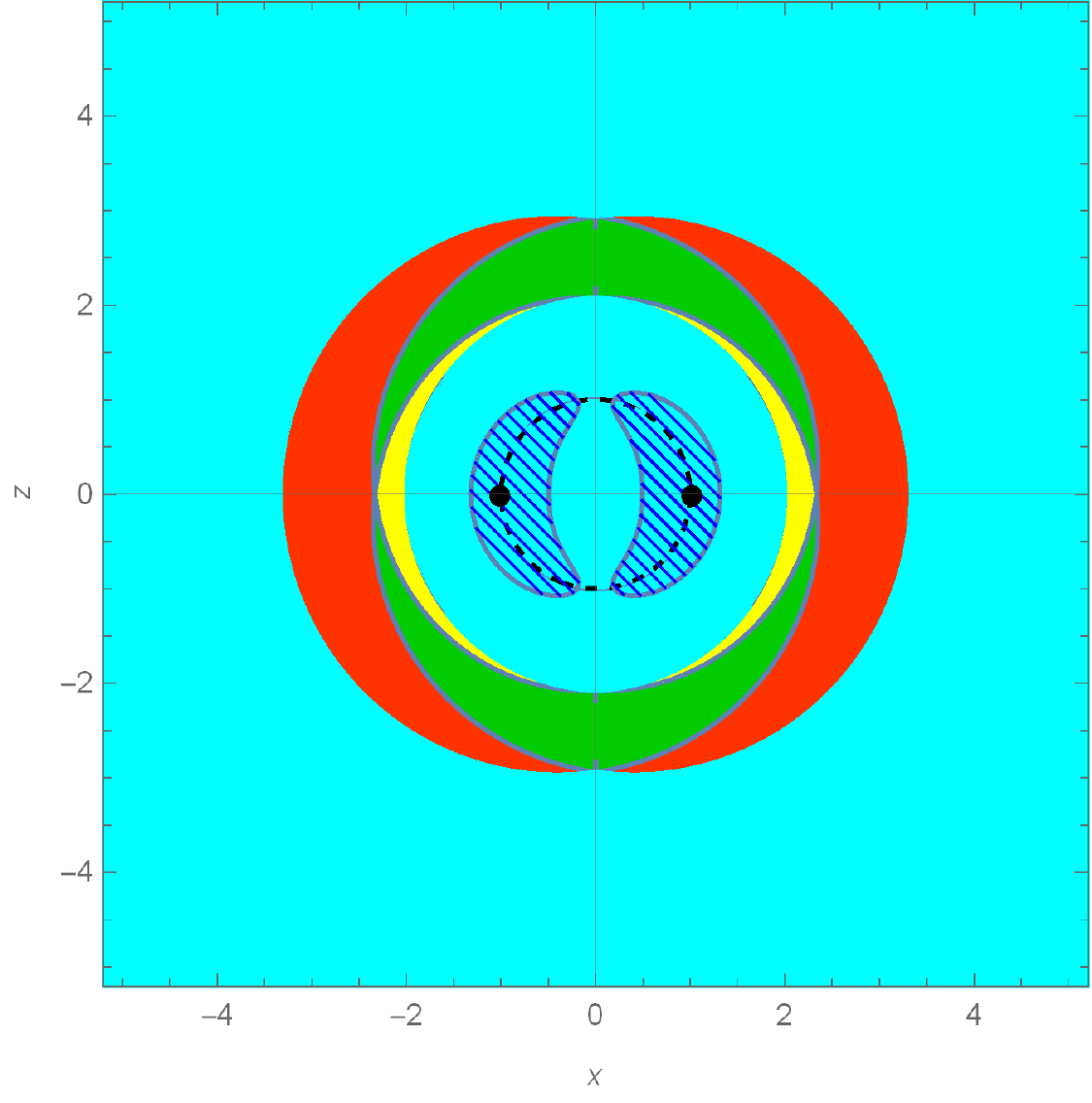}
	\label{KerrN34a}	
 }
 \quad
 \subfloat[][$a=a_c$, $q=q_e$]{
		\includegraphics[scale=0.32]{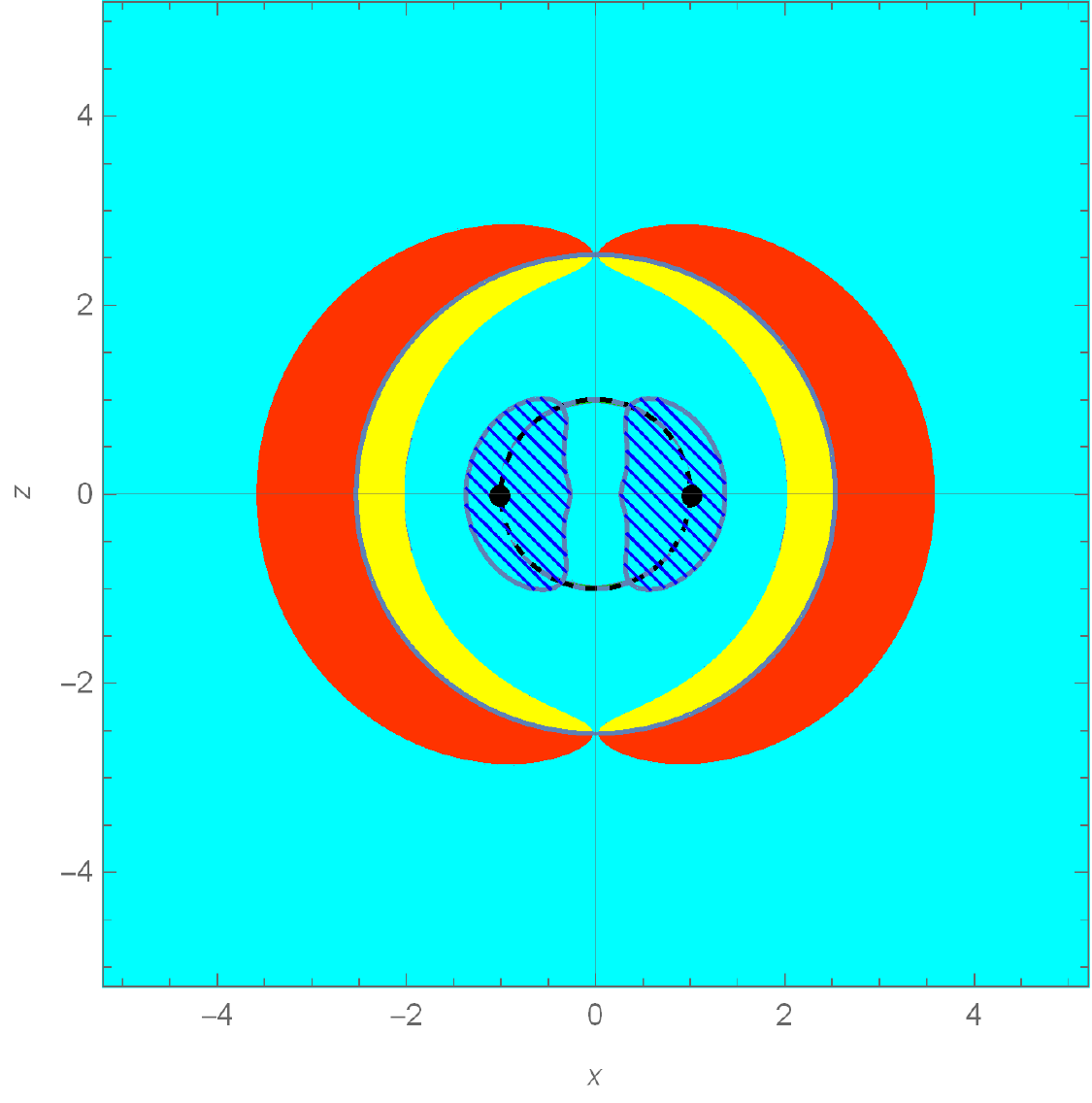}
		\label{KerrN35a}	
 }
  \\
 \subfloat[][$a=0$, $q>q_e$]{
  			\includegraphics[scale=0.32]{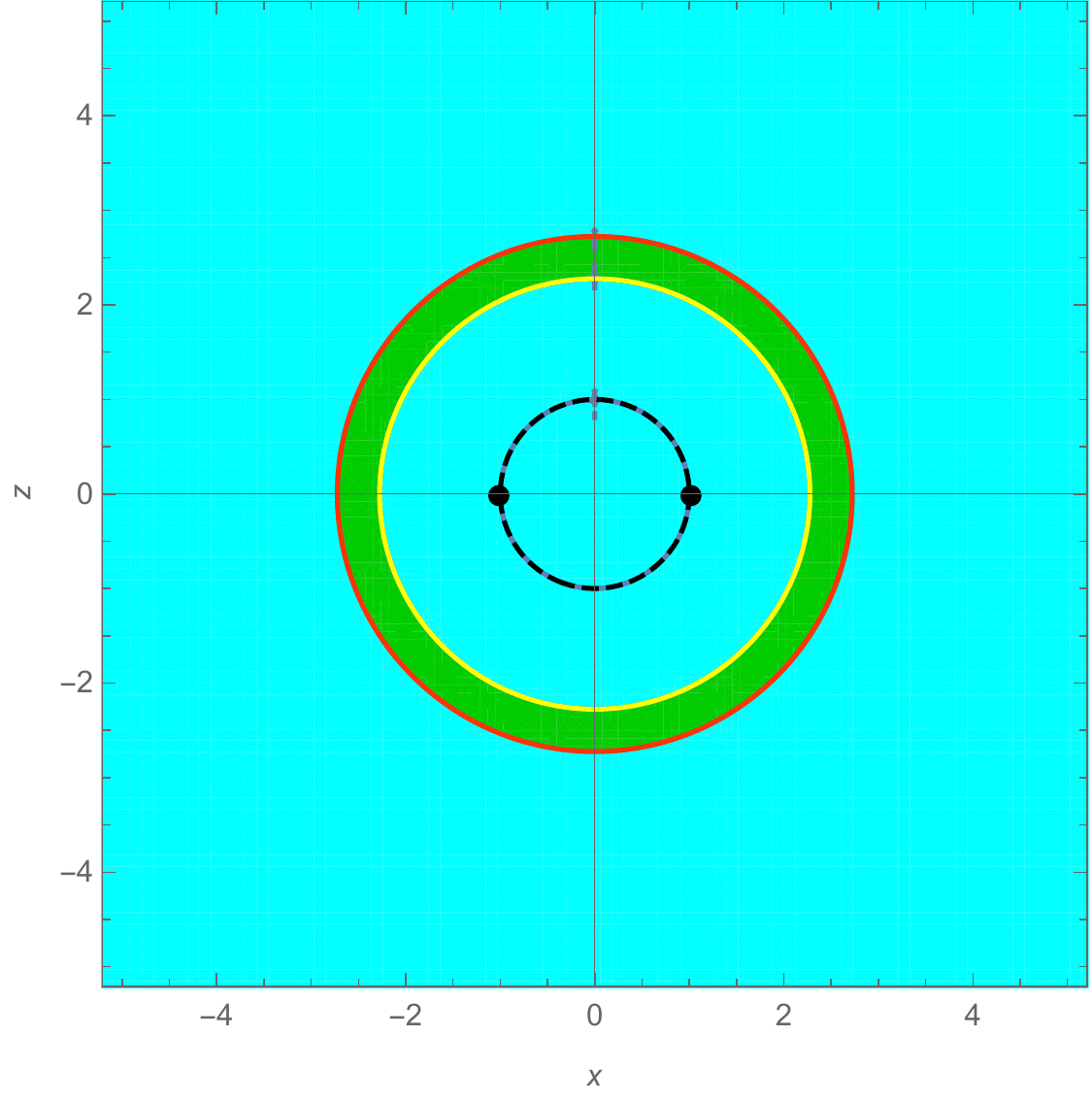}
	\label{ShN41a}	
 }
 \quad
\subfloat[][$a=a_C$, $q>q_e$]{
		\includegraphics[scale=0.32]{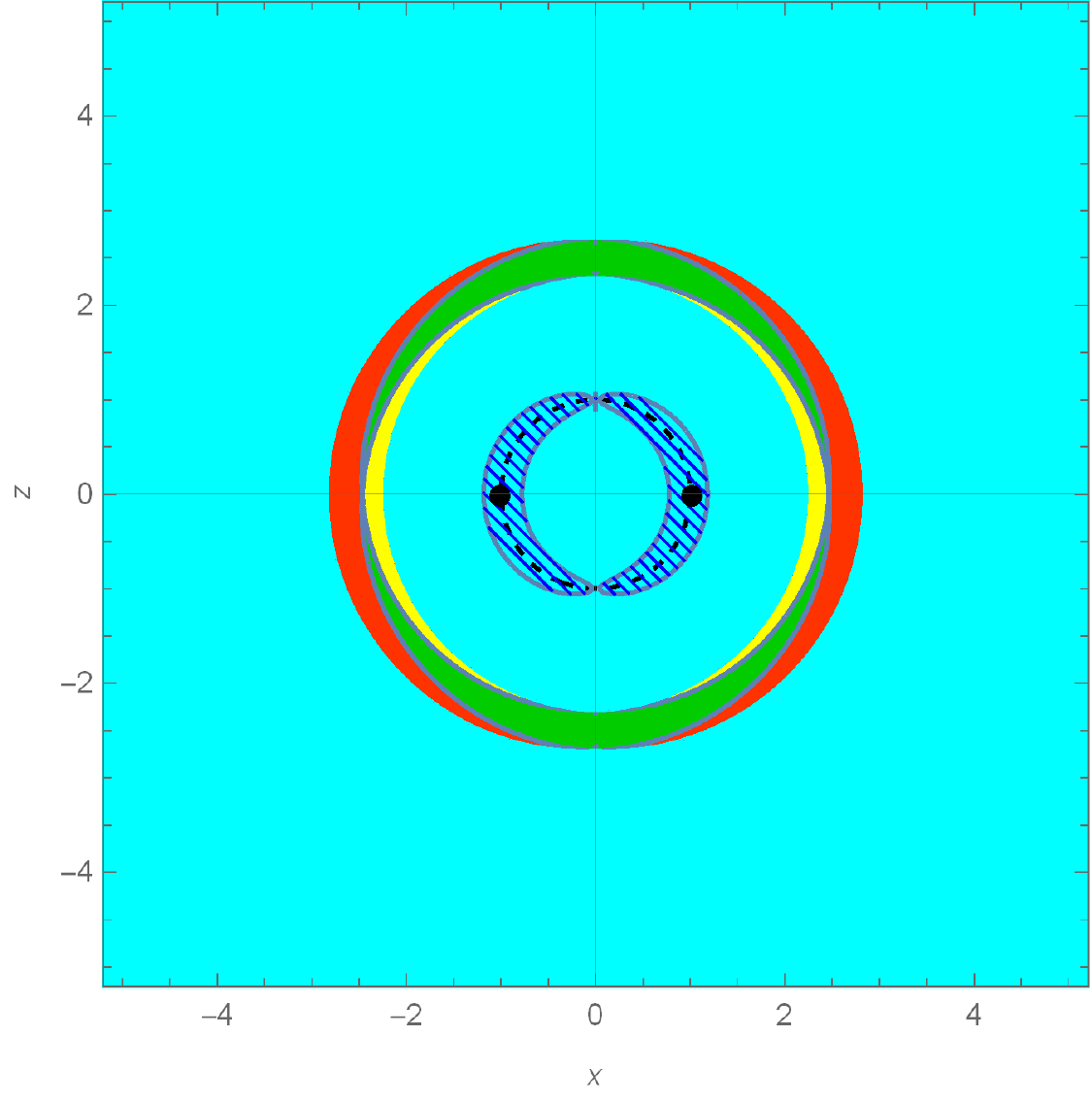}
		\label{KerrN44a}
 }
 \quad
 \subfloat[][$a=a_c$, $q>q_e$]{
		\includegraphics[scale=0.32]{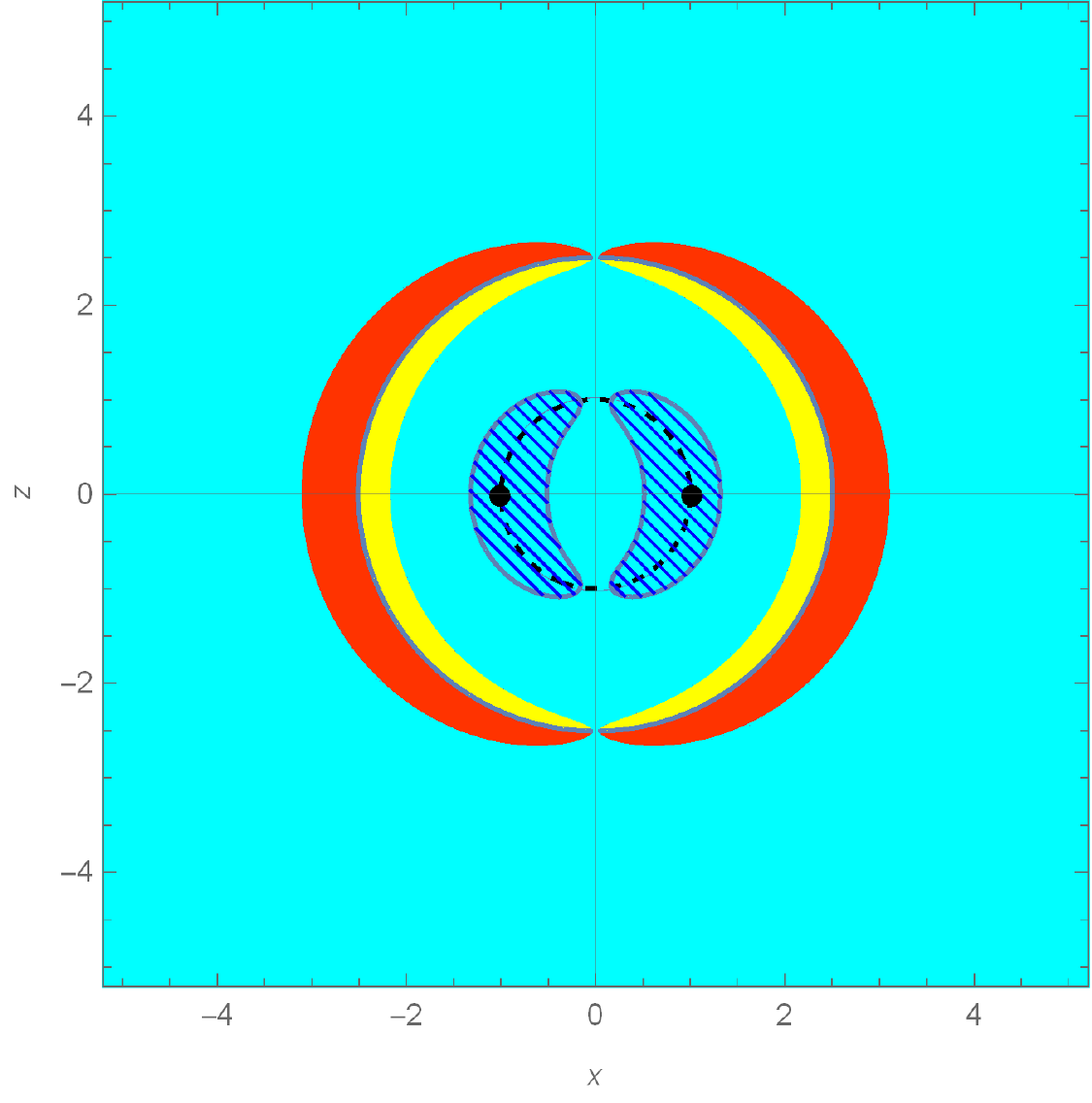}
		\label{KerrN45a}
 }
\caption{Optical types of Kerr and Kerr-Newman spaces.  Red color -- unstable photon region, yellow -- stable photon region, dark blue -- the region   $\Delta_r\leq0$, dashed -- the throat at $r = 0$, mesh -- the ergoregion, blue mesh -- the causality violating region,  green -- (P)TTR, aqua - A(P)TTR.}
\label{Kerr3}
\end{figure}

In order to illustrate the above classification, we use the visualization method proposed in  \cite{Grenzebach}. The Figs. (\ref{Kerr1}-\ref{Kerr3}) represents slices $y=0$ of Cartesian type coordinates where each region of space is filled with a certain color depending on the properties of the corresponding photon surface $r={\rm const}$. Schematic representation of the geodetic behavior for each area is shown in the Fig. (\ref{Kerr}).  

Let's start with the Kerr metric ($q=0$). In absence of rotation ($a=0$), there is a unique photon sphere and  TTSs filling the region between the photon sphere and the horizon (\ref{Sha}). The inner region (\ref{Shb}) behind the singularity is filled by ATTSs, which correspond  to weak gravity or antigravity. The solution obviously belongs to an optical type $I$ in our classification. 

For weaker rotation,  $|a|<a_e$, the photon sphere   spreads into the photon region. There is an area with PTTSs and ATTSs in the vicinity of the poles and the  singularities, respectively (\ref{Kerr2a},\ref{Kerr3a}). The presence of the latter in these cases corresponds to  antigravity effects occurring in the vicinity of the singularity, while the PTTSs characterize  some properties of the strong gravitational lensing. The closed TTR persists also and decreases with increasing rotation, as shown in the Fig. (\ref{EXT1}). A stable photon region emerges directly from the ring singularity and is always located inside the inner horizon. Behind the throat, a new PTTR appears, which characterizes   strong gravitational field in this accessible region, separated from the ATTR by an unstable photon region (\ref{Kerr2b}, \ref{Kerr3b}). These solutions also belong to the type $I$. 

In the extreme case $|a|=a_e$, the closed TTSs and ATTSs completely disappear, while the PTTSs survives in the vicinity of the poles. The APTTS region passes exactly into the stable photon region (\ref{Kerr4a}). In the overextreme case, the stable photon region is no longer hidden by the horizon. As usual for the horizonless ultracompact objects, this indicates on their instability \cite{Cunha:2017qtt}. These solutions belong to the type $II$.  At critical rotation $a_c$, even the PTTSs disappear  completely and the  ATTR merges with the region of a weak gravitational field (\ref{KerrN53a}) which is typical for the type $III$ solutions.

The difference from Kerr in the case of the Kerr-Newman metric ($q<q_C$) can already be seen from the Fig. (\ref{EXT2}). First, there is a large antigravity area characterized by closed ATTSs. Second, the region of stable photons is separated from the singularity and arises as small islands in APTTR (\ref{KerrN13b}) only for certain values of rotation ($|a|> a_C $)  (see  \ref{Kerr2b}, \ref{KerrN12b} and \ref{Kerr3b},\ref{KerrN13b}). In the extreme case,  closed TTSs completely disappear again,   but the stable photon region is separated from the singularity by the ATTR (\ref{KerrN14a}). However, as in the case of Kerr, the solution is unstable in an overextreme mode and falls into the type $II$.

In contrast to this, when $q>q_C$, i.e. even in the overextreme regime, the TTSs still exist provided $|a|<a_C$ (\ref{EXT3}). At the same time, the stable photon regions completely disappear and reappear only in the superextreme space, which indicates on their instability in the same way, but now this region is hidden by  closed TTSs (\ref{KerrN24a}, \ref{KerrN25a}) right up to the rotation value $a_C$, which may indicate some improvement in the  observability properties as being type $I$. From the observational point of view, this solution mimics a subextreme one.

With an extreme value of the charge $q_e$, the horizon disappears at any value of rotation, at the same time a stable photon region appears (\ref{KerrN34a}) and it is still hidden by the TTS area right up to the rotation $a_C$.

In the overextreme case $q>q_e$, there are both photon and antiphoton spheres that limit the TTR (\ref{ShN41a}). Rotation leads to smearing of these surfaces into the corresponding stable and unstable photon regions separated by (P)TTS (\ref{KerrN44a}). Full classification is presented in the  Table \ref{Ch}  and Figs. (\ref{Kerr1}-\ref{Kerr3}). We will provide more motivation for the above classification in the next section.

\begin{table}[h!!]
\caption{Different optical types  of Kerr-Newman }
\begin{center}
\begin{tabular}{|c|c|c|c|c|c|c|c|}
\hline
$q$ & $a$ & Horizon & TTS & PTTS & UPR &  SPR & Type \\
\hline
$q<q_C$ &  $a=0$ & $+$ & $+$ & $-$ & $+$ & $-$ & $I$\\
 &  $0<|a|<a_C$ & $+$ & $+$ & $+$ & $+$ & $-$ & $I$\\
&  $a_C<|a|<a_e$ & $+$ & $+$ & $+$ & $+$ & $+$ & $I$\\
 &  $a_e<|a|<a_c$ & $-$ & $-$ & $+$ & $+$ & $+$ & $II$\\
 &  $|a|>a_c$ & $-$ & $-$ & $-$ & $+$ & $+$ & $III$\\
\hline
$q_C<q<q_e$ &  $a=0$ & $+$ & $+$ & $-$ & $+$ & $-$ & $I$\\
 &  $0<|a|\leq a_e$ & $+$ & $+$ & $+$ & $+$ & $-$ & $I$\\
 &  $a_e<|a|\leq a_C$ & $-$ & $+$ & $+$ & $+$ & $+$ & $I$\\
 &  $a_C<|a|<a_c$ & $-$ & $-$ & $+$ & $+$ & $+$ & $II$\\
 &  $|a|>a_c$ & $-$ & $-$ & $-$ & $+$ & $+$ & $III$\\
\hline
$q_e<q\leq q_m$ &  $a=0$ & $-$ & $+$ & $-$ & $+$ & $-$ & $I$\\
 &  $0<|a|\leq a_C$ & $-$ & $+$ & $+$ & $+$ & $+$ & $I$\\
 &  $a_C<|a|<a_c$ & $-$ & $-$ & $+$ & $+$ & $+$ & $II$\\
 &  $|a|>a_c$ & $-$ & $-$ & $-$ & $+$ & $+$ & $III$\\
\hline
$q> q_m$ &  & $-$ & $-$ & $-$ & $-$ & $-$ & $IV$\\
\hline
\end{tabular}
\end{center}
\label{Ch}
\end{table}

\setcounter{equation}{0}

\section{Photon escape}

To clarify physical meaning of the  geometric structures introduced above, we consider  escape  to infinity of the photons emitted from  different types of characteristic surfaces  in the Kerr-Newman spacetime, generalizing the work of Synge \cite{Synge}. A detailed analysis of  null geodesics behavior in the Kerr-Newman metric as well as its shadow can be found, e.g., in \cite{A.deVries}. The goal here is to relate the escape properties to our set of characteristic surfaces including (P)TTS. We will restrict   by the outer region $r>r_{h+}$ ($\Delta_r>0$) of the (sub)extreme Kerr-Newman metric. 

Let's start by listing the equations of motion in an explicit form \cite{Grenzebach}:
\begin{align}
&\Sigma\dot{\phi}=\frac{L-Ea\sin^2\theta}{\sin^2\theta}+\frac{a((r^2+a^2)E-aL)}{\Delta_r}, \label{f1a} \\
&\Sigma\dot{t}=a(L-Ea\sin^2\theta)+\frac{(r^2+a^2)((r^2+a^2)E-aL)}{\Delta_r}\label{f1b} \\
&(\Sigma\dot{r})^2=((r^2+a^2)E-aL)^2-K\Delta_r\equiv R(r), \label{f1e} \\
&(\Sigma\dot{\theta})^2=K-\left(\frac{L-Ea\sin^2\theta}{\sin\theta}\right)^2\equiv \Theta(\theta), \label{f1d}
\end{align} 
where $K$, $L$ and $E$ are the Carter constant, the azimuthal orbital moment and the energy, respectively.

We want to find the conditions under which a photon emitted at a certain point $ r_o,\theta_o$ on a spherical surface with a certain angle of inclination $\psi$ relative to an outward normal vector goes to infinity. 
The angular parameter $\psi$ is defined following Synge \cite{Synge}:
\begin{align}
\cot^2\psi=\frac{1}{\Delta_r}\left(\frac{dr}{d\theta}\right)^2,
\label{f3a}
\end{align}
so that $0\leq\psi\leq\pi/2$, with $\psi=0$ for the normally emitted photons and $\psi=\pi/2$ for the tangentially emitted ones. Actually each orbit depends on the ratio $L/E=\rho$ which has a meaning of the azimuthal impact parameter. Assuming that the orbit starts at $r_o,\theta_o$, one can express $\psi$ via the Eqs.
(\ref{f1e}), (\ref{f1d}) and (\ref{f3a}) through $\rho, K, r_o, \theta_o$. Alternatively, the Carter constant $K$ can be expressed in terms on $\psi$, defining the function $K_o$:
\begin{equation}
K_o/E^2=(\rho-a \sin^2 \theta_o)^2\csc^2\theta_o\cos^2\psi+\frac{(a^2+r^2_o-a \rho)^2\sin^2\psi}{\Delta_r}.
\label{f3}
\end{equation}
Substituting this (and the impact parameter) into the radial potential divided by $E^2$, we obtain the function 
\begin{equation}
R_o(r)=E^{-2}R(r,K_o,\rho),
\label{f3b}
\end{equation}
which depends on the initial point on the sphere $r_o,\,\theta_o$ and the inclination angle $\psi$ as parameters. The specified initial conditions are suitable if and only if
\begin{align}
R_o(r_o)\geq0, \quad \Theta_o(\theta_o)\geq0.
\label{f3c}
\end{align}
Now we can formulate the escape condition as the absence of a turning point of the radial variable for all $r>r_o$, which means:  
\begin{align}
R_o(r)>0, \quad \forall r>r_o.
\label{f2}
\end{align}
Note that instead of $\psi$ we could use the second (polar) impact parameter as in \cite{A.deVries}.

One can find the following {\em sufficient} condition for photons escape in terms of the initial conditions:
\begin{align}
R'_o(r_o)>0, 
\label{f4}
\end{align}
where prime denotes a derivative and we keep (\ref{f3c}) in mind. 
Indeed, the right-hand side of the Eq. (\ref{f1e}) is the difference of two polynomials of the forth and the second degree. Both of these polynomials monotonously increase for values of $ r>m $ if additionally $ a \rho <a^2 + r^2 $.   Moreover, the polynomial of the fourth degree grows faster at large $r$. Accordingly, if initially $R'_o(r_o)>0$, the condition $R'_o(r)>0$ will be fulfilled for all $r>r_o$, which means that $R_o(r)$  grows monotonously. Since $R_o(r_o)\geq0$, this would mean that the condition (\ref{f2}) is satisfied.  

For tangentially emitted photons ($\psi=\pi/2$), this condition is also a necessary one. Indeed, in this  case the expression (\ref{f3a}) implies that $R_o(r_o)=0$ and the condition (\ref{f2}) is violated already in a small neighborhood of $r_o$ if $R_o(r_o)<0$.   

For the non-tangent photons $R_o(r_o)>0$, and we can find another sufficient condition. Indeed, suppose the condition (\ref{f4}) is violated. This means that the second-degree polynomial grows faster than the fourth-degree one, and $R_o(r)$ monotonously decreases. Then at some point $r'_o>r_o$ the growth rate will be the same. If at this point still $R_o(r'_o)>0$, there will be no turning point and the photon can escape to infinity. In this way we obtain the second sufficient condition:
\begin{align}
R'_o(r'_o)=0, \quad  R_o(r'_o)>0, \quad R_o(r_o)>0, \quad r'_o>r_o, 
\label{f5}
\end{align}
If the condition (\ref{f5}) is violated, the light cannot leave the compact area and will be trapped. Thus for the validity of the Eq. (\ref{f2}), it is necessary that at least one of the conditions (\ref{f4}, \ref{f5}) holds. In fact, the first sufficient conditions (\ref{f4}) are stronger and gives an overestimated escape angle.

Using the Eqs. (\ref{f1e}), (\ref{f3}) and (\ref{f3b}) it is easy to find an explicit form of the inequalities (\ref{f3c}) and (\ref{f4}):
\begin{align}
&(a^2+r^2_o-a\rho)^2-\Delta_r(\rho-a \sin^2\theta_o)^2\csc^2\theta_o\geq0, \label{f6b}\\
&4r_o(a^2+r^2_o-a \rho)\Delta_r-(a^2+r^2_o-a \rho)^2\Delta'_r\nonumber\\
&+\Delta'_r\left((a^2+r^2_o-a \rho)^2-\Delta_r(\rho-a \sin^2 \theta_o)^2\csc^2\theta_o\right)\cos^2\psi>0, \label{f6a}
\end{align} 
accurate to the non-negative multipliers $\cos^2 \psi$,  $\sin^2 \psi/\Delta_r$ and  $1/\Delta_r$ respectively. In what follows, we will omit the index $o$. The roots of the equality (\ref{f6b}) determine the allowable values of the impact parameter for a given  emission point. For example, if we are outside the ergosphere, the coefficient at $\rho^2$ will be negative and the expression (\ref{f6b}) is non-negative only for $\rho_{min}\leq\rho\leq\rho_{max}$, where:
\begin{align}
\rho_{max,min}=\frac{\sin\theta(a^2+r^2\pm a\sin\theta\sqrt{\Delta_r})}{a\sin\theta \pm \sqrt{\Delta_r}}.
\label{f7}
\end{align}
From the condition (\ref{f6b}) it is clear that the coefficient at $\cos^2\psi$ in (\ref{f6a}) is positive for  $r>m$ and therefore the escape area monotonously decreases with $\psi$  and is maximal for the normally emitted photons, as could be expected. For a photon emitted tangentially to the sphere $\psi=\pi/2$, the first condition will be independent of the angle $\theta$ and is simply given by  
\begin{align}
4r\Delta_r-(a^2+r^2-a \rho)\Delta'_r>0,
\label{f8}
\end{align} 
implying
\begin{equation}
r(r^2-3mr+2q^2)+a^2(r+m)+a(r-m)\rho>0.
\label{f9}
\end{equation} 
This differs from the expression (\ref{e6}) in one term proportional to $\rho$.

The simplest case is that of a photon emitted at the pole, in which case   $\rho=0$. It is easy to see that for  $\psi=0$ (normal photons) the escape region reaches the horizon, while for  $\psi=\pi/2$ (tangent photons) it extends to the boundary of the PTTR. 

To find the boundary of the escape region  one has to solve the corresponding equality with respect to $\rho$ which defines the boundary of the escape zone, i.e. the photon orbits of the constant radius \cite{A.deVries}:
\begin{equation}
\rho_c=\frac{(r^2+a^2)\Delta_r'-4r\Delta_r}{a\Delta_r'},
\label{f10}
\end{equation} 
 and substitute the solution into the Eq. (\ref{f6b}):
\begin{equation}
\left(16 r^2a^2\Delta_r\sin^2\theta-(4r\Delta_r-\Sigma\Delta_r')^2\right)/(a^2\Delta'^2_r\sin^2\theta)\geq0.
\label{f11}
\end{equation} 
Thus we get the condition for the photon region. Let us explain the meaning of the conditions we obtained. Set for definiteness $a>0$. Then the condition  (\ref{f8}) is satisfied for all $\rho>\rho_c$. Suppose that the point $P(r_o,\theta_o)$ is located inside the photon region. The condition (\ref{f11}) means   that $\rho_c$ is limited by $\rho_{min}<\rho_c<\rho_{max}$. Respectively, for all $\rho_c<\rho<\rho_{max}$, $P$ is the point of escape for tangent photons. Consequently, the tangent ray can leave the photon region, but only with sufficiently large values of the impact parameter. If the condition of the photon region is violated, there may be two situations. The first corresponds to the case
\begin{align}
4r\Delta_r-(a^2+r^2)\Delta'_r>0,
\label{f12}
\end{align} 
than $\rho_c<0$ and accordingly $\rho_c<\rho_{min}$. So the tangent geodesics with all possible values $\rho_{min}<\rho<\rho_{max}$ escape from the region. This happens in the the weak field domain. It is obvious that the condition (\ref{f12}) corresponds to our previously defined A(P)TTR. 

Another opportunity is
\begin{align}
4r\Delta_r-(a^2+r^2)\Delta'_r<0,
\label{f13}
\end{align} 
than $\rho_c>0$ and accordingly $\rho_c>\rho_{max}$. So tangent geodesics with all possible values $\rho_{min}<\rho<\rho_{max}$ can not escape from the region what corresponds to a strong field area, or (P)TTR. 

Thus, we established an explicit correspondence between the geometric structure of the (P)TTSs and behavior of geodesics with different value of the impact parameter. It is worth noting that in the case of metrics admitting separation of variable in the geodesic equations, there is no big difference between the properties of TTSs and PTTSs for a distant observer. Of course, the tangent light can leave the PTTR across its boundary. However, such geodesics cannot  reach the spatial infinity. Nevertheless, photons escape is possible for other emission angles $\psi$. 

\begin{figure}[tb]
\centering
\subfloat[][$\rho=2m$]{
  	\includegraphics[scale=0.3]{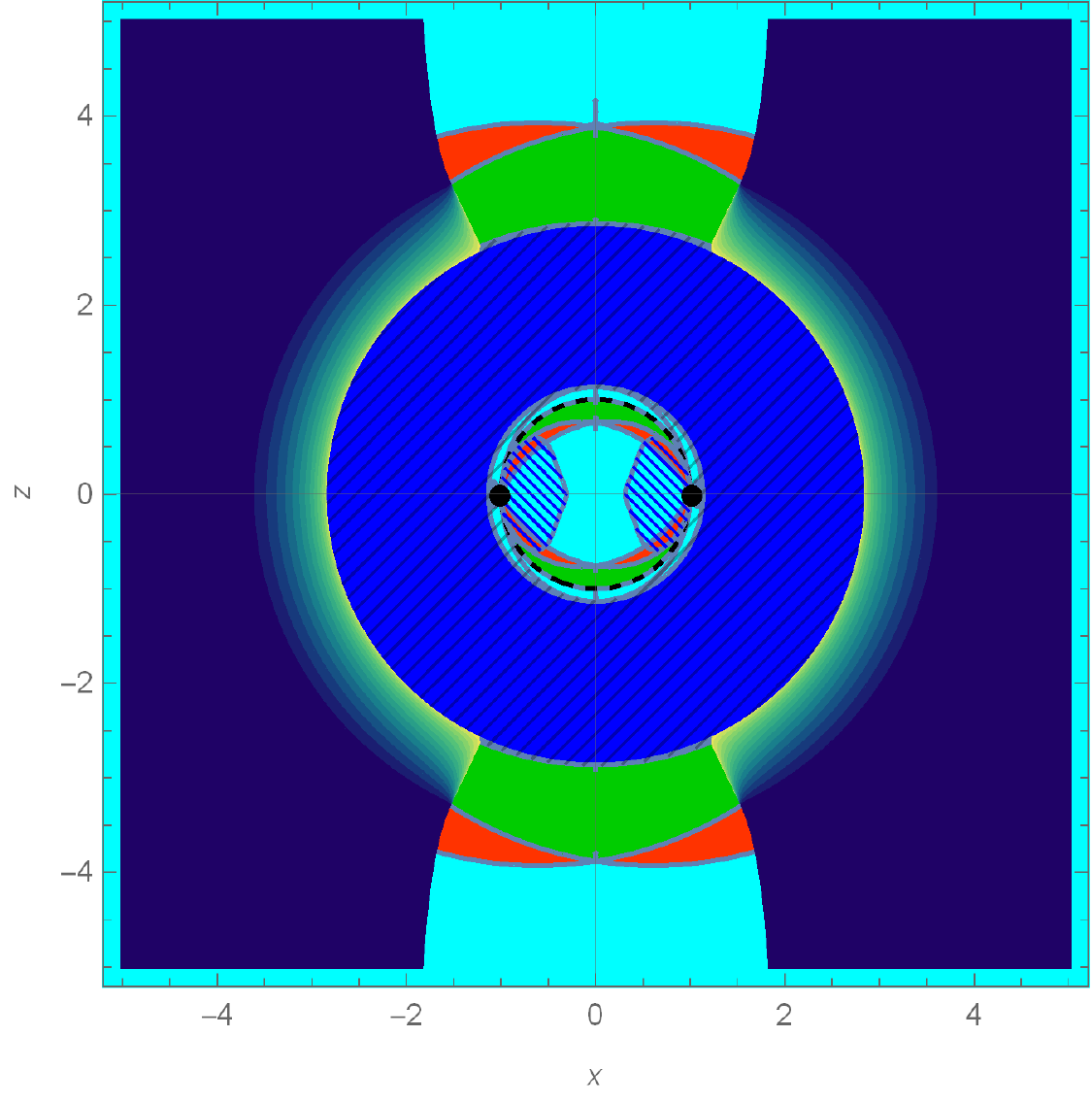}
		\label{ESC4}	
 }
 \quad
\subfloat[][$\rho=0$]{
	\includegraphics[scale=0.3]{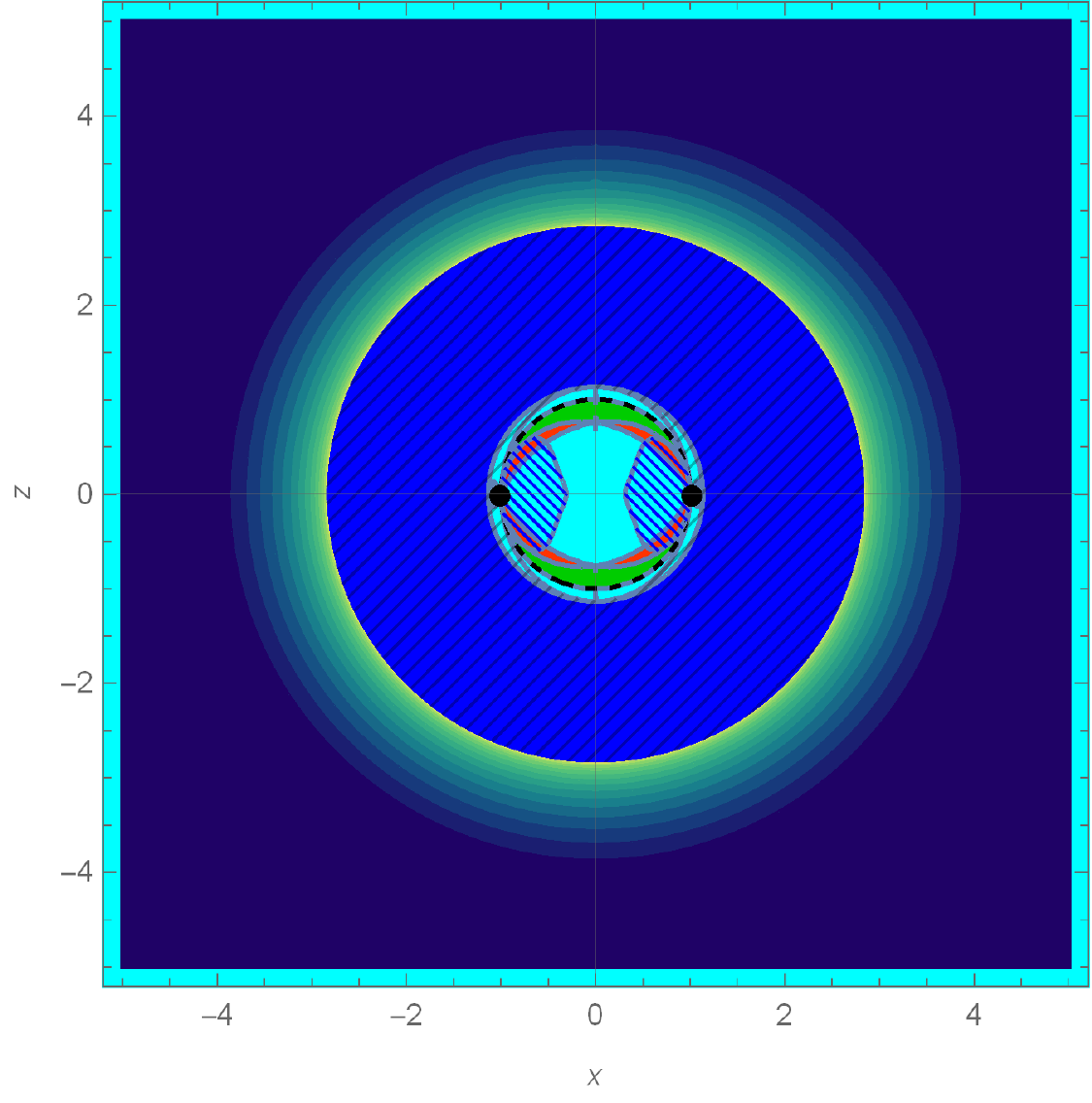}
		\label{ESC0}	
 }
 \quad
 \subfloat[][$\rho=-2m$]{
		\includegraphics[scale=0.3]{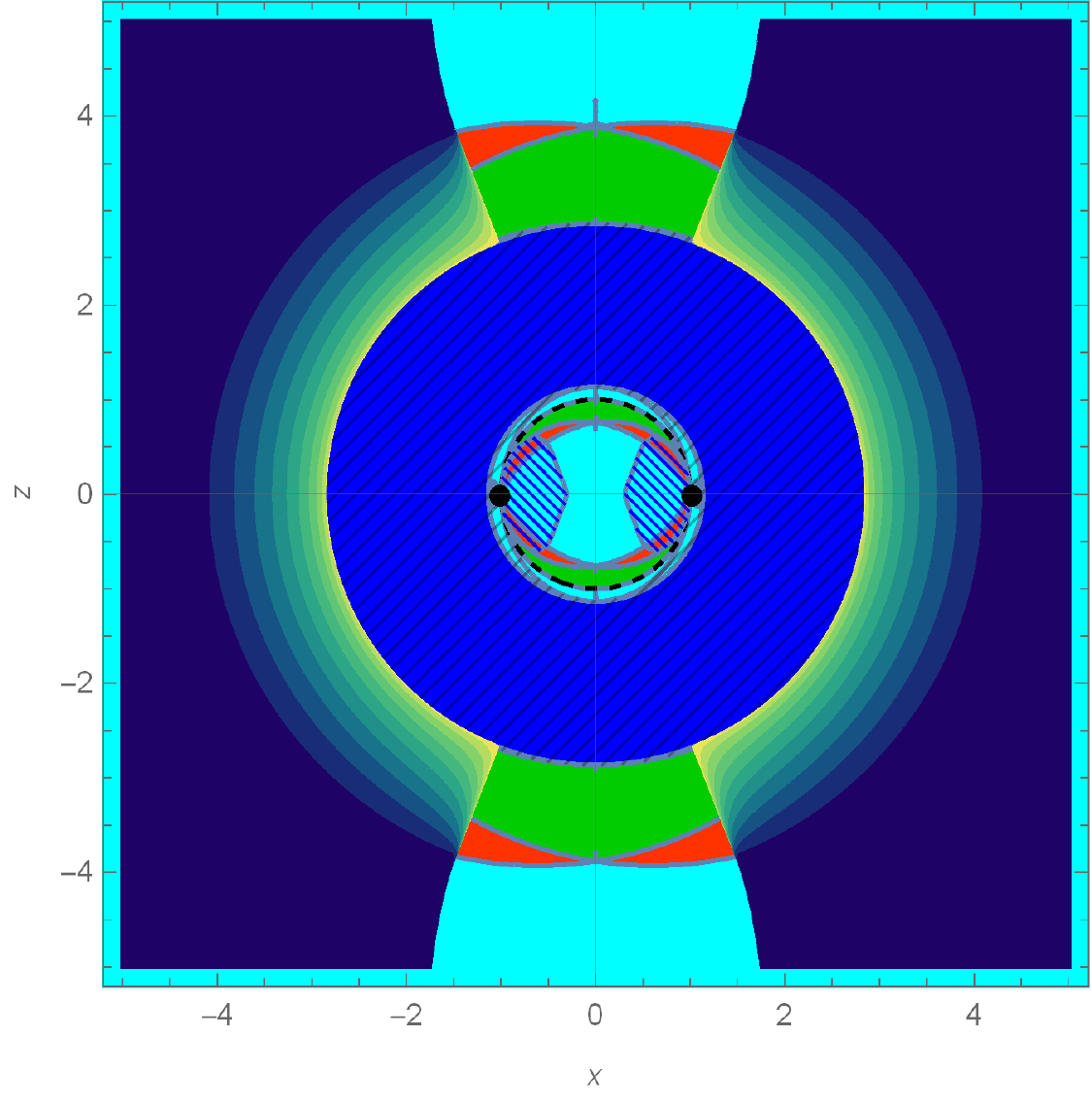}
		\label{ESC3}	
 }
 \\
 \subfloat[][$\rho=4m$]{
  			\includegraphics[scale=0.3]{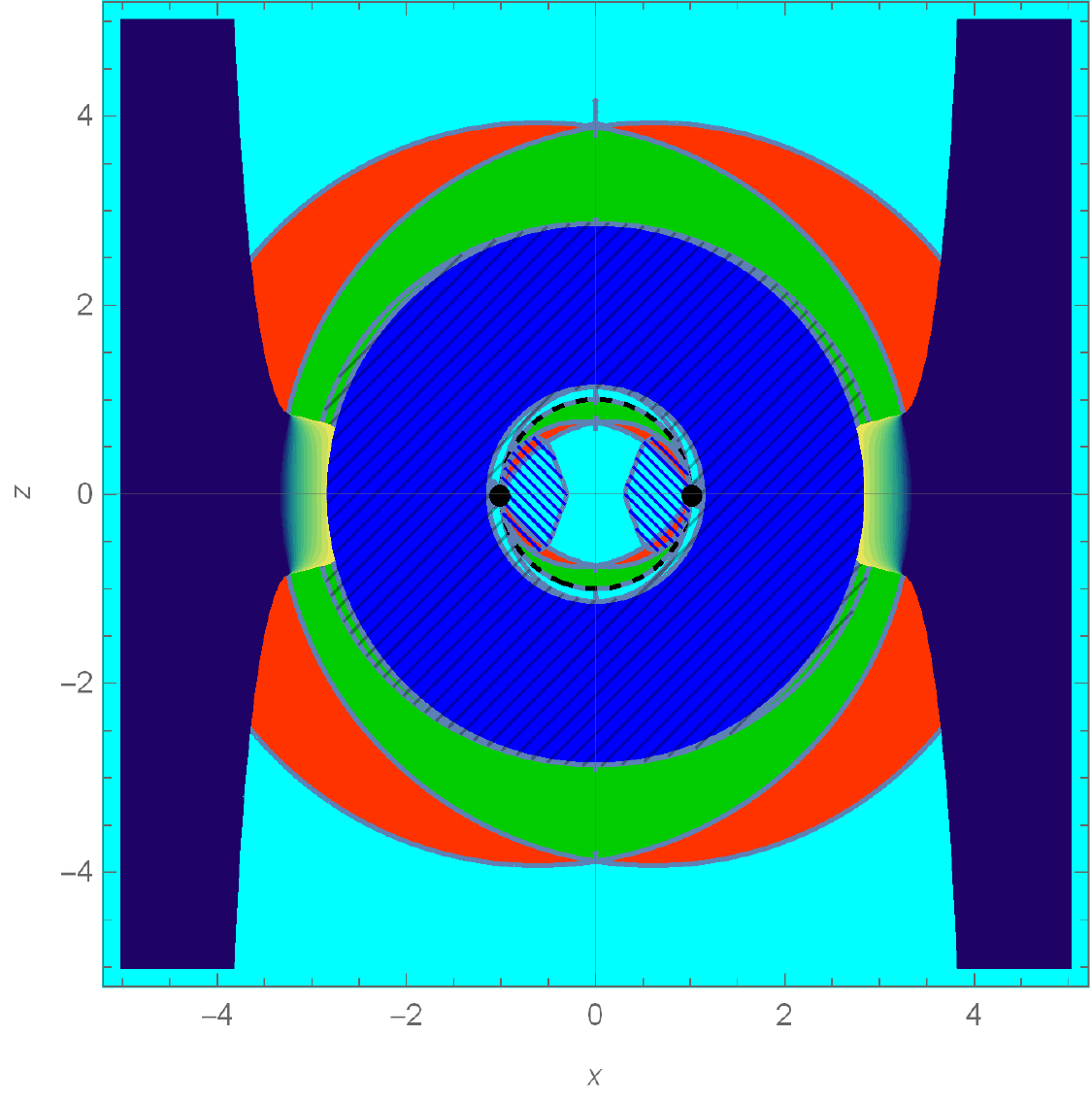}
		\label{ESC2}	
 }
 \quad
\subfloat[][$\rho=-4m$]{
	\includegraphics[scale=0.3]{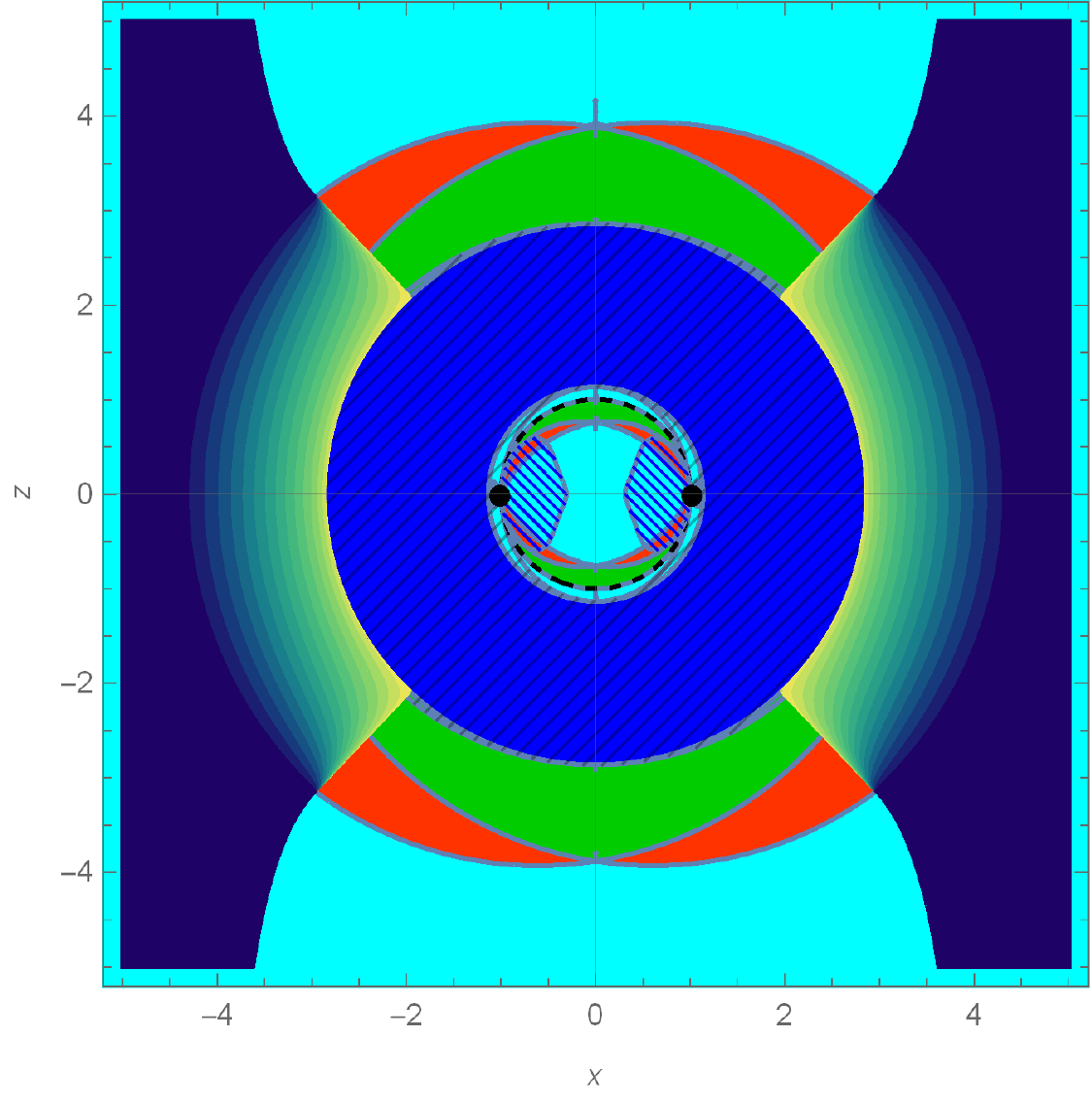}
		\label{ESC1}
 }
 \\
 \subfloat[][$a=0.5a_e$, $\rho\geq0$]{
  	\includegraphics[scale=0.3]{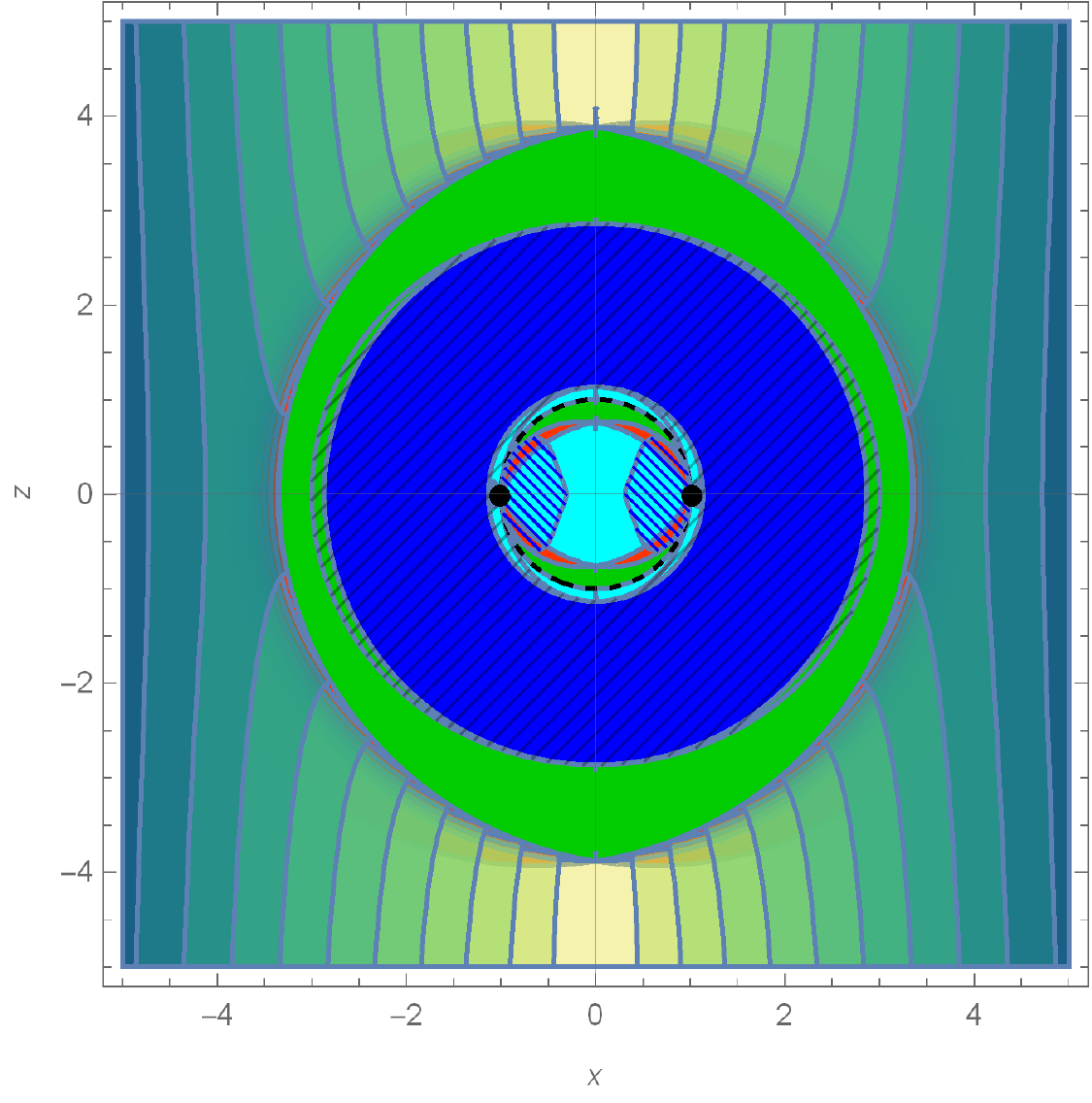}
		\label{ESC0a}
 }
 \quad
\subfloat[][$a=0.5a_e$, $\rho\leq0$]{
		\includegraphics[scale=0.3]{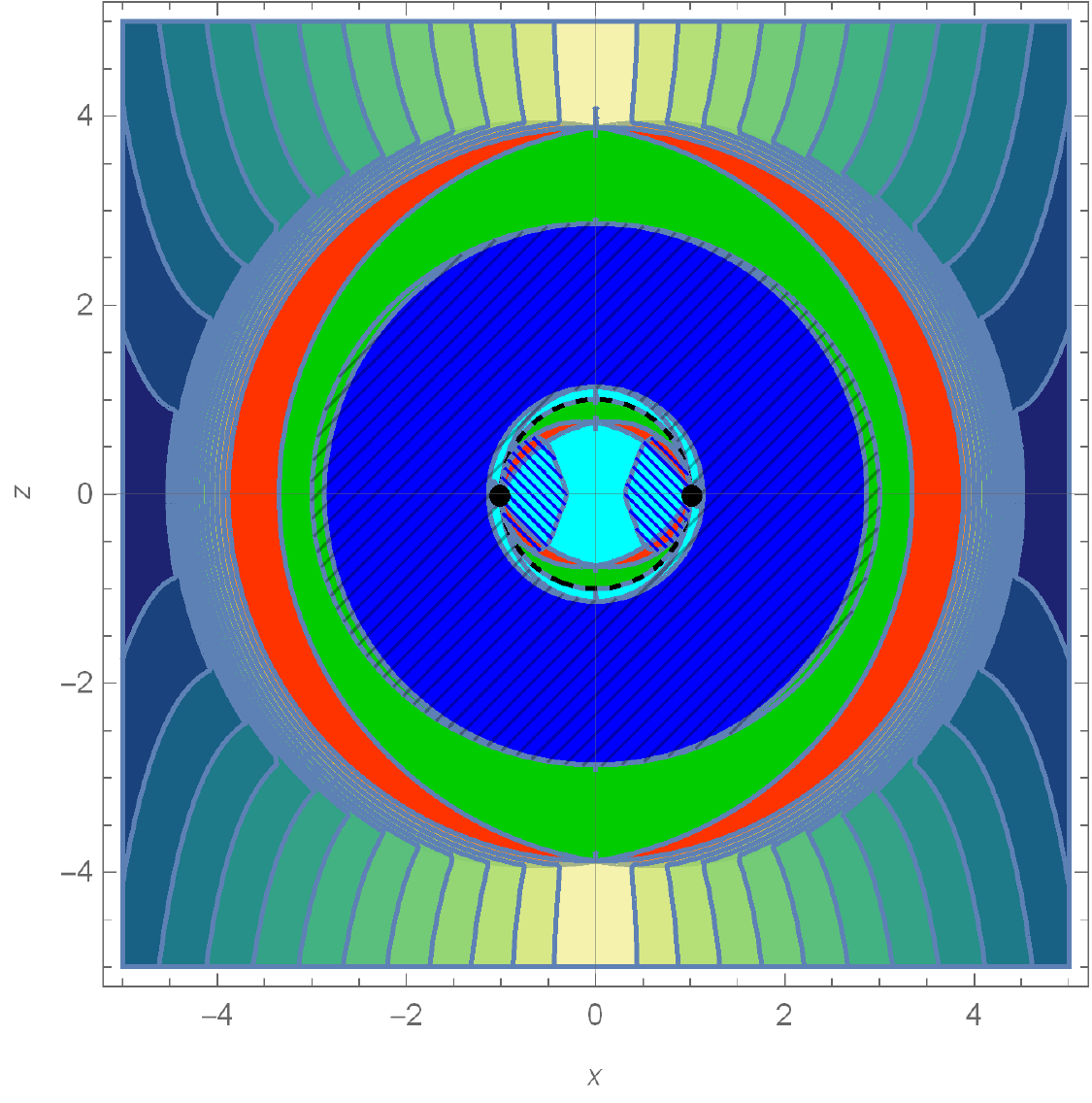}
		\label{ESC0b}
 }
  \\
 \subfloat[][$a=a_e$, $\rho\geq0$]{
  			\includegraphics[scale=0.3]{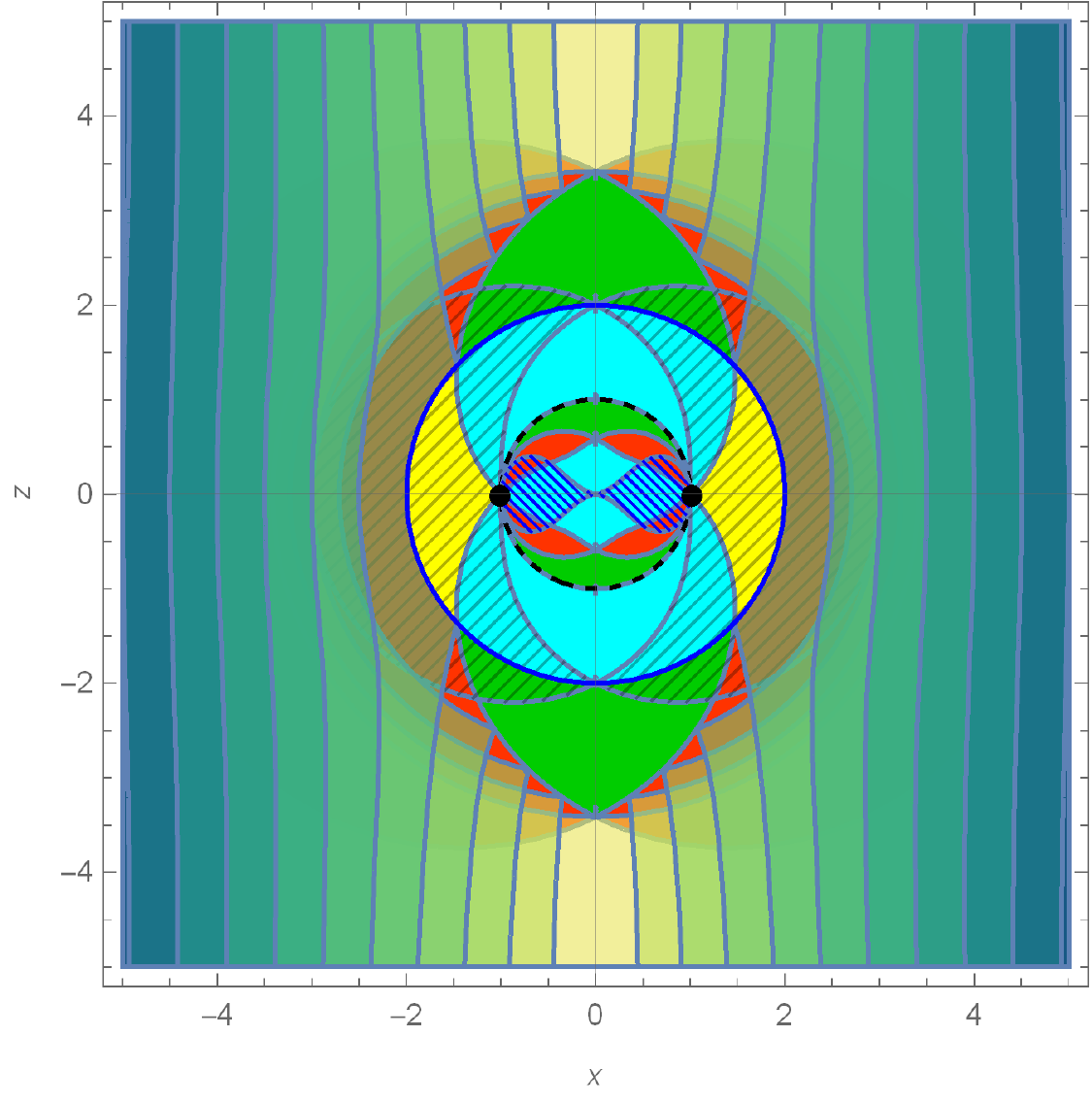}
		\label{ESC0c}	
 }
 \quad
\subfloat[][$a=a_e$, $\rho\leq0$]{
		\includegraphics[scale=0.3]{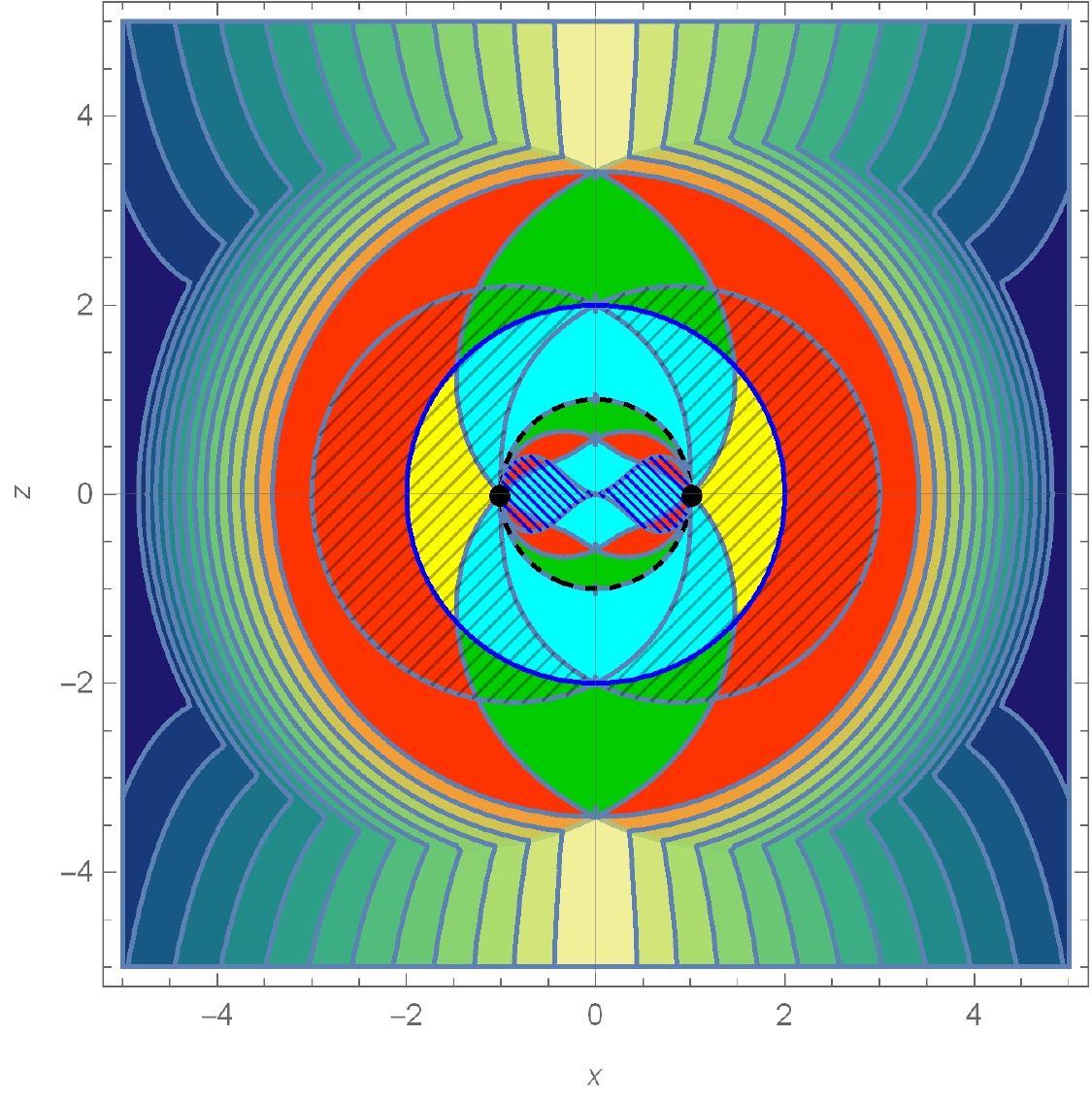}
		\label{ESC0d}	
 }
\caption{Density graphs for the escape angle and the impact parameter. For (\ref{ESC4}-\ref{ESC1}) dark blue corresponds to the escape angle $\psi=\pi/2$ while yellow corresponds to the escape angle $\psi=0$. For (\ref{ESC0a}-\ref{ESC0d}) dark blue corresponds to the $\rho=\pm6m$ while yellow corresponds to the $\rho=0$.  The progressive colours cover the intermediate angles or impact parameters.}
\label{Kerr4}
\end{figure}

From (\ref{f4}) it is easy to get an overestimated escape angle $\psi$:
\begin{align}
\cos^2\psi=\frac{\left(a^2+r^2-a\rho\right)\left([a^2+r^2-a \rho]\Delta_r'-4r\Delta_r\right)}{\Delta_r'\left((a^2+r^2-a \rho)^2-\Delta_r(\rho-a \sin^2 \theta)^2\csc^2\theta\right)}.
\label{f14}
\end{align} 
The exact value of the escape angle can be obtained applying two conditions (\ref{f4}, \ref{f5}) simultaneously. The Fig. (\ref{ESC4}-\ref{ESC1}) represents density graphs for the escape angle for given $\rho$ superimposed on previously obtained images (\ref{Kerr1}) with $a=0.5a_e$  related to the type I. Dark blue color corresponds to the escape angle $\psi=\pi/2$, and  yellow corresponds to the escape angle $\psi=0$. The progressive colours cover the intermediate angles.  As expected, in general, the (P)TTS region becomes permeable for some emission angles that tend to normal as we approach the horizon.

The other  Figs. (\ref{ESC0a}-\ref{ESC0d}) represent discrete density graphs of escape impact parameter ($\rho=0,\pm0.5m,...$) with a fixed angle of emission $\psi=\pi/2$. Dark blue color corresponds to the $\rho=\pm6m$, and yellow corresponds to the $\rho=0$. From the figures it can be derived that the (P)TTR is completely inaccessible for tangent geodesics going from infinity. So the (P)TTSs themselves cut half of all possible emission angles for arbitrary values of the impact parameter, and in this sense are darkest, while the photon region is lighter. The outer region corresponds to the escape of all possible rays as it could be expected \cite{Synge}. The difference between the types I and II is visible from the Figs. (\ref{ESC0a}, \ref{ESC0c}): in the latter case, even the tangent geodesics can leave the vicinity of the horizon.  More detailed analysis of the  properties of null geodesic as well as their relationship to the shadow can be found, e.g., in \cite{Grenzebach,A.deVries}. 

\section{Summary and discussion}
In this paper, we argued that one can reasonably extend the notion of transversely trapped photon surfaces (TTS) in stationary spacetimes originally intended for closed two-sections to the non-closed case. The TTS is defined as a surface such that an initially tangent photon either remains in it forever or leaves it in the inward direction.
In Boyer-Lindquist coordinates, such surfaces have a constant radial coordinate, but their closure is not guaranteed due to the presence of an additional restriction on the polar angle. As a consequence, the conditions on TTSs  do  not necessarily hold for all $\theta$. Such TTSs therefore are not enough to ensure  filling (together with the photon region) of the full space manifold in Kerr-like spacetimes.  For this purpose one has also to consider non-closed TTSs, whose two-sections cover only a spherical cap. The initially tangent photons in principle could leave the inner region  across the boundary of the cap, but, as we have seen, such photons do not escape to spatial infinity in Kerr and Kerr-Newman metrics.

We also found that it is useful to introduce the anti-trapping surfaces (ATTS), so that initially tangent photons can leave them on the outside. For them, the necessary conditions are also $\theta$-dependent, so this concept should be extended to the non-closed surfaces (APTTS).

The (P)TTSs and A(P)TTSs foliate the corresponding volume regions (P)TTR and A(P)TTR, which together with the photon regions (PR) cover the
entire space manifold in Kerr-like spacetimes.
This was illustrated explicitly in the cases of  Kerr and Kerr-Newman metrics.  
Although the separability of geodesic equations in these cases makes it possible to explicitly describe all photon orbits, therefore our characteristic of the null geodesic structure may seem excessive, but this does not apply to metrics (for example, the Weyl class) for which the geodetic equations are non-separable.
Our method does not appeal to solving the geodesic equations while it still  remains a purely analytic tool.
 
Using the Kerr-Newman solution as an example, we have demonstrated how this description can be  applied to classify optical properties of strong gravitational fields. 
The critical values of the charge parameter $q^2_C=3/4m^2$, $q^2_m=9/8m^2$  and the rotation parameter $a_C, a_c$, marking the qualitative changes in the optical structure, are found. In particular, it was shown that for $q>q_C$, the singularity  in an overextreme regime can be weakly naked in the sense of the existence of a closed TTS, and it is strongly naked when $q>q_m$, according to classification introduced in \cite{Virbhadra:2002ju}.
 
To clarify  physical significance of different regions, we examined their escape properties, extending the work of Synge \cite{Synge, A.deVries}. In this description, the rays starting at  certain angles  to a normal to a chosen characteristic surface are traced to infinity.
The escape angle values are visualized associating to them different colors on the density graphs. We found that, as expected, the TTSs capture all tangential geodesics for any values of the impact parameter, while the photon regions absorb only a fraction of such geodesics. The (P)TTSs are still permeable to some non-tangential ray starting angles. However, now the angle is different for different values of the impact parameter.
 
In the case of the Kerr-Newman metric, the system admits separation of variables, and the difference between  properties of TTS and PTTS is reduced.  But the situation may be different in the case of non-separable spaces and  non-integrable geodesic equations \cite{Lukes}. 
It should be noted that in such cases the geometric structure of the transversely trapping  surfaces is preserved, while other characteristics like the photon region are not applicable, which can be  seen, for example, for Zipoy-Voorhees or Tomimatsu-Sato \cite{Kodama:2003ch} solutions. But even such spaces can be   still  classified with the help of (P)TTS.    
It turns out that in these cases the differences between the TTSs and the PTTSs  become significant. A ray that starts  at a certain angle in the region bounded by  PTTS may leave this area at a different angle. In any case, TTSs will retain their main feature - to characterize the compact trapping region of a strong gravitational field, while the PTTSs   characterize  the  strong lensing \cite{Perlick:2004tq,Bozza:2002zj,Cunha:2018acu}.
 
\section*{Acknowledgement}
We thank G\'erard Cl\'ement for careful reading of the manuscript and valuable comments. The work was supported by the Russian Foundation for Basic Research on the project 17-02-01299a, and by the Government program of competitive growth of the Kazan Federal University. 

\section*{APPENDIX A}

Here we give derivation of the conditions listed in Table \ref{Con} from the inequalities (\ref{b3c}, \ref{b3d}). First, the roots of a polynomial
\begin{align} 
P(\xi)=\left(b-c\right)\xi^2-2d\xi+\left(a-b\right)=0
\end{align} 
read
\begin{align} 
\xi_\pm=\frac{d\pm\sqrt{d^2-(a-b)(b-c)}}{b-c}.
\label{AP2}
\end{align} 
Let's start with the PTTS conditions $P(\xi)\geq0$, $\xi^2\leq1$. The simplest opportunity is
\begin{align} 
b>c, \quad d^2\leq(a-b)(b-c),
\end{align}  
meaning that $P(\xi)$ is non-negative everywhere and corresponds to the first condition in Table \ref{Con}. Another possibility is that the both roots (\ref{AP2})  are located to the left or to the right of the interval $\xi^2\leq1$.
\begin{align} 
b>c, \quad \xi_-\geq1 \quad or  \quad \xi_+\leq-1.
\end{align}  
This condition reduces to the following
\begin{align} 
b>c, \quad a-c\geq 2 |d|, \quad |d|>(b-c).
\end{align}  
Squaring this we get the  second conditions in   Table \ref{Con}. The last opportunities are: 
\begin{align} 
b<c \quad or \quad b=c,  \quad P(\pm 1)\geq0, 
\end{align}  
which means that the roots  (\ref{AP2})  of the polynomial $P(\xi)$ are located on the opposite sides of the interval $\xi^2\leq1$ or the line segment $P(\xi)|_{b=c}$ is above the axis $\xi$ respectively. Explicitly, it is the third condition in   Table \ref{Con}.    

Inside the ergoregion, the  PTTS conditions become $P(\xi)\leq0$, $\xi^2\geq1$. The simplest opportunity is again
\begin{align} 
b<c, \quad d^2\leq(a-b)(b-c).
\end{align}  
Another nontrivial possibility is that the both roots are enclosed in the interval $\xi^2\leq1$:
\begin{align} 
b<c, \quad -1 \leq \xi_\pm\leq 1.
\end{align}
This condition reduces to the following
\begin{align} 
c>b, \quad a-c\leq -2 |d|, \quad |d|<(b-c).
\end{align} 
Squaring this, we get the required second conditions in   Table \ref{Con} for the ergoregion. There are no other possibilities. For A(P)TTS the  calculations are  similar. Note that we impose additionally that $\alpha$ and $\gamma$  have the same sign and  $\beta>0$ always. This is true for the metrics we used, but may be wrong in more general cases which has to be considered independently.

\begin{table}[h]
\caption{Parameters table}
\begin{center}
\begin{tabular}{|c|c|c|c|c|c|c|c|}
\hline  
($a,q$) & 0  &  $0.5q_e$ &  $0.9q_e$ & $9.5q_e$ & $q_C$ & $q_e$ & $\sqrt{11/10}q_e$ \\
\hline
0 &  (\ref{Sha}) &  (\ref{ShN11a}) & (\ref{SHN21a})& (\ref{ShN51a}) & (\ref{ShN53a}) & (\ref{ShN31a}) &  (\ref{ShN41a})  \\
$0.7a_e$ &  (\ref{Kerr2a}) & (\ref{KerrN12a}) & (\ref{KerrN22a})& - & - & - & - \\
$0.9a_e$ &  (\ref{Kerr3a}) & (\ref{KerrN13a})  & (\ref{KerrN23a}) & - & - & - & - \\
$a_e$ & (\ref{Kerr4a})  &  (\ref{KerrN14a}) & (\ref{KerrN24a})& - & - & - & -\\
$a_C$ & - &  - & -& (\ref{KerrN25a}) &  (\ref{KerrN52a}) & (\ref{KerrN34a}) & (\ref{KerrN44a}) \\
$a_c$ & - &  - & -& (\ref{KerrN26a}) & (\ref{KerrN53a}) & (\ref{KerrN35a}) & (\ref{KerrN45a})  \\
\hline
\end{tabular}
\end{center}
\label{Ch}
\end{table}

\end{document}